\documentclass[12pt]{article}
\usepackage{ragged2e}
\usepackage[colorlinks=true,urlcolor=blue,anchorcolor=blue,citecolor=blue,filecolor=blue,linkcolor=blue,menucolor=blue,linktocpage=true]{hyperref}
\usepackage{makeidx}
\usepackage{amsmath}
\usepackage{amsfonts}
\usepackage{amssymb}
\usepackage{graphicx}
\usepackage{cite}
\usepackage{xcolor}
\usepackage{amsthm}

\makeatletter
\g@addto@macro\th@plain{\thm@headpunct{:}}
\makeatother
\theoremstyle{plain}
\newtheorem*{acknowledgement*}{Acknowledgement}
\input epsf.sty
\makeatletter \@addtoreset{equation}{section}

\input{tcilatex}
\begin{document}

\title{\textbf{Partial Breaking in Rigid Limit of }$\mathcal{N}=2$\textbf{\
Gauged Supergravity }}
\author{R. Ahl Laamara$^{1,2,3}$, E.H Saidi$^{1,3}$, M. Vall$^{1,3}$ \\
%EndAName
{\small 1. LPHE-MS, Science Faculty}, {\small Mohammed V University, Rabat,
Morocco}\\
{\small 2. Centre R\'{e}gional des Metiers de L'Education et de La
Formation, F\`{e}s-Mekn\`{e}s, Morocco}\\
{\small 3. Centre of Physics and Mathematics, CPM- Morocco}}
\maketitle

\begin{abstract}
Using a new manner to rescale fields in $\mathcal{N}=2$ gauged supergravity
with n$_{V}$ vector multiplets and n$_{H}$ hypermultiplets, we develop the
explicit derivation of the rigid limit of quaternionic isometry Ward
identities agreeing with known results. We show that the rigid limit can be
achieved, amongst others, by performing two successive transformations on
the covariantly holomorphic sections $V^{M}\left( z,\bar{z}\right) $ of the
special Kahler manifold: a particular symplectic change followed by a
particular Kahler transformation. We also give a geometric interpretation of
the $\eta _{i}$ parameters used in \href{http://arxiv.org/abs/1508.01474}{{\
arXiv:1508.01474}} to deal with the expansion of the holomorphic
prepotential $\mathcal{F}\left( z\right) $ of the $\mathcal{N}=2$ theory. We
give as well a D- brane realisation of gauged quaternionic isometries and an
interpretation of the embedding tensor $\vartheta _{M}^{u}$ in terms of type
IIA/IIB mirror symmetry. Moreover, we construct explicit metrics for a new
family of $4r$- dimensional quaternionic manifolds $\boldsymbol{M}%
_{QK}^{\left( n_{H}\right) }$ classified by ADE Lie algebras generalising
the $SO\left( 1,4\right) /SO\left( 4\right) $ geometry which corresponds to $%
A_{1}\sim su\left( 2\right) $. The conditions of the partial breaking of $%
\mathcal{N}=2$ supersymmetry in the rigid limit are also derived for both
the observable and the hidden sectors. Other features are also studied.
\end{abstract}

%\tableofcontents

%\newpage

%\keywords{$\mathcal{N}=2$ gauged supergravity, gauging isometries, embedding tensor and mirror symmetry, rigid limit, partial supersymmetry breaking.}

\section{Introduction}

It has been shown since a long time that the no-go theorem of \textrm{\cite%
{Witten:1981nf, Cecotti:1984rk,Cecotti:1985mx}}, which forbids the partial
breaking of extended supersymmetries, can be overcome by turning on
appropriate fluxes \textrm{\cite%
{Hughes:1986dn,Hughes:1986fa,Antoniadis:1995vb,Antoniadis:1996ki,Partouche:1996yp,Ivanov:1997mt,Taylor:1999ii,Fujiwara:2006ze,Itoyama:2007kk,Antoniadis:2012cg,Laamara:2015rbq}%
}, or by using non linear realisations of extended supersymmetry \textrm{%
\cite{Bagger:1996wp,Bagger:1997pi,Rocek:1997hi}}, or also by taking rigid
limits of extended gauged supergravities \textrm{\cite%
{Ferrara:1995gu,Porrati:1996xu}}; see also \textrm{\cite%
{Ferrara:1995xi,Fre:1996js,Fre:1996xw,Louis:2002vy,Itoyama:2006ef,Louis:2009xd,Louis:2010ui,Hansen:2013dda,Fujiwara:2005kf,Kuzenko:2015rfx}
} for the local case. For the 4d $\mathcal{N}=2$ extended supersymmetric
theories in rigid limit, the study partial breaking involves two breaking
scales $\Lambda _{susy}$ and $\Lambda _{susy}^{\prime }$, one for each
supersymmetry ($\Lambda _{susy}<\Lambda _{susy}^{\prime })$,\ and leads to
interesting phenomenological implications as well as formal ones. If the two 
$\Lambda _{susy}$'s are widely separated as proposed in{\normalsize \ 
\textrm{\cite{Antoniadis:2012cg}}}, the fermion chirality needed by
phenomenology is recovered and we are with a new window for the building of
quasi-realistic particle models and for the study of effective low energy
stringy inspired prototypes. Also, the infrared of partial breaking in the
rigid limit of $\mathcal{N}=2$ supergravity gives a way to think about the
supergravitational origin of the BI theory{\normalsize \  \textrm{\cite%
{Andrianopoli:2015rpa,Andrianopoli:2014mia}}}. Recall as well that in%
{\normalsize \ }$\mathcal{N}=2${\normalsize \ }partial breaking, the no-go
theorem is understood as a problem which concerns only the $\mathcal{N}=2$
superalgebra of supersymmetric charges $Q_{\alpha }^{A},$ $\bar{Q}_{A}^{\dot{%
\alpha}}$, and can be avoided by using the $\mathcal{N}=2$ supersymmetric
current algebra generated by local supercurrents $\mathcal{J}_{\mu \alpha
}^{A},$ $\mathcal{\bar{J}}_{\mu B}^{\dot{\alpha}}$ which allows, in addition
to a diagonal energy momentum tensor $\delta _{B}^{A}\mathcal{T}_{\mu \nu }$%
, the introduction of a non- diagonal deformation $C_{B}^{A}$ of the scalar
potential. This deformation can be different from zero if one introduces
magnetic Fayet-Iliopoulos (FI) terms besides the electric ones with a
particular $SU(2)_{R}$ rotation of each of them. In the local theory, the
partial breaking of $\mathcal{N}=2$ of supersymmetry is due to the negative
contribution of the two gravitini to the scalar potential $\mathcal{V}_{%
\text{sugra}}$ which allows to bypass naturally the no-go theorem. As first
shown in \textrm{\cite{Ferrara:1995xi},} the global supersymmetry can be
derived as a rigid limit of the electrically gauged supergravity for which
the avoiding of the no-go theorem restricts the rigid limit to be performed
in a particular symplectic frame where the holomorphic prepotential $%
\mathcal{F}$ of the special Kahler manifold $\boldsymbol{M}_{SK}$ vanishes.
In this frame, the non-diagonal deformation in the rigid theory corresponds
to the contribution of the hidden sector of the $\mathcal{N}=2$ gauged
supergravity to the induced scalar potential $\mathcal{V}_{\text{sugra}}$
given by the underlying Ward identities \textrm{\cite%
{Castellani:1985ka,Cecotti:1984wn}}. \  \  \  \  \newline
Recently, a new approach has been done in \textrm{\cite{Andrianopoli:2015wqa}%
} to deal with the rigid Ward identities of $\mathcal{N}=2$ global
supersymmetry. In this study, FI couplings are implemented into an SP$\left(
2n_{V}\right) $ symplectic system of $SU(2)_{R}$ isotriplets $\left( 
\mathcal{P}^{a}\right) ^{M}=\mathcal{P}^{aM}$, interpreted in terms of
moment maps living on the hypermultiplet manifold, and partial breaking of
supersymmetry is induced by the symplectic invariant isovector $\zeta
_{a}\sim \varepsilon _{abc}\mathcal{P}^{Mb}\mathcal{C}_{MN}\mathcal{P}^{Nc}$
pointing into a particular direction of the $SU(2)_{R}$ isospin space. The
supergravity origin of this generalized partially broken $\mathcal{N}=2$
global supersymmetric model has been derived in \textrm{\cite%
{Andrianopoli:2015rpa} where }$\zeta _{a}$ is interpreted as an anomaly of
the $\mathcal{N}=2$ supercurrent algebra in 4- dimensions coming from the
hidden sector of the parent $\mathcal{N}=2$ gauged supergravity. The authors
of \textrm{\cite{Andrianopoli:2015rpa}} showed also that the restriction to
the symplectic frame in which the prepotential of the special Kahler
manifold vanishes, can be overcome by generalizing the electric gauging of 
\textrm{\cite{Ferrara:1995xi}} to a dyonic gauging nicely encoded in the
embedding tensor $\vartheta _{M}^{u}=(\theta _{\Lambda }^{u},\tilde{\theta}%
^{u\Lambda })$ of \textrm{\cite{deWit:2005ub}} with electric $\theta
_{\Lambda }^{u}$ and magnetic $\tilde{\theta}^{u\Lambda }$ components\textrm{%
\ }which can be interpreted as fluxes from higher energy theories \textrm{%
\cite%
{deWit:2003hq,Angelantonj:2003rq,D'Auria:2003jk,Samtleben:2008pe,Trigiante:2016mnt}%
}.

\  \  \  \  \newline
In this paper, we contribute to the study of partial breaking of $\mathcal{N}%
=2$ supersymmetry along the approach of \textrm{\cite%
{Andrianopoli:2015wqa,Andrianopoli:2015rpa}. }We\textrm{\ }develop the
explicit derivation of the rigid limit of Ward identities of $\mathcal{N}=2$
gauged supergravity by focusing on gauging isometries of particular class of
quaternionic manifolds $\boldsymbol{M}_{QK}^{\left( n_{H}\right) }$ in the
hypermultiplet branch. We show, amongst others, that the splitting of the
graviphoton and Coulomb branch contributions of the covariantly holomorphic
symplectic sections $V^{M}\left( z,\bar{z}\right) $ can be achieved by
performing two successive transformations; a particular symplectic frame
change followed by a particular Kahler transformation. We also give another
manner to rescale the fields of the $\mathcal{N}=2$ theory leading to a
natural interpretation of the components $\theta _{\Lambda }^{u},$ $\tilde{%
\theta}^{u\Lambda }$ of the embedding tensor as FI couplings scaling like
mass$^{2}$. We give as well a geometric interpretation of the $\eta _{i}$
parameters of \textrm{\cite{Andrianopoli:2015rpa} }as the leading term of the%
\textrm{\ }$\frac{m_{3/2}}{M_{pl}}$- expansion of the 1- form Kahler
connection on $\boldsymbol{M}_{SK}$; we also\textrm{\ }complete partial
results and we recover other ones of literature as particular cases.
Moreover, we construct the metrics of a new class of quaternionic manifolds $%
\boldsymbol{M}_{QK}^{\left( ADE\right) }$ classified by the finite
dimensional ADE Lie algebras; the rank $r$ of the ADE algebras is given by
the number $n_{H}$ of hypermultiplets where the leading $\boldsymbol{M}%
_{QK}^{\left( su_{2}\right) }$ is precisely the $SO\left( 1,4\right)
/SO\left( 4\right) $ quaternionic manifold. The conditions for partial
breaking of $\mathcal{N}=2$ supersymmetry in observable and hidden sectors
as well as the breaking of SU$\left( 2\right) _{R}$ down to U$\left(
1\right) _{R}$ are also studied.

\  \  \  \  \newline
The organisation of this paper is as follows: \textrm{In \autoref{sec2}}, we
review some useful aspects on $\mathcal{N}=2$ gauged supergravity in 4-
dimensions. \textrm{In \autoref{sec3}}, we study the rigid limit of
symplectic sections on the special Kahler manifold of the Coulomb branch. We
derive also the rigid limit of other basic quantities like the period matrix 
$\mathcal{N}_{\Lambda \Sigma }$ and the rank two symplectic $\mathcal{U}%
^{MN} $ of the local special Kahler geometry involved in gauge and scalar
sectors. \textrm{In \autoref{sec4}}, we study the description of the
quaternionic Kahler manifold in the case $n_{H}=1$ and give a new
generalisation for $n_{H}>1$ inspired from ADE Toda field theories in two
dimensions. We also study the gauging of abelian quaternionic isometries by
focussing on $SO\left( 1,4\right) /SO\left( 4\right) $. \textrm{In sections %
\ref{sec5} and \ref{sec6}}, we use $\mathcal{N}=2$ supersymmetric
representations and a property TAUB-NUT hyperKahler metric to motivate
another manner to rescale the fields of gauged $\mathcal{N}=2$ supergravity;
fields in gravity and matter supermultiplets are rescaled by Planck mass $%
M_{pl}$ while gauge fields in vector supermultiplets are rescaled by $%
\Lambda \sim m_{3/2}$, the mass of the gravitino after partial breaking of
supersymmetry. We also derive the model of \textrm{\cite%
{Andrianopoli:2015wqa}} as a rigid limit and we discuss the partial breaking
in the rigid theory. \textrm{Section 7} is devoted to the conclusion and
discussions. The last sections are devoted to appendices: \textrm{In \autoref%
{appendixA}}, we recall results on $\mathcal{N}=2$ ungauged and gauged
supergravities in 4d as well as on the general constraints obeyed by the
embedding tensor. \textrm{In \autoref{appendixB}}, we describe basic tools
of special Kahler and special quaternionic manifolds; the proof of some
results, that have been used in sections \ref{sec3} and \ref{sec4}, are also
described in this appendix. In {\normalsize \autoref{appendixC}}\textrm{, we
describe the rigid limit of the coupling matrices } $\mathcal{N}_{\Lambda
\Sigma }$ and $\mathcal{U}^{MN} $\textrm{\ and }in \textrm{\autoref%
{appendixD}}, we study the Ward identities and give details on their
solutions for the case of gauging of abelian quaternionic isometries.
Appendix {\normalsize \autoref{appendixE}}\textrm{\ is devoted to rigid
limit of obtained Ward identities.}

\section{Gauged $\mathcal{N}=2$ supergravity}

\label{sec2}

In this section, we describe some useful tools for studying the coupling of
vector and matter supermultiplets in $\mathcal{N}=2$ gauged supergravity in
4d. Of particular interest for us here is the issue concerning the gauging
of isometries of the scalar manifold $\boldsymbol{M}_{scal}=\boldsymbol{M}%
_{SK}\times \boldsymbol{M}_{QK}$ of this theory, the underlying Ward
identities and the induced scalar potential to be used later for the study
of partial supersymmetry breaking. For explicit details regarding this
section, see \emph{appendix A} of this paper on 4d $\mathcal{N}=2$ ungauged
and gauged extended supergravities; further details can be also found in a
recent review given in \textrm{\cite{Trigiante:2016mnt}.}

\subsection{$\mathcal{N}=2$ supergravity and isometries}

We first describe the on shell \ degrees of freedom of $\mathcal{N}=2$
supergravity; then we give the component field lagrangian $\mathcal{L}_{%
\text{sugra}}^{\mathcal{N}=2}$ and Ward identities of gauged isometries of
scalar manifold $\boldsymbol{M}_{scal}$ of this theory.

\subsubsection{Component fields}

$\mathcal{N}=2$ supergravity in four space time dimensions involves three
basic $\mathcal{N}=2$ supersymmetric representations carrying as well
quantum numbers under SU$\left( 2\right) $ R- symmetry of the underlying
supersymmetric algebra; these are the $\mathcal{N}=2$ gravity
supermultiplet, the $\mathcal{N}=2$ vector supermultiplet and the $\mathcal{N%
}=2$ matter supermultiplet respectively denoted as follows%
\begin{equation}
\boldsymbol{G}_{\mathcal{N}=2}\qquad ,\qquad \boldsymbol{V}_{\mathcal{N}%
=2}\qquad ,\qquad \boldsymbol{H}_{\mathcal{N}=2}  \label{21}
\end{equation}%
In building 4D $\mathcal{N}=2$ local models, we use one $\boldsymbol{G}_{%
\mathcal{N}=2}$ supermultiplet; but we may have several copies of vector
supermultiplets $\left \{ \boldsymbol{V}_{\mathcal{N}=2}^{i}\right \}
_{1\leq i\leq n_{V}}$, whose number $n_{V}$ define the rank of the compact
gauge symmetry $\mathcal{G}$ of the model. We may also have several
hypermultiplets $\left \{ \boldsymbol{H}_{\mathcal{N}=2}^{I}\right \}
_{1\leq I\leq n_{H}}$ that interact with $\boldsymbol{G}_{\mathcal{N}=2}$
and $\boldsymbol{V}_{\mathcal{N}=2}^{i}$. The interaction between $%
\boldsymbol{V}_{\mathcal{N}=2}^{i}$ and $\boldsymbol{H}_{\mathcal{N}=2}^{I}$
requires hypermultiplets carrying charges under the gauge symmetry group $%
\mathcal{G}$; this means that hypermultiplets $\boldsymbol{H}_{\mathcal{N}%
=2}^{I}$ have to transform in some non trivial representations of the gauge
symmetry to be taken in this study as just $\mathcal{G}=U\left( 1\right)
^{n} $. However, in dealing with the scalar component fields in $\boldsymbol{%
H}_{\mathcal{N}=2}$ that capture information on the underlying geometry of
the quaternionic Kahler submanifold $\boldsymbol{M}_{QK}$ of the scalar
manifold $\boldsymbol{M}_{scal}$ of the $\mathcal{N}=2$ supergravity model,
we have to worry about the SU$\left( 2\right) $ R-symmetry representations
hosting the real 4n$_{H}$ scalars of $\boldsymbol{H}_{\mathcal{N}=2}^{I}$
namely%
\begin{equation}
\left( s_{1},s_{2},s_{3},s_{4}\right) ^{I}=\left(
s_{1}^{I},s_{2}^{I},s_{3}^{I},s_{4}^{I}\right) \qquad ,\qquad I=1,...,n_{H}
\end{equation}%
Indeed, recall that from SU$\left( 2\right) _{R}$ view, the four scalar
components in a given hypermultiplet $\boldsymbol{H}_{\mathcal{N}=2}$ can be
described in different manners; in particular by the two following ones to
be used later on: $\left( i\right) $ either by using a complex field doublet 
$f^{A}$ and its complex conjugate $\bar{f}_{A}$ expressed in terms of the $%
s_{k}$'s as follows.%
\begin{equation}
f^{A}=\left( 
\begin{array}{c}
s_{1}+is_{2} \\ 
s_{3}+is_{4}%
\end{array}%
\right) \qquad ,\qquad \bar{f}_{A}=\left( s_{1}-is_{2},s_{3}-is_{4}\right)
\label{ssc}
\end{equation}%
or $\left( ii\right) $ a real isotriplet $\vec{\phi}$ and a real isosinglet $%
\varphi $ given by 
\begin{equation}
\phi ^{a}=\left( 
\begin{array}{c}
s_{1} \\ 
s_{2} \\ 
s_{3}%
\end{array}%
\right) \qquad ,\qquad \varphi =s_{4}  \label{re}
\end{equation}%
In this study, we will mainly use (\ref{re}) described by real SU$\left(
2\right) $ representations to parameterise the scalars of hypermultiplets $%
\boldsymbol{H}_{\mathcal{N}=2}^{I}$; the action of the gauge symmetry $%
\mathcal{G}=U\left( 1\right) ^{n_{V}}$ on these hypermultiplets will be
realised in terms of local isometries of the quaternionic Kahler submanifold 
$\boldsymbol{M}_{QK}$ parameterised by scalar fields%
\begin{equation}
Q^{uI}=\left( \varphi ^{I},\phi ^{aI}\right) \qquad ,\qquad u=1,2,3,4\qquad
,\qquad I=1,...,n_{H}
\end{equation}%
As we have also scalar fields coming from $\boldsymbol{V}_{\mathcal{N}%
=2}^{i} $, let us recall rapidly here below the component fields of the
supersymmetric multiplets (\ref{21}) and their quantum numbers under SU$%
\left( 2\right) _{R}$.

\begin{description}
\item[$\left( 1\right) $] Gravity supermultiplet $\boldsymbol{G}_{\mathcal{N}%
=2}:$ the spins of its on shell degrees of freedom are as $\left( 2,\frac{3}{%
2}^{2},1\right) $; in addition to the spin 2 graviton $e_{\mu }^{m}$, it has
a graviphoton $A_{\mu }^{0}$, that plays an important role in present study,
and two spin $\frac{3}{2}$ gravitini $\psi _{\mu }^{\hat{\alpha}1}$ and $%
\psi _{\mu }^{\hat{\alpha}2}$ forming an SU$\left( 2\right) $ doublet
described by 
\begin{equation}
\psi _{\mu }^{\hat{\alpha}A}=\left( 
\begin{array}{c}
\psi _{\mu }^{\alpha A} \\ 
\bar{\psi}_{\mu \dot{\alpha}A}%
\end{array}%
\right) \qquad ,\qquad \alpha ,\dot{\alpha}=1,2  \label{gt}
\end{equation}%
As far as $\mathcal{N}=2$ supersymmetry is preserved, these two gravitini
are massless; but if it is partially broken, half of the degrees of freedom
in (\ref{gt}) become massive with some mass $m_{3/2}$ thought of below as
the scale of partial supersymmetry breaking.

\item[$\left( 2\right) $] Vector supermultiplet $\boldsymbol{V}_{\mathcal{N}%
=2}:$ its on shell degrees of freedom are as $(1,\frac{1}{2}^{2},0^{2});$ in
addition to the gauge field $A_{\mu }$ which will also play an important
role in the present analysis, it has two scalars $z$ and $\bar{z}$
parameterising $\boldsymbol{M}_{SK}$ as well as two Majorana gauginos $%
\lambda ^{\hat{\alpha}1}$, $\lambda ^{\hat{\alpha}2}$ forming a doublet as 
\begin{equation}
\lambda ^{\hat{\alpha}A}=\left( 
\begin{array}{c}
\lambda ^{\alpha A} \\ 
\bar{\lambda}_{\dot{\alpha}A}%
\end{array}%
\right) 
\end{equation}%
In the U$\left( 1\right) ^{n_{V}}$ Coulomb branch, we have a set of $n_{V}$
vector supermultiplets indexed like $\boldsymbol{V}_{\mathcal{N}=2}^{i}$
with on shell degrees described by the component fields 
\begin{equation}
A_{\mu }^{i}\qquad ,\qquad z^{i}\qquad ;\qquad \lambda ^{\hat{\alpha}Ai}
\label{zi}
\end{equation}%
The gauginos are organised into $n_{V}$ doublets of SU$\left( 2\right) _{R}$%
; and the real 2$n_{V}$ scalars are described by $n_{V}$ complex scalars $%
z^{i}$ and their complex conjugates $\bar{z}^{i}$; they are singlets under
the global SU$\left( 2\right) _{R}$ and the local $U\left( 1\right) ^{n_{V}}$%
. Notice that along with the $n_{V}$\ abelian potential\textrm{\ }vectors%
\textrm{\ }$A_{\mu }^{i}$ of the Coulomb branch\textrm{, }we moreover have
the\ $A_{\mu }^{0}$ graviphoton of the\textrm{\ }$\boldsymbol{G}_{\mathcal{N}%
=2}$ multiplet which is also an abelian gauge field; \textrm{these electric
gauge potentials will be denoted collectively as }$A_{\mu }^{\Lambda
}=\left( A_{\mu }^{0},A_{\mu }^{i}\right) $ with an upper index $\Lambda $
running from $0$ to $n_{V}$. \textrm{Thus, the total electric abelian gauge
group of the low energy theory is }$U\left( 1\right) ^{1+n_{V}}$. \textrm{By
implementing the magnetic duals of the electric vector potentials }$A_{\mu
}^{\Lambda }$\textrm{, which for convenience we denote them like }$A_{\mu
\Lambda }=\left( A_{\mu 0},A_{\mu i}\right) $\textrm{\ with index }$\Lambda $%
\textrm{\ down stairs, the resulting }$\mathcal{N}=2$\textrm{\ dyonic theory
has now an }$SP\left( 2n+2\right) $\textrm{\ symplectic symmetry and still a 
}$U\left( 1\right) ^{1+n_{V}}$\textrm{\ gauge symmetry\footnote{
\ This description has a nice interpretation in type IIB string compactified
on Calabi-Yau threefolds $\left( CY3\right) $ with symplectic homology three
cycle basis $\left( \alpha ^{\Lambda },\beta _{\Lambda }\right) $. The
4-form gauge potenntial $\mathbf{C}_{4}$ of the dyonic D3 brane gets reduced
after compactification down to the space-time 1-form gauge potentials $%
\mathbf{A}_{1}^{\Lambda }=A_{\mu }^{\Lambda }dx^{\mu }$ and $\mathbf{A}%
_{1\Lambda }=A_{\mu \Lambda }dx^{\mu }$ related to the 4-form potantial like 
$\mathbf{A}_{1}^{\Lambda }=\int_{\alpha ^{\Lambda }}\mathbf{C}_{4}$ and $%
\mathbf{A}_{1\Lambda }=\int_{\beta _{\Lambda }}\mathbf{C}_{4}$ and to each
other by the symplectic symmetry inherited from $\left( \alpha ^{\Lambda
},\beta _{\Lambda }\right) $.}.} As was first argued in\textrm{\normalsize \ 
}\textrm{{\cite{Seiberg:1994rs,Seiberg:1988ur}}}, the low energy properties
of the electric- magnetic theory are described by one function, the
prepotential\textrm{\ }$\mathbf{F}$, which is a holomorphic function of the
scalar field\textrm{\ }$\mathbf{F}=\mathbf{F}(z)$; more details on this
description are reported in section 3.

\item[$\left( 3\right) $] Matter hypermultiplet $\boldsymbol{H}_{\mathcal{N}%
=2}:$ its on shell degrees of freedom are given by $(\frac{1}{2}^{2},0^{4});$
it has four real scalars $\left( s_{1},s_{2},s_{3},s_{4}\right) $ and two
complex Fermi fields which can be hosted by different representations of SU$%
\left( 2\right) _{R}$; in particular as follows%
\begin{equation}
\begin{tabular}{l|lll||lll}
$\boldsymbol{H}_{\mathcal{N}=2}$ & \multicolumn{3}{|l}{\small \  \  \  \  \  \  \
\  \  \  \  \ scalars} & \multicolumn{3}{||l}{\small \  \  \  \  \  \  \  \  \  \  \  \  \  \
\  \  \ fermions} \\ \hline
{\small Rep I} & {\small complex} & : & $f^{A}=\left( f^{1},f^{2}\right) $ & 
{\small \ Dirac } & : & $\Psi ^{\hat{\alpha}}=\left( \psi ^{\alpha },\bar{%
\chi}_{\dot{\alpha}}\right) $ \\ \hline
{\small Rep II} & {\small real} & : & $Q^{u}=\left( \varphi ,\phi
^{a}\right) $ & {\small \ Majorana} & : & $\xi ^{\hat{\alpha}A}=\left( \xi
^{\alpha A},\bar{\xi}_{\dot{\alpha}A}\right) $%
\end{tabular}
\label{tb1}
\end{equation}%
\begin{equation*}
\end{equation*}%
In the first representation (Rep I), the $f^{A}$ is a complex isodoublet of
scalar fields as in (\ref{ssc}), and $\Psi ^{\hat{\alpha}}$ is a Dirac
spinor; while in Rep II, the hyperini $\xi ^{\hat{\alpha}A}$ form an
isodoublet of Majorana fermions. The scalar field doublet $f^{A}$ has the
usual canonical dimension, that is scaling as mass; and a priori $Q^{u}$
should be treated in a similar manner. However this is not the case in
present study since we will use quantities like $e^{2\varphi }$ as in eq(\ref%
{t}). In \textrm{\autoref{rsp}}, we will comment on this issue and give a
proposal for the scaling dimensions of Rep II inspired from the structure of
TAUB-NUT metric preserving manifestly the SU$\left( 2\right) $ R- symmetry;
see eqs(\ref{fh}-\ref{hf}) and (\ref{tb}-\ref{bt}).
\end{description}

$\  \  \ $

$\bullet $ \emph{A comment on scalars of }$\boldsymbol{H}_{\mathcal{N}=2}$%
\newline
The two manners Rep I and Rep II in describing the four real scalars of the
hypermultiplet $\boldsymbol{H}_{\mathcal{N}=2}$ given by (\ref{tb1}) are
very remarkable and suggestive. The point is that for the complex
representation, the SU$\left( 2\right) $ doublet $f^{A}$ and its complex
conjugate $\bar{f}_{A}$ allow U$\left( 1\right) $ phase change preserving SU$%
\left( 2\right) $ R- symmetry like%
\begin{equation}
\begin{tabular}{lllllll}
$f^{A}$ & $\rightarrow $ & $f^{^{\prime }A}=e^{iqw}f^{A}$ & $\qquad ,\qquad $
& $J_{0}f^{A}$ & $=$ & $qf^{A}$ \\ 
$\bar{f}_{A}$ & $\rightarrow $ & $\bar{f}_{A}^{^{\prime }}=e^{-iqw}\bar{f}%
_{A}$ & $\qquad ,\qquad $ & $J_{0}\bar{f}_{A}$ & $=$ & $-q\bar{f}_{A}$%
\end{tabular}%
\end{equation}%
where the real $w$ stands for the parameter of the abelian U$\left( 1\right) 
$ group and the hermitian $J_{0}$ is its generator which can be thought of
as 
\begin{equation}
J_{0}=q\left( f^{A}\frac{\partial }{\partial f^{A}}-\bar{f}_{A}\frac{%
\partial }{\partial \bar{f}_{A}}\right) \qquad ,\qquad \left( \frac{\partial 
}{\partial f^{A}}\right) ^{\dagger }=-\frac{\partial }{\partial \bar{f}_{A}}
\end{equation}%
The same transformations are valid for the gradient $\partial _{\mu }f^{A}$
of the field doublet which transform like $\partial _{\mu }f^{A}\rightarrow
e^{iqw}\partial _{\mu }f^{A}$ for global parameter w. By gauging the phase
change, $w=w\left( x\right) $, one needs implementation of an abelian gauge
field $\mathfrak{A}_{\mu }^{0}$ with gauge transformation $\mathfrak{A}_{\mu
}^{0}\rightarrow \mathfrak{A}_{\mu }^{0}-\partial _{\mu }w$; and replace $%
\partial _{\mu }f^{A}$ by the gauge covariant derivative 
\begin{equation}
D_{\mu }f^{A}=\left( \partial _{\mu }+i\mathfrak{A}_{\mu }^{0}J_{0}\right)
f^{A}
\end{equation}%
in order to have covariance $D_{\mu }f^{A}\rightarrow e^{iqw}D_{\mu }f^{A}$.
Such abelian phase change is not possible for the real representation $%
Q^{u}=\left( \varphi ,\phi ^{a}\right) $ since the fields are real\ scalars.
To engineer an abelian phase symmetry, we have to complexify the fields $%
\varphi $ and $\vec{\phi}=\left( \phi ^{a}\right) $ by doubling the number
of degrees of freedom which in general corresponds to having an even number
of hypermultiplets. However, one can still engineer abelian symmetries by
using field translations like 
\begin{equation}
\varphi \rightarrow \varphi ^{\prime }=\varphi +c_{0}\qquad ,\qquad \vec{\phi%
}\rightarrow \vec{\phi}^{\prime }=\vec{\phi}+\vec{c}
\end{equation}%
respectively generated by 
\begin{equation}
T_{0}=\frac{\partial }{\partial \varphi }\qquad ,\qquad T_{a}=\frac{\partial 
}{\partial \phi ^{a}}
\end{equation}%
These kind of translations will be encountered later on when studying
gauging abelian quaternionic isometries of the scalar manifold $\boldsymbol{M%
}_{QK}$ associated with the hypermultiplet sector. There, we will give more
details on the gauging of these translations. \newline
Notice by the way that for the generic case where there are $n_{H}$
hypermultiplets $\boldsymbol{H}_{\mathcal{N}=2}^{r}$, the component fields
of the hypermultiplets carry also the index $r=1,...,n_{H}$; for the example
of the real representation we have $Q^{ur}=\left( \varphi ^{r},\phi
^{ar}\right) $ for scalars and $\xi _{\hat{\alpha}}^{Ar}$ for Majorana
fermions. A similar indexing is valid as well for the complex scalars $%
f^{Ar} $ and their Dirac spinor partners $\Psi ^{\hat{\alpha}r}=\left( \psi
^{\alpha r},\bar{\chi}_{\dot{\alpha}}^{r}\right) $.\  \  \  \newline
To conclude this paragraph, observe that scalar fields in $\mathcal{N}=2$
supergravity come from two sectors: $\left( i\right) $ these are the $n_{V}$
complex scalars $z^{i}$ parameterising a complex $n_{V}$ dimension special
Kahler manifold $\boldsymbol{M}_{SK}$; and $\left( ii\right) $ the 4$n_{H}$
real scalars $Q^{ur}$ parameterising a quaternionic Kahler manifold $%
\boldsymbol{M}_{QK}$. Altogether, these scalars parameterise a manifold $%
\boldsymbol{M}_{scal}=\boldsymbol{M}_{SK}\times \boldsymbol{M}_{QK}$ with
real $2n_{V}+4n_{H}$ dimensions and a block diagonal metric of the form 
\begin{equation}
\mathcal{\tilde{G}}=\left( 
\begin{array}{cc}
\mathcal{G}_{i\bar{j}} & 0 \\ 
0 & h_{\pi \pi ^{\prime }}%
\end{array}%
\right)  \label{dm}
\end{equation}%
with $\mathcal{G}_{i\bar{j}}$ is the metric of $\boldsymbol{M}_{SK}$ and $%
h_{\pi \pi ^{\prime }}$ is the metric of $\boldsymbol{M}_{QK}$ with $4n_{H}$
field coordinates $q^{\pi }\equiv Q^{ur}$. The special Kahler and the
quaternionic Kahler submanifolds of $\boldsymbol{M}_{scal}$ are given by the
coset group manifolds\textrm{\footnote{%
In the embedding of this $\mathcal{N}=2$ QFT$_{4}$ in type II string theory
on CY3, we have two extra scalars parameterising the particular Kahler
submanifold $SU\left( 1,1\right) /U\left( 1\right) .$}} 
\begin{equation}
\boldsymbol{M}_{SK}= \frac{SO\left( 2,n_{V}\right) }{SO\left( 2\right)
\times SO\left( n_{V}\right) }
\end{equation}%
and%
\begin{equation}
\boldsymbol{M}_{QK}=\frac{SO\left( 4,n_{H}\right) }{SO\left( 4\right) \times
SO\left( n_{H}\right) }
\end{equation}%
In what follows, we shall set $n_{V}\equiv n$, and keep the notation $n_{H}$
for the number of hypermultiplets with scalars as $q^{\pi }$. Later on, we
will also consider the particular and interesting case where $n_{H}=1$ by
using the real representation $Q^{u}=\left( \varphi ,\phi ^{a}\right) $.

\subsubsection{Gauging isometries and Lagrangian density}

The only known way to introduce a scalar potential in the $\mathcal{N}=2$
supergravity theory which is compatible with $\mathcal{N}=2$ supersymmetry
is through the procedure of gauging isometries of the scalar manifold of the
theory \textrm{\cite{Gallerati:2016oyo, Trigiante:2016mnt}}. This procedure,
to be detailed on a particular example \textrm{in \autoref{ssec43}}, is
nicely described by using the embedding tensor $\vartheta _{M}^{\gamma }$,
carrying two different kinds of indices, with components as%
\begin{equation}
\vartheta _{M}^{\gamma }=\left( \theta _{\Lambda }^{\gamma },\tilde{\theta}%
^{\Lambda \gamma }\right) \qquad ,\qquad \vartheta ^{\gamma M}=\mathcal{C}%
^{MN}\vartheta _{N}^{\gamma }=\left( 
\begin{array}{c}
\tilde{\theta}_{\Lambda }^{\gamma } \\ 
-\theta ^{\Lambda \gamma }%
\end{array}%
\right)  \label{eg}
\end{equation}%
with $\Lambda =1,...,n_{V}$ and $\gamma =1,...,n_{H}$ respectively indexing
the number of gauge and matter multiplets. This symplectic object has been
first introduced in \textrm{\cite{deWit:2005ub}}, and can be of two types: $%
\left( i\right) $ either having only electric components $\vartheta
_{M}^{\gamma }=\left( \theta _{\Lambda }^{\gamma },0\right) $; in this case
the gauging is called \textrm{electric gauging;} or $\left( ii\right) $ 
\textrm{dyonic gauging where }$\vartheta _{M}^{\gamma }$\textrm{\ }has both
electric and magnetic components $\left( \theta _{\Lambda }^{\gamma },\tilde{%
\theta}^{\Lambda \gamma }\right) $ \textrm{\cite{deWit:2005ub,
deVroome:2007qu}}. In this regards, notice that the rigid limit of an
electric gauged $\mathcal{N}=2$ supergravity was constructed a long time ago
in \textrm{\cite{Ferrara:1995xi}}; there the authors showed that to have, in
the rigid theory, a partially broken $\mathcal{N}=2$ supersymmetry one must
choose a symplectic frame in which the prepotential $\mathcal{F}$\ of the
special Kahler geometry does not exist. This constraint is overcome in
dyonic $\mathcal{N}=2$ gauged supergravity; there the rigid limit of dyonic $%
\mathcal{N}=2$ theory has been recently constructed in \textrm{\cite%
{Andrianopoli:2015rpa}} where the limit of $\mathcal{N}=2$ supergravity is
achieved in \emph{any} symplectic frame and more importantly in a frame in
which the\emph{\ }prepotential\emph{\ }$\mathcal{F}$\emph{\ }exists.

\  \  \  \  \ 

$\bullet $ \emph{Gauging isometries}\newline
Following \textrm{\cite{deWit:2005ub, Andrianopoli:2015rpa},} the gauging of
a subgroup \b{H} of the global isometry group \b{G} of the scalar manifold $%
\boldsymbol{M}_{scal}=\boldsymbol{M}_{SK}\times \boldsymbol{M}_{QK}$ of the $%
\mathcal{N}=2$ gauged supergravity is nicely encoded in the embedding tensor 
$\vartheta _{M}^{\gamma }$ \textrm{of (\ref{eg}). This tensor }allows to
express objects valued in the Lie algebra of \b{H}, like the gauge field $%
C_{\mu }$, in two equivalent manners: $\left( \mathit{1}\right) $ either as $%
C_{\mu }=\mathcal{A}_{\mu }^{M}X_{M}$ where the $X_{M}$'s are the generators
of \b{H} and where the $\mathcal{A}_{\mu }^{M}$'s are given by $(A_{\mu
}^{\Lambda },\tilde{A}_{\mu \Lambda })$ transforming as an \textrm{SP}$%
\left( 2n_{V}+2,\mathbb{R}\right) $\textrm{\ symplectic vector; or }$\left( 
\mathit{2}\right) $\textrm{\ like }$C_{\mu }=\mathcal{A}_{\mu }^{M}\vartheta
_{M}^{\gamma }\boldsymbol{T}_{\gamma }$ where now the $\boldsymbol{T}%
_{\gamma }$'s are the generators of \b{G}. To fix ideas, let us consider the
general picture regarding the gauging of isometries in both $\boldsymbol{M}%
_{SK}$ and $\boldsymbol{M}_{QK}$ factors of $\boldsymbol{M}_{scal}$; the
restriction to the gauging of isometries in $\boldsymbol{M}_{QK}$ is
directly obtained by dropping out the part concerning $\boldsymbol{M}_{SK}$;
particular examples concerning the values of the $n_{V}$ integer will be
also considered in section 4. To that purpose, let us denote \textrm{by }$%
\boldsymbol{T}_{\gamma }=\left( T_{\underline{a}},T_{\underline{m}}\right) $%
\textrm{\ the set of generators of \b{G} with }$T_{\underline{a}}$\textrm{'s
refering to isometry generators in }$\boldsymbol{M}_{SK}$\textrm{\ and }$T_{%
\underline{m}}$\textrm{'s to those in }$\boldsymbol{M}_{QK}$\textrm{. The
electric gauging of a part of \b{G} is achieved by considering a subgroup \b{%
H}} \textrm{generated by some linear combinations of }$T_{\underline{a}}$
and $T_{\underline{m}}$\textrm{\ that we define as }$X_{\Lambda }=\theta
_{\Lambda }^{\underline{a}}T_{\underline{a}}+\theta _{\Lambda }^{\underline{m%
}}T_{\underline{m}}$\textrm{\ and which can be rewritten in a condensed
manner like }$X_{\Lambda }=\theta _{\Lambda }^{\gamma }\boldsymbol{T}%
_{\gamma }$\textrm{\ where the coefficients }$\theta _{\Lambda }^{\gamma }$%
\textrm{\ are as in the first entry block of eq(\ref{eg}). To these electric 
}$X_{\Lambda }$\textrm{\ generators, it is associated the }$A_{\mu
}^{\Lambda }$ \textrm{electric gauge fields of the supergravity theory and
so we can think about the }gauge field valued in the Lie algebra of \b{H}%
\textrm{\ as }$C_{\mu }=A_{\mu }^{\Lambda }X_{\Lambda }$. By substituting $%
X_{\Lambda }$ in terms of $\boldsymbol{T}_{\gamma }$ we can also imagine
this vector potential as valued in \b{G} like $C_{\mu }=C_{\mu }^{\gamma }%
\boldsymbol{T}_{\gamma }$ but with components as $C_{\mu }^{\gamma }$\textrm{%
\ }$=\theta _{\Lambda }^{\gamma }A_{\mu }^{\Lambda }$.\textrm{\ In the
dyonic gauging, one has in addition to the electric }$X_{\Lambda }$\textrm{,
magnetic duals defined as} $X^{\Lambda }=\tilde{\theta}^{\gamma \Lambda }%
\boldsymbol{T}_{\gamma }$ \textrm{with }$\tilde{\theta}^{\gamma \Lambda }$%
\textrm{\ given by the second entry block of (\ref{eg}). In this case, the
previous expression of the }$C_{\mu }^{\gamma }$\textrm{\ gauge potentials
gets promoted to a general one involving the }$\tilde{A}_{\mu \Lambda }$%
\textrm{'s as well; so we have }$C_{\mu }^{\gamma }=\theta _{\Lambda
}^{\gamma }A_{\mu }^{\Lambda }+\tilde{\theta}^{\gamma \Lambda }\tilde{A}%
_{\mu \Lambda }$. \textrm{The symplectic structure of these kinds of dyonic
quantities can be manifestly exhibited by using SP}$\left( 2n_{V}+2,\mathbb{R%
}\right) $\textrm{\ symplectic representations. For the case of the gauge
field, this may be done as follows: first consider the potential form }$%
C_{\mu }=C_{\mu }^{\gamma }\boldsymbol{T}_{\gamma }$,\textrm{\ substitute }$%
C_{\mu }^{\gamma }$ in terms of $A_{\mu }^{\Lambda },$ $\tilde{A}_{\mu
\Lambda }$ and use the symplectic symmetry of the dyonic theory to put it in
the form $C_{\mu }=\mathcal{A}_{\mu }^{M}X_{M}$ from which we learn that%
\textrm{\ }$X_{M}=\vartheta _{M}^{\gamma }\boldsymbol{T}_{\gamma }$ which
expands explicitly like 
\begin{equation}
X_{M}=\vartheta _{M}^{\underline{a}}T_{\underline{a}}+\vartheta _{M}^{%
\underline{m}}T_{\underline{m}}
\end{equation}%
\textrm{In the case where the gauging of isometries of the scalar manifold
is restricted to quaternionic isometries in} $\boldsymbol{M}_{QK}$, the
embedding tensor has no $\vartheta _{M}^{\underline{a}}$ components, and so
the above expression reduces to $X_{M}=\vartheta _{M}^{\underline{m}}T_{%
\underline{m}}$. Notice also that consistency of the gauging of the \b{H}
subgroup of \b{G} requires a set of constraint relations; in particular the
following ones: \newline
$\left( i\right) $ conditions on the embedding tensor components%
\begin{equation}
\begin{tabular}{lll}
$\vartheta _{M}^{\underline{a}}\vartheta _{N}^{\underline{b}}f_{\underline{a}%
\underline{b}}^{\underline{c}}+\left( X_{M}\right) _{N}^{P}\vartheta _{P}^{%
\underline{c}}$ & $=$ & $0$ \\ 
$\vartheta _{M}^{\underline{m}}\vartheta _{N}^{\underline{n}}f_{\underline{m}%
\underline{n}}^{\underline{p}}+\left( X_{M}\right) _{N}^{P}\vartheta _{P}^{%
\underline{p}}$ & $=$ & $0$%
\end{tabular}
\label{lc}
\end{equation}%
and%
\begin{equation}
\vartheta _{M}^{\underline{a}}\mathcal{C}^{MN}\vartheta _{N}^{\underline{b}%
}=\vartheta _{M}^{\underline{a}}\mathcal{C}^{MN}\vartheta _{N}^{\underline{n}%
}=\vartheta _{M}^{\underline{m}}\mathcal{C}^{MN}\vartheta _{N}^{\underline{n}%
}=0
\end{equation}%
where $f_{\underline{a}b}^{\underline{c}}$, $f_{\underline{m}\underline{n}}^{%
\underline{p}}$ are structure constants of the underlying Lie algebra of \b{G%
}, $\left( X_{M}\right) _{N}^{P}$ the matrix elements of $X_{M}$; and where
the antisymmetric $\mathcal{C}^{MN}$ is the $Sp(2n+2)$ invariant matrix
given by 
\begin{equation}
\mathcal{C}^{MN}=%
\begin{pmatrix}
0 & \mathcal{I}_{n+1} \\ 
-\mathcal{I}_{n+1} & 0%
\end{pmatrix}%
\end{equation}%
$\left( ii\right) $ the dualisation of the part of the four real scalars $%
Q^{u}$ of the matter sector, involved in the gauging, in terms of rank 2-
antisymmetric tensors. This dualisation is needed in order to avoid the
introduction of new degrees of freedom in the theory associated with the
gauge fields $A_{\mu }^{\Lambda }=\left( A_{\mu }^{0},A_{\mu }^{i}\right) $ 
\textrm{\cite{Andrianopoli:2015rpa, D'Auria:2004yi, Sommovigo:2004vj,
Dall'Agata:2003yr, Louis:2002ny, deWit:2005ub}}. If referring to the part of
scalars in question scalars as $\phi ^{\underline{m}}$ and to the
antisymmetric tensors like $B_{\mu \nu }^{\underline{m}}$, the dualisation
can be described by help of the coupling $\mathcal{L}_{d}\sim \varepsilon
^{\mu \nu \rho \sigma }\partial _{\mu }\phi ^{\underline{m}}H_{\underline{m}%
\nu \rho \sigma }+\frac{1}{2}H_{\underline{m}\nu \rho \sigma }H^{\underline{m%
}\nu \rho \sigma }$ with $H_{\nu \rho \sigma }^{\underline{m}}$ standing for
the field strengths of $B_{\mu \nu }^{\underline{m}}$ given by%
\begin{equation}
H_{\nu \rho \sigma }^{\underline{m}}=\frac{1}{3!}\partial _{\lbrack \nu
}B_{\rho \sigma ]}^{\underline{m}}  \label{hb}
\end{equation}%
After the dualisation, one can interpret the gauge fields $A_{\mu }^{%
\underline{m}}$ as the degrees of freedom needed for the antisymmetric $%
B_{\mu \nu }^{\underline{m}}$ \textrm{to become a massive tensor field \cite%
{Andrianopoli:2007ep,Cecotti:1987qr}. Recall that in a generic d-
dimensional space time, a massless gauge field }$\mathfrak{A}_{\mu }$\textrm{%
\ has }$\left( d-2\right) $ degrees of freedom while a massless $\mathfrak{B}%
_{\mu \nu }$ has $\left( d-2\right) \left( d-3\right) /2$; in our case $d=4$%
. By eating $\mathfrak{A}_{\mu }$, the number of the degrees of freedom of
the resulting antisymmetric field $\mathfrak{\tilde{B}}_{\mu \nu }$ is given
by the sum of degrees of $\mathfrak{B}_{\mu \nu }$ and degrees of $\mathfrak{%
A}_{\mu }$ namely\textrm{\ }%
\begin{equation}
\frac{1}{2}\left( d-2\right) \left( d-3\right) +\left( d-2\right) =\left(
d-2\right) \left( d-1\right)
\end{equation}%
This number is exactly the total degrees of freedom of a massive rank- 2
antisymmetric field $\mathfrak{\tilde{B}}_{\mu \nu }$. \newline
$\left( iii\right) $ To preserve $\mathcal{N}=2$ supersymmetry, one must add
a scalar potential $\mathcal{V}_{scal}^{\mathcal{N}=2}=\mathcal{V}\left( z,%
\bar{z},Q\right) $ to the Lagrangian density. We notice that after the
dualisation, the $Q^{u}$ scalars of the hypermultiplet split as follows%
\begin{equation}
Q^{u}=\left( \phi ^{\hat{a}},B_{\mu \nu }^{\underline{m}}\right)
\end{equation}%
and the metric $h_{uv}$ of $\boldsymbol{M}_{QK}$ becomes \textrm{\cite%
{Dall'Agata:2003yr,Andrianopoli:2015rpa}}%
\begin{equation}
h_{uv}=%
\begin{pmatrix}
g_{\hat{a}\hat{b}} & A_{\hat{a}}^{\underline{m}} \\ 
A_{\hat{a}}^{\underline{m}} & M^{\underline{m}\underline{n}}%
\end{pmatrix}
\label{dmet}
\end{equation}%
Notice that for $n_{H}=1$, the indices $\hat{a}$ and $\underline{m}$ take
respectively the values $\hat{a}=1,2$ and $\underline{m}=3,4$.

\  \  \  \  \  \ 

$\bullet $ \emph{Lagrangian density}\newline
The component field Lagrangian $\sqrt{g}\mathcal{L}_{\mathcal{N}=2}$ of the
4d $\mathcal{N}=2$ gauged supergravity depends on the geometry of $%
\boldsymbol{M}_{scal}$; the explicit expression of its bosonic part,
describing the interacting dynamics of the above mentioned degrees of
freedom, is given by \textrm{\cite{D'Auria:2004yi,Andrianopoli:2015rpa,
Dall'Agata:2003yr}} 
\begin{equation}
\mathcal{L}_{N=2}=\mathcal{R}+\mathcal{L}-\mathcal{V}_{scal}^{\mathcal{N}=2}
\label{act}
\end{equation}%
with 
\begin{equation}
\begin{tabular}{lll}
$\mathcal{L}$ & $=$ & $\mathcal{G}_{i\bar{j}}\partial ^{\mu }z^{i}\partial
_{\mu }\bar{z}^{\bar{j}}+g_{\hat{a}\hat{b}}\partial _{\mu }\phi ^{\hat{a}%
}\partial ^{\mu }\phi ^{\hat{b}}+2\varepsilon ^{\mu \nu \rho \sigma }A_{\hat{%
a}}^{\underline{m}}H_{\underline{m}\nu \rho \sigma }\partial _{\mu }\phi ^{%
\hat{a}}$ \\ 
&  & $-2\varepsilon {^{\mu \nu \lambda \sigma }}B_{\underline{m}\mu \nu
}\vartheta _{\Lambda }^{\underline{m}}\left( \mathcal{\hat{F}}_{\rho \sigma
}^{\Lambda }-\vartheta ^{\Lambda \underline{n}}B_{\underline{n}\rho \sigma
}\right) $ \\ 
&  & $+i\left( \bar{\mathcal{N}}_{\Lambda \Sigma }\mathcal{\hat{F}}_{\mu \nu
}^{-\Lambda }\mathcal{\hat{F}}^{-\Sigma \mu \nu }-\mathcal{N}_{\Lambda
\Sigma }\mathcal{\hat{F}}_{\mu \nu }^{+\Lambda }\mathcal{\hat{F}}^{+\Sigma {%
\mu \nu }}\right) $ \\ 
&  & $+6M^{\underline{m}\underline{n}}H_{\underline{m}\nu \rho \sigma }H_{%
\underline{n}}^{\nu \rho \sigma }$%
\end{tabular}
\label{tca}
\end{equation}%
where ${\mathcal{F}}_{\mu \nu }^{\Lambda }$ are the field strengths of the
vectors $A_{\mu }^{\Lambda }$ defined as%
\begin{equation}
\mathcal{\hat{F}}_{\mu \nu }^{\Lambda }={\mathcal{F}}_{\mu \nu }^{\Lambda
}+2\vartheta ^{\Lambda m}B_{m\mu \nu }
\end{equation}%
and%
\begin{equation}
{\mathcal{F}}_{\mu \nu }^{\pm \Lambda }=\frac{1}{2}({\mathcal{F}}_{\mu \nu
}^{\Lambda }\pm {\frac{i}{2}}\epsilon _{\mu \nu \rho \sigma }{\mathcal{F}}%
^{\Lambda \rho \sigma })
\end{equation}%
In (\ref{tca}), the $\mathcal{N}_{\Lambda \Sigma }$ is the period matrix
that we will define later on, see (\ref{permat}); the metric components $g_{%
\hat{a}\hat{b}}$, $A_{\hat{a}}^{\underline{m}}$, $M^{\underline{m}\underline{%
n}}$ are as in (\ref{dmet}) and the $H_{\nu \rho \sigma }^{\underline{m}}$'s
are the field strength tensors of the $B_{\mu \nu }^{\underline{m}}$'s given
by (\ref{hb}). As we will show in \textrm{\autoref{sec5}}, the scalar
potential reads \cite{Andrianopoli:2015rpa}%
\begin{equation}
\mathcal{V}_{scal}^{\mathcal{N}=2}=\left( U_{i}^{M}\mathcal{G}^{i\bar{j}}%
\bar{U}_{\bar{j}}^{N}\right) \mathcal{P}_{M}^{a}\mathcal{P}_{N}^{a}-V^{M}%
\bar{V}^{N}\mathcal{P}_{M}^{a}\mathcal{P}_{N}^{a}  \label{vscc}
\end{equation}%
where $V^{M}$ is the covariantly holomorphic section on $\boldsymbol{M}_{SK}$
with metric $\mathcal{G}_{i\bar{j}}$, $U_{i}^{M}$ is the covariant
derivative of $V^{M}$ and $\mathcal{P}_{M}^{a}$ are moment maps on $%
\boldsymbol{M}_{QK}$; all these objects and their rigid expressions will
described with details later on.

\subsection{Isometries and Ward identities}

Under the gauging of isometries of the scalar manifold $\boldsymbol{M}%
_{scal}=\boldsymbol{M}_{SK}\times \boldsymbol{M}_{QK}$ of $\mathcal{N}=2$
supergravity, the usual supersymmetric transformations $\delta _{\epsilon
}^{\left( 0\right) }\chi $ of the fermion fields $\chi =\left( \lambda
^{iA},\zeta ^{\gamma },\psi _{A\mu }\right) $ of the theory get extra
contributions $\delta _{\epsilon }^{\left( \vartheta \right) }\chi $
required by supersymmetric invariance. These extra transformations read in
terms of the supersymmetric parameters $\epsilon ^{B}$ and auxiliary
functional fields $\left( W^{i}\right) _{B}^{A}$, $N_{A}^{\gamma }$ and $%
S_{B}^{A}$ as follows%
\begin{equation}
\begin{tabular}{lll}
$\delta _{\epsilon }^{\left( \vartheta \right) }\lambda ^{iA}$ & $=$ & $%
\left( W^{i}\right) _{B}^{A}\epsilon ^{B}$ \\ 
$\delta _{\epsilon }^{\left( \vartheta \right) }\zeta ^{\gamma }$ & $=$ & $%
N_{A}^{\gamma }\epsilon ^{A}$ \\ 
$\delta _{\epsilon }^{\left( \vartheta \right) }\psi _{A\mu }$ & $=$ & $%
i\gamma _{\mu }\epsilon ^{B}S_{B}^{A}$%
\end{tabular}
\label{stt}
\end{equation}%
The various fermion shift matrices $\left( W^{i}\right) _{B}^{A}$, $%
N_{A}^{\gamma }$ and $S_{B}^{A}$ involved in these transformations are not
arbitrary; they are related to each others by Ward identities given by the 2$%
\times $2 matrix equation \textrm{\cite{Ferrara:1985gj,Cecotti:1985mx}}%
\begin{equation}
\mathcal{G}_{i\bar{j}}\left( W^{i}\right) _{C}^{A}(\bar{W}^{\bar{j}%
})_{B}^{C}+2\bar{N}_{\alpha }^{A}N_{B}^{\alpha }-12\bar{S}%
_{C}^{A}S_{B}^{C}=\delta _{B}^{A}\mathcal{V}_{\text{sugra}}  \label{wid1}
\end{equation}%
where $\mathcal{V}_{\text{sugra}}$ is the scalar potential of the
supergravity theory; this formula may be naively compared to the usual
expression of the scalar potential $\left \vert F\right \vert ^{2}+\frac{1}{2%
}D^{2}$\ of global supersymmetry where F and D are the auxiliary fields of
the $\mathcal{N}=2$ U$\left( 1\right) $ vector multiplet in the Wess- Zumino
gauge. Because of the particular form of the right hand side of above matrix
equation, these identities can be split by help of SU$\left( 2\right) $
R-symmetry representations as follows 
\begin{equation}
\begin{tabular}{lll}
$Tr\left( \tau ^{a}\left[ \mathcal{G}_{i\bar{j}}\left( W^{i}\right) _{C}^{A}(%
\bar{W}^{\bar{j}})_{B}^{C}+2\bar{N}_{\alpha }^{A}N_{B}^{\alpha }-12\bar{S}%
_{C}^{A}S_{B}^{C}\right] \right) $ & $=$ & $0$ \\ 
$Tr\left( \left[ \mathcal{G}_{i\bar{j}}\left( W^{i}\right) _{C}^{A}(\bar{W}^{%
\bar{j}})_{B}^{C}+2\bar{N}_{\alpha }^{A}N_{B}^{\alpha }-12\bar{S}%
_{C}^{A}S_{B}^{C}\right] \right) $ & $=$ & $2\mathcal{V}_{\text{sugra}}$%
\end{tabular}
\label{cst}
\end{equation}%
The projection along the isosinglet dimension gives the induced scalar
potential $\mathcal{V}_{\text{sugra}}$ in $\mathcal{N}=2$ gauged
supergravity and whose rigid limit is given \textrm{\autoref{ssec52}}. If we
restrict ourselves to the case of gauging only isometries of the
quaternionic Kahler manifold $\boldsymbol{M}_{QK}\subset \boldsymbol{M}%
_{scal}$, the constraint eqs(\ref{cst}) are solved in terms of Killing
vectors $\kappa _{M}^{u}$, isovectors of moment maps $\mathcal{P}%
_{M}^{a}=\left( \mathcal{P}_{M}^{x},\mathcal{P}_{M}^{y},\mathcal{P}%
_{M}^{z}\right) $ and quaternionic vielbeins $\mathcal{E}_{u\alpha }^{A}$ as
follows \textrm{\cite{D'Auria:1990fj}}%
\begin{equation}
\begin{tabular}{lll}
$\left( W^{i}\right) _{B}^{A}$ & $=$ & $-i\mathcal{G}^{i\bar{j}}\bar{U}_{%
\bar{j}}^{M}\left( \mathcal{P}_{M}^{a}\tau _{a}\right) _{B}^{A}$ \\ 
$N_{\alpha }^{A}$ & $=$ & $2\mathcal{E}_{u\alpha }^{A}\kappa _{M}^{u}\bar{V}%
^{M}$ \\ 
$S_{AB}$ & $=$ & $\frac{i}{2}\left( \tau _{a}\right) _{A}^{C}\varepsilon
_{BC}\mathcal{P}_{M}^{a}V^{M}$%
\end{tabular}
\label{wb1}
\end{equation}%
with complex conjugates as%
\begin{equation}
\begin{tabular}{lll}
$\left( \bar{W}^{\bar{j}}\right) _{BC}$ & $=$ & $i\left( \tau _{a}\right)
_{B}^{D}\varepsilon _{DC}\mathcal{P}_{M}^{a}\mathcal{G}^{i\bar{j}}U_{i}^{M}$
\\ 
$\bar{N}_{A}^{\alpha }$ & $=$ & $-2\mathcal{E}_{uA}^{\alpha }\kappa
_{M}^{u}V^{M}$ \\ 
$\bar{S}^{AB}$ & $=$ & $\frac{i}{2}\left( \tau _{a}\right)
_{C}^{A}\varepsilon ^{CB}\mathcal{P}_{M}^{a}\bar{V}^{M}$%
\end{tabular}%
\end{equation}%
In this solution, $\mathcal{E}_{u\alpha }^{A}$ is the vielbein of the
quaternionic Kahler manifold $\boldsymbol{M}_{QK}$; see eq(\ref{14}) and eqs(%
\ref{mf}-\ref{fm}) to fix the ideas.

\section{ Kahler sector in $\mathcal{N}=2$ supergravity}

\label{sec3}

In this section, we develop the study of the rigid limit of $\mathcal{N}=2$
supergravity and completes partial results obtained in \textrm{\cite%
{Andrianopoli:2015rpa}.} In particular, we give the explicit derivation of
the rigid limits of the symplectic sections $\Omega ^{M},V^{M}$ and $%
U_{i}^{M}$, living in the Kahler sector of $\mathcal{N}=2$ supergravity, as
well as the $\mathcal{N}_{\Lambda \Sigma }$ and $\mathcal{U}^{MN}$ tensors
involved in eqs(\ref{act},\ref{tca}) and (\ref{vscc}). These rigid limits,
which are needed for the computation of the scalar potential $\mathcal{V}%
_{kah}^{\mathcal{N}=2}$ and the matrix anomaly $\mathring{C}_{B}^{A}$ in the
observable sector of the gauged supergravity, were first considered in 
\textrm{\cite{Ferrara:1995xi}} for the case of one $\mathcal{N}=2$ vector
supermultiplet $\boldsymbol{V}_{\mathcal{N}=2}$; an extension of \textrm{%
\cite{Ferrara:1995xi}} to an arbitrary number $n_{V}$ of $\mathcal{N}=2$
vector supermultiplets $\left \{ \boldsymbol{V}_{\mathcal{N}=2}^{i}\right \}
_{1\leq i\leq n}$ was studied in \textrm{\cite{David:2003dh}}, and a new
extension, using extra parameters $\eta _{i}$ {\normalsize interpreted as
charges} associated with the gauging procedure, has been proposed recently
in \textrm{\cite{Andrianopoli:2015rpa}}. \newline
In the present study, the aforementioned $\eta _{i}$'s will be also given a
geometric interpretation; they are the rigid limit of the gauge connection
components $\omega _{i}^{0}\left( z,\bar{z}\right) $ of the U$\left(
1\right) $ bundle on the special Kahler manifold $\boldsymbol{M}_{SK}$. We
also show that the splitting of the $V^{M}$ covariantly holomorphic sections
on the special Kahler manifold as the sum $V_{grav}^{M}+V_{rigid}^{M}$ can
be achieved by using two successive and particular symmetry transformations
namely a particular symplectic change mapping $V^{M}$ to $V^{\prime M}$
followed by a particular Kahler transformation mapping $V^{\prime M}$ to $%
V^{\prime \prime M}$. In both these transformations, the $\eta _{i}$
parameters \textrm{play an important role.}

\subsection{$\frac{1}{\mathrm{\protect \mu }}$- expansion of symplectic
sections}

\label{chs}

We begin by recalling that in special Kahler geometry of the Coulomb branch
of $\mathcal{N}=2$ supergravity, one distinguishes three kinds of symplectic
sections $\Omega ^{M},V^{M}$ and $U_{i}^{M}$ related to each others: First,
we have the holomorphic $\Omega ^{M}$ satisfying $\frac{\partial \Omega ^{M}%
}{\partial \bar{z}^{\bar{\imath}}}=0$ with complex field coordinates $z^{i}$
as in eq(\ref{zi}). Second, we also have the covariantly holomorphic
sections $V^{M}$ related to $\Omega ^{M}$ as follows 
\begin{equation}
V^{M}=e^{\frac{\mathcal{K}}{2}}\Omega ^{M}\qquad ,\qquad \bar{\nabla}_{\bar{%
\imath}}V^{M}=0  \label{mv}
\end{equation}%
with $\mathcal{K}=\mathcal{K}\left( z,\bar{z}\right) $ standing for the
special Kahler potential obeying the usual Kahler transformation symmetry 
\begin{equation}
\mathcal{K}^{\prime }\left( z,\bar{z}\right) =\mathcal{K}\left( z,\bar{z}%
\right) -f\left( z\right) -\bar{f}\left( \bar{z}\right)  \label{kt}
\end{equation}%
where $f\left( z\right) $ is an arbitrary holomorphic function. $\bar{\nabla}%
_{\bar{\imath}}$ is the covariant derivative induced by the Kahler
transformation, it is given $\bar{\nabla}_{\bar{k}}=\frac{\partial }{%
\partial \bar{z}^{k}}-\frac{1}{2}\frac{\partial \mathcal{K}}{\partial \bar{z}%
^{\bar{k}}}$. Both $\Omega ^{M}$ and $V^{M}$ are symplectic vector of SP$%
\left( 2n+2\right) $. Third, we have as well the covariant gradient $%
U_{i}^{M}$ defined as 
\begin{equation}
U_{i}^{M}=\nabla _{i}V^{M}\qquad ,\qquad \nabla _{i}=\frac{\partial }{%
\partial z^{k}}+\frac{1}{2}\frac{\partial \mathcal{K}}{\partial z^{k}}
\label{mi}
\end{equation}%
Our objective here is to determine the $\frac{1}{\mathrm{\mu }}$- expansions
of these sections; for that purpose, we need to introduce extra tools like
the Kahler 1-form $\omega _{1}^{0}$ and the holomorphic prepotential $%
\mathcal{F}$ as described below.

\subsubsection{Kahler 1-form $\protect \omega _{1}^{0}$ and holomorphic
prepotential $\mathcal{F}$}

The covariant derivatives $\nabla _{i}$ and $\bar{\nabla}_{\bar{\imath}}$ in
eqs(\ref{mv},\ref{mi}) are associated with the 1-form gauge connection 
\begin{equation}
\omega _{1}^{0}=\frac{i}{2}\left( \partial \mathcal{K}-\bar{\partial}%
\mathcal{K}\right)
\end{equation}%
of the U$\left( 1\right) $ bundle living on $\boldsymbol{M}_{SK}$; here $%
\partial =dz^{k}\frac{\partial }{\partial z^{k}}$ and $\bar{\partial}=d\bar{z%
}^{k}\frac{\partial }{\partial \bar{z}^{\bar{k}}}$. This hermitian 1-form is
an isosinglet of SU$\left( 2\right) _{R}$; and the index $0$ on the
hermitian $\omega _{1}^{0}$ is to distinguish it from an analogous
isotriplet of 1-forms denoted like 
\begin{equation}
\omega _{1}^{a}=\left( 
\begin{array}{c}
\omega _{1}^{x} \\ 
\omega _{1}^{y} \\ 
\omega _{1}^{z}%
\end{array}%
\right)
\end{equation}%
and which will be introduced later on when studying the geometry of the
quaternionic Kahler manifold $\boldsymbol{M}_{QK}$ of hypermultiplets. By
expressing the 1-form $\omega _{1}^{0}$ as%
\begin{equation}
\omega _{1}^{0}=i\left( dz^{k}\omega _{k}^{0}-d\bar{z}^{\bar{k}}\bar{\omega}%
_{\bar{k}}^{0}\right)  \label{gc}
\end{equation}%
we have%
\begin{equation}
\omega _{k}^{0}=\frac{1}{2}\frac{\partial \mathcal{K}}{\partial z^{k}}\qquad
,\qquad \bar{\omega}_{\bar{k}}^{0}=\frac{1}{2}\frac{\partial \mathcal{K}}{%
\partial \bar{z}^{\bar{k}}}
\end{equation}%
Notice that under the Kahler transformation (\ref{kt}), the $\omega _{k}^{0}$
and $\bar{\omega}_{\bar{k}}^{0}$ transform like%
\begin{equation}
\omega _{k}^{\prime 0}=\omega _{k}^{0}-\frac{1}{2}\frac{\partial f}{\partial
z^{k}}\qquad ,\qquad \bar{\omega}_{\bar{k}}^{\prime 0}=\bar{\omega}_{\bar{k}%
}^{0}-\frac{1}{2}\frac{\partial \bar{f}}{\partial \bar{z}^{\bar{k}}}
\label{tg}
\end{equation}%
Following \textrm{\cite{Andrianopoli:2015rpa}}, the rigid limit of $\mathcal{%
N}=2$ supergravity, in particular of $\Omega ^{M},$ $V^{M},$ $U_{i}^{M}$ and 
$\omega _{1}^{0}$, can be approached by considering the holomorphic
prepotential $\boldsymbol{F}\left( z\right) $ of the $\mathcal{N}=2$ theory
and using perturbation techniques with a perturbation parameter \underline{$%
\lambda $}$=\frac{1}{\mathrm{\mu }}$. The general procedure is as summarised
below:

\begin{itemize}
\item Introduce a real expansion parameter $\frac{1}{\mathrm{\mu }}<<1$
where the dimensionless $\mathrm{\mu }$ is given by the reduced mass scale $%
\mathrm{\mu }=\frac{M_{pl}}{\Lambda }$; with $M_{pl}$ is the Planck scale
and $\Lambda $ a mass scale of order of the supersymmetric breaking scale;
say the mass $m_{3/2}$ of a $\mathcal{N}=1$ \emph{massive} gravitino
supermultiplet $\boldsymbol{G}_{\frac{3}{2}}^{\mathcal{N}=1}$ with spins as 
\begin{equation}
\boldsymbol{G}_{\frac{3}{2}}^{\mathcal{N}=1}=\left( \frac{3}{2},1^{2},\frac{1%
}{2}\right)
\end{equation}%
The use of this supermultiplet is motivated by partial supersymmetry
breaking assumed to take place at scale $\Lambda $.

\item Expand the usual holomorphic prepotential $\mathcal{F}\left( X\right) $
of the $U\left( 1\right) ^{n}$ Coulomb branch in power series of $\frac{1}{%
\mathrm{\mu }}$. In this regards, recall that the holomorphic $\mathcal{F}%
\left( X\right) $ is homogeneous function of degree 2; it can be factorised
like 
\begin{equation}
\mathcal{F}\left( X\right) =-i\left( X^{0}\right) ^{2}\boldsymbol{F}\left( 
\frac{X^{I}}{X^{0}}\right)  \label{10}
\end{equation}
\end{itemize}

Thinking of the $X$- homogenous coordinates as $X^{\Lambda }=\left(
X^{0},X^{I}\right) $ and working in the special frame where $\frac{X^{I}}{%
X^{0}}=\delta _{i}^{I}z^{i}$ (conveniently by setting $X^{0}=1)$, a general
form of the expansion of $\boldsymbol{F}(X^{I}/X^{0})$ into $\frac{1}{%
\mathrm{\mu }}$ power series reads, up to third order, as follows 
\begin{equation}
\boldsymbol{F}\left( z\right) =\frac{1}{4}+\frac{1}{2\mathrm{\mu }}%
\sum_{i=1}^{n}\eta _{i}z^{i}+\frac{1}{2\mathrm{\mu }^{2}}\Phi \left(
z\right) +\mathfrak{O}\left( \frac{1}{\mathrm{\mu }^{3}}\right)  \label{ef}
\end{equation}%
\textrm{This parametrisation choice of }$\boldsymbol{F}\left( z\right) $ 
\textrm{reproduces the model of Antoniadis- Partouche-Taylor; a simple
version \ of }$\boldsymbol{F}\left( z\right) $\textrm{\ was first introduced
by Ferrara- Girardelo- Poratti for a single vector multiplet \cite%
{Ferrara:1995xi} and generalized to the form (\ref{ef}) by Andranopoli \emph{%
et al} for the case of several multiplets. }In this development, the
constant parameters $\eta _{1},...,\eta _{n}$ are real numbers interpreted
in \textrm{\cite{Andrianopoli:2015rpa}} as charges of gauged abelian
isometries of the scalar manifold of $\mathcal{N}=2$ supergravity; the
reality condition of the $\eta _{i}$'s is a requirement of special Kahler
geometry. \textrm{For later use, notice that the terms }$\eta _{i}\times
z^{i}$\textrm{\ in above expansion is scale invariant as it behaves l}ike
the leading constant $\frac{1}{4}$ in (\ref{ef}); so the $\eta _{i}$
parameters can be also interpreted as a kind of dual parameters to $z^{i}$
in a similar way as space time coordinates $x^{\mu }$ and energy momentum
vector $P_{\mu }$. Notice also that in addition to the complex field
variables $z^{i}$, the prepotential $\boldsymbol{F}\left( z\right) $ depends
as well on the $\eta _{i}$'s and the perturbation parameter $\frac{1}{%
\mathrm{\mu }}$; i.e: $\boldsymbol{F}\left( z\right) \equiv \boldsymbol{F}%
\left( z,\eta ,\frac{1}{\mathrm{\mu }}\right) $. These extra parameters will
be often hidden below.

\subsubsection{Special Kahler potential $\mathcal{K}$}

Using the above $\frac{1}{\mathrm{\mu }}$- expansion (\ref{ef}) of the
holomorphic prepotential $\mathcal{F}\left( X\right) $, one can write down
the $\frac{1}{\mathrm{\mu }}$- development of the holomorphic sections $%
\Omega ^{M}=\Omega ^{M}\left( z\right) $ living on the special Kahler
subspace $\boldsymbol{M}_{SK}$ of the scalar manifold $\boldsymbol{M}%
_{SK}\times \boldsymbol{M}_{QK}$ of $\mathcal{N}=2$ supergravity. By using (%
\ref{10}), we can express the symplectic vector $\Omega ^{M}$ as follows%
\begin{equation}
\Omega ^{M}=\left( 
\begin{array}{c}
X^{\Lambda } \\ 
F_{\Lambda }%
\end{array}%
\right)  \label{rg}
\end{equation}%
with%
\begin{equation}
X^{\Lambda }=\left( 
\begin{array}{c}
1 \\ 
z^{k}%
\end{array}%
\right) \qquad ,\qquad F_{\Lambda }=\left( 
\begin{array}{c}
F_{0} \\ 
F_{k}%
\end{array}%
\right)  \label{rh}
\end{equation}%
and 
\begin{equation}
\begin{tabular}{lll}
$\mathcal{F}_{0}$ & $=$ & $\frac{1}{i}\left( 2\boldsymbol{F}-z^{k}\partial
_{k}\boldsymbol{F}\right) $ \\ 
$\mathcal{F}_{k}$ & $=$ & $\frac{1}{i}\partial _{k}\boldsymbol{F}$%
\end{tabular}%
\end{equation}%
then using (\ref{ef}), we obtain the following behaviour up to third order
in $\frac{1}{\mathrm{\mu }},$%
\begin{equation}
\begin{tabular}{lll}
$\mathcal{F}_{0}$ & $=$ & $\frac{1}{i}\left[ \frac{1}{2}+\frac{1}{2\mathrm{%
\mu }}\eta _{k}z^{k}+\frac{1}{\mathrm{\mu }^{2}}\left( \Phi -\frac{1}{2}%
z^{k}\partial _{k}\Phi \right) \right] +\mathfrak{O}\left( \frac{1}{\mathrm{%
\mu }^{3}}\right) $ \\ 
$\mathcal{F}_{k}$ & $=$ & $\frac{1}{i}\left( \frac{\eta _{k}}{2\mathrm{\mu }}%
+\frac{1}{2\mathrm{\mu }^{2}}\partial _{k}\Phi \right) +\mathfrak{O}\left( 
\frac{1}{\mathrm{\mu }^{3}}\right) $%
\end{tabular}
\label{f}
\end{equation}%
Substituting these expansions back into (\ref{rg}), we can split the
holomorphic $\Omega ^{M}$ as a sum of a power series of $\frac{1}{\mathrm{%
\mu }}$ with leading term corresponding to the limit $\frac{1}{\mathrm{\mu }}%
\rightarrow 0$. Expressing this expansion like 
\begin{equation}
\Omega ^{M}=\Omega _{\left( 0\right) }^{M}+\frac{1}{\mathrm{\mu }}\Omega
_{\left( 1\right) }^{M}+\frac{1}{\mathrm{\mu }^{2}}\Omega _{\left( 2\right)
}^{M}+\mathfrak{O}(\frac{1}{\mathrm{\mu }^{3}})
\end{equation}%
we learn that 
\begin{equation}
\Omega _{\left( 0\right) }^{M}=\left( 
\begin{array}{c}
1 \\ 
z^{i} \\ 
\frac{1}{2i} \\ 
0%
\end{array}%
\right) \quad ,\quad \Omega _{\left( 1\right) }^{M}=\left( 
\begin{array}{c}
0 \\ 
0 \\ 
\frac{\eta _{i}z^{i}}{2i} \\ 
\frac{\eta _{i}}{2i}%
\end{array}%
\right) \quad ,\quad \Omega _{\left( 2\right) }^{M}=\left( 
\begin{array}{c}
0 \\ 
0 \\ 
\frac{2\Phi -z^{i}\partial _{i}\Phi }{2i} \\ 
\frac{\partial _{i}\Phi }{2i}%
\end{array}%
\right)  \label{in}
\end{equation}%
But this splitting is not interesting for our analysis as it does not
separate the pure gravity sector, associated with graviphoton direction,
from the rigid limit of the supergravity theory. To overcome this
difficulty, we use the symmetries of the supergravity theory; in particular
the symplectic and Kahler invariance, to sit in a particular coordinate
frame on $\boldsymbol{M}_{SK}$ where gravity and the rigid limit appear as
two "orthogonal" sectors at the first order of the limit $\frac{1}{\mathrm{%
\mu }}\rightarrow 0$. Indeed, as the holomorphic section $\Omega ^{M}$ is
not unique since it is defined up to a symplectic transformation 
\begin{equation}
\Omega ^{M}\rightarrow \Omega ^{\prime M}=\mathbb{S}_{N}^{M}\Omega ^{N}
\label{so}
\end{equation}%
one can work with $\Omega ^{\prime M}$ instead of $\Omega ^{M}$ without
affecting physical properties. In this regards, recall that the above
symplectic transformation leaves invariant the special Kahler potential of
the $\boldsymbol{M}_{SK}$ 
\begin{equation}
\mathcal{K}=-\ln \left[ -i\Omega ^{M}\mathcal{C}_{MN}\bar{\Omega}^{N}\right]
\end{equation}%
Recall also that the matrix $\mathbb{S}_{N}^{M}$ in (\ref{so}) is a real
symplectic $\left( 2n+2\right) \times \left( 2n+2\right) $ matrix satisfying
the usual property $S_{P}^{M}\mathcal{C}^{PQ}S_{Q}^{N}=\mathcal{C}^{MN}$ and
can be defined in terms of four submatrices as follows%
\begin{equation}
\mathbb{S}=%
\begin{pmatrix}
\mathbb{A} & \mathbb{B} \\ 
\mathbb{C} & \mathbb{D}%
\end{pmatrix}
\label{sym}
\end{equation}%
Here the antisymmetric $\mathcal{C}^{MN}$ is the symplectic metric of SP$%
\left( 2n+2\right) $, and the $\mathbb{A},$ $\mathbb{B}$, $\mathbb{C}$ and $%
\mathbb{D}$ are $\left( n+1\right) \times \left( n+1\right) $ submatrices
related to each others like 
\begin{equation}
\mathbb{A}^{T}\mathbb{D-C}^{T}\mathbb{B}=\mathbb{I}_{n+1}\hspace{0.6cm},%
\hspace{0.6cm}\mathbb{A}^{T}\mathbb{C=C}^{T}\mathbb{A}\hspace{0.6cm},\hspace{%
0.6cm}\mathbb{B}^{T}\mathbb{D=D}^{T}\mathbb{B}
\end{equation}%
By using (\ref{rg}) and substituting the expressions (\ref{rh}-\ref{f}) back
into the special Kahler potential $\mathcal{K}$ given by 
\begin{equation}
\mathcal{K}=-\ln \left( i\left[ \bar{X}^{\Lambda }F_{\Lambda }-X^{\Lambda }%
\bar{F}_{\Lambda }\right] \right)  \label{kp}
\end{equation}%
we get the following $\frac{1}{\mathrm{\mu }}$- development of the special
Kahler potential 
\begin{equation}
\mathcal{K}=-\ln \left[ 1+\frac{1}{\mathrm{\mu }}\eta _{i}\left( z^{i}+\bar{z%
}^{i}\right) +\frac{1}{\mathrm{\mu }^{2}}\left( \Phi +\bar{\Phi}\right) -%
\frac{1}{2\mathrm{\mu }^{2}}\left( z-\bar{z}\right) ^{i}\left( \partial
_{i}\Phi -\bar{\partial}_{i}\bar{\Phi}\right) +\mathfrak{O}\left( \frac{1}{%
\mathrm{\mu }^{3}}\right) \right]  \label{k}
\end{equation}%
Expanding the logarithm up to third order in $\frac{1}{\mathrm{\mu }}$, we
end with the following expression%
\begin{equation}
\begin{tabular}{lll}
$\mathcal{K}$ & $=$ & $-\frac{1}{\mathrm{\mu }}\eta _{i}\left( z^{i}+\bar{z}%
^{i}\right) $ \\ 
&  & $-\frac{1}{\mathrm{\mu }^{2}}\left[ \left( \Phi +\bar{\Phi}\right) -%
\frac{1}{2}\left( z-\bar{z}\right) ^{i}\left( \partial _{i}\Phi -\bar{%
\partial}_{i}\bar{\Phi}\right) -\frac{1}{2}\left( \eta _{i}z^{i}+\eta _{i}%
\bar{z}^{i}\right) ^{2}\right] $ \\ 
&  & $+\mathfrak{O}\left( \frac{1}{\mathrm{\mu }^{3}}\right) $%
\end{tabular}
\label{ka}
\end{equation}%
from which we learn several information; in particular the $\frac{1}{\mathrm{%
\mu }}$- expansion of the $U\left( 1\right) $ Kahler gauge components $%
\omega _{i}^{0}=\frac{1}{2}\partial _{i}\mathcal{K}$ and $\bar{\omega}_{\bar{%
j}}^{0}=\frac{1}{2}\partial _{\bar{j}}\mathcal{K}$ as well as the special
Kahler metric $\mathcal{G}_{i\bar{j}}=\partial _{i}\partial _{\bar{j}}%
\mathcal{K}$. For the gauge components we have%
\begin{equation}
\begin{tabular}{lll}
$\omega _{i}^{0}$ & $=$ & $-\frac{1}{2\mathrm{\mu }}\eta _{i}-\frac{1}{2%
\mathrm{\mu }^{2}}\mathfrak{a}_{i}+\mathfrak{O}\left( \frac{1}{\mathrm{\mu }%
^{3}}\right) $ \\ 
$\bar{\omega}_{\bar{j}}^{0}$ & $=$ & $-\frac{1}{2\mathrm{\mu }}\eta _{j}-%
\frac{1}{2\mathrm{\mu }^{2}}\mathfrak{\bar{a}}_{\bar{j}}+\mathfrak{O}\left( 
\frac{1}{\mathrm{\mu }^{3}}\right) $%
\end{tabular}
\label{om}
\end{equation}%
with 
\begin{equation}
\begin{tabular}{lll}
$\mathfrak{a}_{i}$ & $=$ & $\frac{1}{2}\partial _{i}\Phi -\frac{1}{2}\left(
z-\bar{z}\right) ^{k}\partial _{i}\partial _{k}\Phi -\eta _{i}\left( \eta
_{k}z^{k}+\eta _{k}\bar{z}^{k}\right) $ \\ 
$\mathfrak{\bar{a}}_{\bar{j}}$ & $=$ & $\frac{1}{2}\partial _{\bar{j}}\bar{%
\Phi}+\frac{1}{2}\left( z-\bar{z}\right) ^{k}\partial _{\bar{j}}\partial _{%
\bar{k}}\bar{\Phi}-\eta _{j}\left( \eta _{k}z^{k}+\eta _{k}\bar{z}%
^{k}\right) $%
\end{tabular}
\label{aa}
\end{equation}%
From the expansions (\ref{om}), we observe in the limit $\mathrm{\mu }%
\rightarrow \infty ,$ the Kahler 1-form $\omega _{1}^{0}$ (\ref{gc})
vanishes; and that the constants $\frac{1}{2}\eta _{k}$ appear as just the
leading components of the U$\left( 1\right) $ Kahler gauge fields $\omega
_{k}^{0}$ and $\bar{\omega}_{\bar{k}}^{0}$ in the $\frac{1}{\mathrm{\mu }}$
expansion; a property which can be stated like%
\begin{equation}
\eta _{k}=2\lim_{\mathrm{\mu }\rightarrow \infty }\left( \mathrm{\mu }\omega
_{k}^{0}\right) \qquad ,\qquad \bar{\eta}_{k}=2\lim_{\mathrm{\mu }%
\rightarrow \infty }\left( \mathrm{\mu }\bar{\omega}_{\bar{k}}^{0}\right)
\end{equation}%
So the $\eta _{k}$'s can be interpreted as just the zero modes in the
expansion of the components of the scaled 1-form connection $\left( \mathrm{%
\mu }\omega _{1}^{0}\right) $ on $\boldsymbol{M}_{SK}$,%
\begin{equation}
\mathrm{\mu }\omega _{1}^{0}=-\frac{i}{2}\eta _{k}\left( dz^{k}-d\bar{z}%
^{k}\right) +\mathfrak{O}\left( \frac{1}{\mathrm{\mu }}\right)
\end{equation}%
More comments on these $\eta _{k}$'s will be given later on; see eqs(\ref%
{344}). Regarding the Kahler metric, we have%
\begin{equation}
\mathcal{G}_{i\bar{j}}=\frac{1}{\mathrm{\mu }^{2}}\mathcal{\mathring{G}}_{i%
\bar{j}}+\mathfrak{O}\left( \frac{1}{\mathrm{\mu }^{3}}\right) \qquad
,\qquad \mathcal{\mathring{G}}_{i\bar{j}}=\eta _{i}\eta _{j}-\frac{1}{2}%
\left( \partial _{i}\partial _{j}\Phi +cc\right)  \label{met}
\end{equation}%
where $\mathcal{\mathring{G}}_{i\bar{j}}$ is the metric of the rigid special
Kahler manifold $\boldsymbol{\mathring{M}}_{SK}$; thus 
\begin{equation}
\mathcal{\mathring{G}}_{i\bar{j}}=\lim_{\mu \rightarrow \infty }\left( 
\mathrm{\mu }^{2}\mathcal{G}_{i\bar{j}}\right)
\end{equation}%
Following \textrm{\cite{Craps:1997gp, Fre:1995dw, Andrianopoli:1996cm}}, the
metric $\mathcal{G}_{i\bar{j}}$ is related to the covariant gradients $%
U_{i}^{M}=\nabla _{i}V^{M}$ and $\bar{U}_{\bar{\imath}}^{M}=\bar{\nabla}_{%
\bar{\imath}}\bar{V}^{M}$ as 
\begin{equation}
\mathcal{G}_{k\bar{l}}=-iU_{k}^{M}\mathcal{C}_{MN}\bar{U}_{\bar{l}}^{N}
\label{-1}
\end{equation}%
with the metric $\mathcal{C}_{MN}$ as before. In the rigid theory, we also
have 
\begin{equation}
\mathcal{\mathring{G}}_{k\bar{l}}=-i\mathring{U}_{k}^{\underline{M}}\mathcal{%
C}_{\underline{M}\underline{N}}\overline{\mathring{U}}_{\bar{l}}^{\underline{%
N}}  \label{0}
\end{equation}%
where $\mathcal{C}_{\underline{M}\underline{N}}$ is now the metric of the SP$%
\left( 2n\right) $ symmetry along the Coulomb branch directions. Its
underlined indices $\underline{M}$, $\underline{N}$ run from 1 to n; and
where 
\begin{equation}
\mathring{U}_{k}^{\underline{M}}=\delta _{M}^{\underline{M}}\frac{\partial 
\mathring{\Omega}^{M}}{\partial z^{k}}
\end{equation}%
with $\mathring{\Omega}^{M}$ the holomorphic section of the rigid theory;
see eq(\ref{rl}) for its explicit expression. Moreover, by thinking of $%
\mathcal{G}_{k\bar{l}}$ of (\ref{-1}) in terms of the approximation (\ref%
{met}), we learn that the $\mathring{U}_{k}^{\underline{M}}$ is nothing but
the leading term in the $\frac{1}{\mathrm{\mu }}$- expansion of $U_{k}^{M}$.
Notice that in eq(\ref{0}), we have used the convention notation $\Delta ^{%
\underline{M}}=(\delta ^{i},\tilde{\delta}_{i})$ for SP$\left( 2n\right) $
vectors with symplectic metric 
\begin{equation}
\mathcal{C}_{\underline{M}\underline{N}}=\delta _{\underline{M}}^{M}\mathcal{%
C}_{MN}\delta _{\underline{N}}^{N}
\end{equation}%
Recall as well that for SP$\left( 2n+2\right) $ of the special Kahler
geometry of $\boldsymbol{M}_{SK}$, the metric is denoted like $\mathcal{C}%
_{MN}$ and vectors 
\begin{equation}
\Delta ^{M}=(\delta ^{\Lambda },\tilde{\delta}_{\Lambda })
\end{equation}%
have $2n+2$ components with index $\Lambda =\left( 0,i\right) $ and $%
i=1,...,n$. To complete our notations, we denote by 
\begin{equation}
\Delta ^{\underline{\tau }}=(\delta ^{0},\tilde{\delta}_{0})
\end{equation}%
the SP$\left( 2\right) $ vectors along the graviphoton direction with 2$%
\times $2 symplectic metric as $\mathcal{C}_{\underline{\tau }\underline{%
\sigma }}=\delta _{\underline{\tau }}^{M}\mathcal{C}_{MN}\delta _{\underline{%
\sigma }}^{N}$. With these notations, an SP$\left( 2n+2\right) $ vector like
the holomorphic section $\Omega ^{M}$ can be written as%
\begin{equation}
\Omega ^{M}=\delta _{\underline{\tau }}^{M}\Omega ^{\underline{\tau }%
}+\delta _{\underline{N}}^{M}\Omega ^{\underline{N}}
\end{equation}%
with 
\begin{equation}
\delta _{\underline{\tau }}^{M}\Omega ^{\underline{\tau }}=\left( 
\begin{array}{c}
1 \\ 
0^{k} \\ 
F_{0} \\ 
0_{k}%
\end{array}%
\right) \qquad ,\qquad \delta _{\underline{N}}^{M}\Omega ^{\underline{N}%
}=\left( 
\begin{array}{c}
0 \\ 
z^{k} \\ 
0 \\ 
F_{k}%
\end{array}%
\right)  \label{tn}
\end{equation}

\subsection{Rigid limit of symplectic sections}

Here we give the rigid limits of several useful quantities; in particular
the holomorphic sections $\Omega ^{M}$ on the special Kahler manifold $%
\boldsymbol{M}_{SK}$, the corresponding covariantly holomorphic $V^{M}$ and
their covariant gradient $U_{i}^{M}=\left( \partial _{i}+\frac{1}{2}\partial
_{i}\mathcal{K}\right) V^{M}$. We also study the splitting of pure gravity
contribution of these quantities from their rigid limits viewed from the
side of the Coulomb branch.

\  \  \  \ 

$\bullet $ \emph{Rigid limit of holomorphic section }$\Omega ^{M}$\newline
First notice that in eq(\ref{met}), the rigid metric $\mathcal{\mathring{G}}%
_{i\bar{j}}$ appears as the leading term of the $\frac{1}{\mathrm{\mu }}$-
expansion of the special Kahler metric $\mathcal{G}_{i\bar{j}}$ of $%
\boldsymbol{M}_{SK}$ subspace of the scalar manifold of $\mathcal{N}=2$
supergravity. The limit $\mathcal{\mathring{G}}_{i\bar{j}}$ can be thought
of as descending from the following "rigid" holomorphic prepotential%
\begin{equation}
\mathcal{\mathring{F}}=\frac{i}{4}\left[ \left( \eta _{i}z^{i}\right)
^{2}-2\Phi \right]
\end{equation}%
By calculating the second derivative of above expression, we have%
\begin{equation}
\partial _{i}\partial _{j}\mathcal{\mathring{F}}=\frac{i}{2}\left( \eta
_{i}\eta _{j}-\partial _{j}\partial _{i}\Phi \right)
\end{equation}%
from which we read the real and imaginary parts which are given by 
\begin{equation}
\begin{tabular}{lll}
$\func{Im}\partial _{i}\partial _{j}\mathcal{\mathring{F}}$ & $=$ & $\frac{1%
}{2}\mathcal{\mathring{G}}_{i\bar{j}}$ \\ 
$\func{Re}\partial _{i}\partial _{j}\mathcal{\mathring{F}}$ & $=$ & $-\frac{i%
}{4}\left( \partial _{i}\partial _{j}\Phi -cc\right) $%
\end{tabular}%
\end{equation}%
The holomorphic section $\mathring{\Omega}^{M}$ of the rigid limit of the $%
\mathcal{N}=2$ supergravity theory can be then imagined as given by 
\begin{equation}
\mathring{\Omega}^{M}=\left( 
\begin{array}{c}
0 \\ 
z^{i} \\ 
0 \\ 
\frac{i}{2}\left( \eta _{i}\eta _{j}z^{j}-\partial _{i}\Phi \right)%
\end{array}%
\right)  \label{rl}
\end{equation}%
with no $\mathring{X}^{0}$ nor $\mathring{F}_{0}$\ components compared to (%
\ref{rg}); so the 2n components $\mathring{\Omega}^{M}$'s behave as an $%
SP\left( 2n\right) $ symplectic vector $\Omega ^{\underline{M}}$ as in (\ref%
{tn}) in contrast to the $\Omega ^{M}$ of eq(\ref{rg}) which form a vector
of $SP\left( 2n+2\right) $. In other words, we have $\mathring{\Omega}%
^{M}=\delta _{\underline{M}}^{M}\mathring{\Omega}^{\underline{M}}$ with%
\begin{equation}
\mathring{\Omega}^{\underline{M}}=\left( 
\begin{array}{c}
z^{i} \\ 
\frac{i}{2}\left( \eta _{i}\eta _{j}z^{j}-\partial _{i}\Phi \right)%
\end{array}%
\right) \qquad ,\qquad i=1,...,n
\end{equation}%
This reduction of the symplectic symmetry in the rigid limit of $\mathcal{N}%
=2$ supergravity is a remarkable property; it will be studied in detail in a
moment; before that notice that $\mathring{U}_{k}^{\underline{M}}=\delta
_{M}^{\underline{M}}\mathring{U}_{k}^{M}$ with 
\begin{equation}
\mathring{U}_{k}^{M}=\frac{\partial \mathring{\Omega}^{M}}{\partial z^{k}}
\end{equation}%
reads as%
\begin{equation}
\mathring{U}_{k}^{M}=\left( 
\begin{array}{c}
0 \\ 
\delta _{k}^{i} \\ 
0 \\ 
\frac{i}{2}\left( \eta _{i}\eta _{k}-\partial _{i}\partial _{k}\Phi \right)%
\end{array}%
\right)  \label{um}
\end{equation}%
showing that $\mathring{U}_{k}^{M}$ has no components along the graviphoton
direction; that is: 
\begin{equation}
\delta _{\underline{\tau }}^{M}\mathring{U}_{k}^{\underline{\tau }}=0
\end{equation}

$\bullet $ \emph{Rigid limit of }$V^{M}$ \emph{and }$U_{i}^{M}$\newline
The reduction of the $SP\left( 2n+2\right) $ symplectic symmetry down to the
subsymmetry $SP\left( 2n\right) $ of the rigid limit is an obvious property
as in the supergravity theory there are $n+1$ electric charges and $n+1$
magnetic ones; two of them are associated with the graviphoton $A_{\mu }^{0}$
living in the pure gravity sector; the other $2n$ ones have origin in the
Coulomb branch with gauge field potentials $A_{\mu }^{i}$. Though expected,
the derivation of this reduction property in the rigid limit is not a
trivial feature. As we will see below, it needs using symmetry properties of
the covariantly holomorphic section $V^{M}=e^{\frac{\mathcal{K}}{2}}\Omega
^{M}$ on $\boldsymbol{M}_{SK}$ which allow to put $V^{M}$ into the form 
\begin{equation}
V^{M}=V_{grav}^{M}+\frac{1}{\mathrm{\mu }}\mathring{\Omega}^{M}+\mathfrak{O}(%
\frac{1}{\mathrm{\mu }^{2}})  \label{rd}
\end{equation}%
with $V_{grav}^{M}$ given by%
\begin{equation}
V_{grav}^{M}=\left( 
\begin{array}{c}
X^{0} \\ 
0 \\ 
-\frac{i}{2}X^{0} \\ 
0%
\end{array}%
\right)  \label{vgra}
\end{equation}%
appearing as the zero mode of the $\frac{1}{\mathrm{\mu }}$- expansion and $%
\mathring{\Omega}^{M}$, which is given by (\ref{rl}), as the first order. In
other words%
\begin{equation}
\lim_{\mu \rightarrow \infty }V^{M}=V_{grav}^{M}\qquad ,\qquad \lim_{\mu
\rightarrow \infty }\left[ \mathrm{\mu }\left( V^{M}-V_{grav}^{M}\right) %
\right] =\mathring{\Omega}^{M}
\end{equation}%
The splitting (\ref{rd}) can be derived by performing some manipulations
using symmetry properties of $V^{M}$ and rigid limit approximation as
follows: \newline
First, we start from the holomorphic $\Omega ^{M}$ and perform two
(commuting) successive transformations: $\left( i\right) $ a particular
symplectic change $\tilde{\Omega}^{M}=\mathbb{S}_{N}^{M}\Omega ^{N}$ with
symplectic matrix $\mathbb{S}_{N}^{M}$ as in eq(\ref{sym}) leaving special
Kahler potential invariant; that is $\mathcal{\tilde{K}}=\mathcal{K}$; and $%
\left( ii\right) $ another particular Kahler transformation on $\tilde{\Omega%
}^{M}$ given by 
\begin{equation}
\tilde{\Omega}^{\prime M}=e^{f_{0}}\tilde{\Omega}^{M}  \label{k1}
\end{equation}%
with holomorphic parameter $f_{0}$ taken as 
\begin{equation}
f_{0}=-\frac{1}{\mathrm{\mu }}\eta _{i}z^{i}  \label{1k}
\end{equation}%
The above particular Kahler transformation acts as well non trivially on the 
$\mathcal{K}$ potential and on the components $\omega _{k}^{0},$ $\bar{\omega%
}_{\bar{k}}^{0}$ of the 1-form gauge connection $\omega _{1}$ like 
\begin{equation}
\begin{tabular}{lll}
$\mathcal{K}^{\prime }$ & $=$ & $\mathcal{K}+\frac{1}{\mathrm{\mu }}\left(
\eta _{i}z^{i}+\eta _{i}\bar{z}^{i}\right) $ \\ 
$\omega _{k}^{\prime 0}$ & $=$ & $\omega _{k}^{0}+\frac{1}{2\mathrm{\mu }}%
\eta _{k}$ \\ 
$\bar{\omega}_{\bar{k}}^{\prime 0}$ & $=$ & $\bar{\omega}_{\bar{k}}^{0}+%
\frac{1}{2\mathrm{\mu }}\eta _{k}$%
\end{tabular}%
\end{equation}%
which can be also rewritten like 
\begin{equation}
\begin{tabular}{lll}
$\frac{1}{\mathrm{\mu }}\left( \eta _{i}z^{i}+\eta _{i}\bar{z}^{i}\right) $
& $=$ & $\mathcal{K}^{\prime }-\mathcal{K}$ \\ 
$\frac{1}{2\mathrm{\mu }}\eta _{k}$ & $=$ & $\omega _{k}^{\prime 0}-\omega
_{k}^{0}$ \\ 
$\frac{1}{2\mathrm{\mu }}\eta _{k}$ & $=$ & $\bar{\omega}_{\bar{k}}^{\prime
0}-\bar{\omega}_{\bar{k}}^{0}$%
\end{tabular}
\label{344}
\end{equation}%
giving another way to think of the $\eta _{k}$ parameters and on the
quantity $\frac{1}{\mathrm{\mu }}\left( \eta _{i}z^{i}+\eta _{i}\bar{z}%
^{i}\right) $. The particular Kahler change leads therefore to the following
new quantities 
\begin{equation}
\mathcal{K}^{\prime }=-\frac{1}{\mathrm{\mu }^{2}}\left[ \left( \Phi +\bar{%
\Phi}\right) -\frac{1}{2}\left( z-\bar{z}\right) ^{i}\left( \partial
_{i}\Phi -\bar{\partial}_{i}\bar{\Phi}\right) -\frac{1}{2}\left( \eta
_{i}z^{i}+\eta _{i}\bar{z}^{i}\right) ^{2}\right] +\mathfrak{O}\left( \frac{1%
}{\mathrm{\mu }^{3}}\right)  \label{k3}
\end{equation}%
and%
\begin{equation}
\begin{tabular}{lll}
$\omega _{i}^{\prime 0}$ & $=$ & $-\frac{1}{2\mathrm{\mu }^{2}}\mathfrak{a}%
_{i}+\mathfrak{O}\left( \frac{1}{\mathrm{\mu }^{3}}\right) $ \\ 
$\bar{\omega}_{\bar{j}}^{\prime 0}$ & $=$ & $-\frac{1}{2\mathrm{\mu }^{2}}%
\mathfrak{\bar{a}}_{\bar{j}}+\mathfrak{O}\left( \frac{1}{\mathrm{\mu }^{3}}%
\right) $%
\end{tabular}%
\end{equation}%
showing that $\mathcal{K}^{\prime }$ and $\omega _{i}^{\prime 0},$ $\bar{%
\omega}_{\bar{j}}^{\prime 0}$ are of the order $\frac{1}{\mathrm{\mu }^{2}}$%
. Observe that $\mathcal{K}^{\prime }$ can be also expressed like 
\begin{equation}
\mathcal{K}^{\prime }=\frac{1}{\mathrm{\mu }^{2}}\mathcal{\mathring{K}}+%
\mathfrak{O}\left( \frac{1}{\mathrm{\mu }^{3}}\right)  \label{ke}
\end{equation}%
with leading term $\mathcal{\mathring{K}}$ in the $\frac{1}{\mathrm{\mu }}$-
expansion as%
\begin{equation}
\mathcal{\mathring{K}}=\left( \Phi +\bar{\Phi}\right) -\frac{1}{2}\left( z-%
\bar{z}\right) ^{i}\left( \partial _{i}\Phi -\bar{\partial}_{i}\bar{\Phi}%
\right) -\frac{1}{2}\left( \eta _{i}z^{i}+\eta _{i}\bar{z}^{i}\right) ^{2}
\end{equation}%
This term $\mathcal{\mathring{K}}$ is nothing but the special Kahler
potential of the rigid theory.\newline
The next step is to consider the new covariantly holomorphic section $\tilde{%
V}^{\prime M}=e^{\frac{\mathcal{K}^{\prime }}{2}}\tilde{\Omega}^{\prime M}$
and expand it in $\frac{1}{\mathrm{\mu }}$ power terms like 
\begin{equation}
\tilde{V}^{\prime M}=\mathring{V}_{\left( 0\right) }^{M}+\frac{1}{\mathrm{%
\mu }}\mathring{V}_{\left( 1\right) }^{M}+\mathfrak{O}(\frac{1}{\mathrm{\mu }%
^{2}})
\end{equation}%
To obtain the expression of $\tilde{V}^{\prime M}$, we need to know the
factors $\mathbb{S}_{N}^{M}$ and $\tilde{\Omega}^{\prime M}$ see that the
expression of $\mathcal{K}^{\prime }$ is given by (\ref{ke}). Let us first
determine $\mathbb{S}_{N}^{M}$; and turn after to $\tilde{\Omega}^{\prime M}$%
. By using eqs(\ref{met}-\ref{0}), we learn that the leading term in the $%
\frac{1}{\mathrm{\mu }}$- expansion of $U_{i}^{M}$ (\ref{-1}) should, up to
a symplectic transformation $\tilde{U}_{i}^{M}=\mathbb{S}_{N}^{M}U_{i}^{N}$,
have the typical form 
\begin{equation}
U_{i}^{M}\equiv \frac{1}{\mathrm{\mu }}\mathring{U}_{i}^{M}+\mathfrak{O}(%
\frac{1}{\mathrm{\mu }^{2}})  \label{uurig}
\end{equation}%
This means that there exists a symplectic frame on $\boldsymbol{M}_{SK}$
where we have 
\begin{equation}
\begin{tabular}{lll}
$\mathbb{S}_{N}^{M}U_{i}^{N}$ & $=$ & $\frac{1}{\mathrm{\mu }}\mathring{U}%
_{i}^{M}+\mathfrak{O}(\frac{1}{\mathrm{\mu }^{2}})$ \\ 
& $=$ & $\frac{1}{\mathrm{\mu }}\delta _{\text{\b{M}}}^{M}\mathring{U}_{i}^{%
\text{\b{M}}}+\mathfrak{O}(\frac{1}{\mathrm{\mu }^{2}})$%
\end{tabular}
\label{su}
\end{equation}%
with $\mathring{U}_{i}^{M}$ as in (\ref{um}). To obtain the explicit
expression of the symplectic matrix $\mathbb{S}_{N}^{M}$ at first order in
the expansion parameter $\frac{1}{\mathrm{\mu }}$, we solve the constraint
equation $\mathbb{S}_{N}^{M}U_{i}^{N}=\frac{1}{\mathrm{\mu }}\mathring{U}%
_{i}^{M}$. Using eq(\ref{um}) and the following $\frac{1}{\mathrm{\mu }}$-
expansion of $U_{k}^{M}$ 
\begin{equation}
U_{k}^{M}=\left( 
\begin{array}{c}
-\frac{1}{\mathrm{\mu }}\eta _{k} \\ 
\delta _{k}^{i}\left[ 1-\frac{1}{2\mathrm{\mu }}\left( \mathbf{\eta .z}+%
\mathbf{\eta .\bar{z}}\right) \right] -\frac{1}{2\mathrm{\mu }}\left(
z^{i}\eta _{k}+\eta _{k}z^{i}\right) \\ 
0 \\ 
\frac{i}{2\mathrm{\mu }^{2}}\left[ \eta _{i}\eta _{k}-\partial _{i}\partial
_{k}\Phi \right]%
\end{array}%
\right) +\mathfrak{...}
\end{equation}%
we find that $\mathbb{S}_{N}^{M}$ is given by (\ref{sym}) with the
particular matrices 
\begin{equation}
\mathbb{A}=%
\begin{pmatrix}
1 & \frac{1}{\mathrm{\mu }}\eta _{I} \\ 
0 & \frac{1}{\mathrm{\mu }}\mathcal{I}_{n\times n}%
\end{pmatrix}%
\qquad ,\qquad \mathbb{D}=%
\begin{pmatrix}
1 & 0 \\ 
-\eta _{I} & \mathrm{\mu }\mathcal{I}_{n\times n}%
\end{pmatrix}
\label{spf}
\end{equation}%
and vanishing$\left( n+1\right) \times \left( n+1\right) $ matrices $\mathbb{%
B}$ and $\mathbb{C}$. Having obtained $\mathbb{S}_{N}^{M}$, we can now
determine $\tilde{\Omega}^{\prime M}$. First by using the particular
symplectic change $\tilde{\Omega}^{M}=\mathbb{S}_{N}^{M}\Omega ^{M}$, we
have the new components of the holomorphic sections 
\begin{equation}
\left( 
\begin{array}{c}
\tilde{X}^{0} \\ 
\tilde{X}^{I} \\ 
\tilde{F}_{0} \\ 
\tilde{F}_{I}%
\end{array}%
\right) =%
\begin{pmatrix}
1 & \frac{1}{\mathrm{\mu }}\eta _{I} & 0 & 0 \\ 
0 & \frac{1}{\mathrm{\mu }}\mathcal{I}_{n\times n} & 0 & 0 \\ 
0 & 0 & 1 & 0 \\ 
0 & 0 & -\eta _{I} & \mathrm{\mu }\mathcal{I}_{n\times n}%
\end{pmatrix}%
\left( 
\begin{array}{c}
X^{0} \\ 
X^{I} \\ 
F_{0} \\ 
F_{I}%
\end{array}%
\right)  \label{syt}
\end{equation}%
that read explicitly as follows%
\begin{equation}
\tilde{\Omega}^{M}=\left( 
\begin{array}{c}
X^{0}+\frac{1}{\mathrm{\mu }}X^{0}\mathbf{\eta .z} \\ 
\frac{1}{\mathrm{\mu }}X^{0}z^{i} \\ 
-\frac{i}{2}X^{0}-\frac{i}{2\mu }\eta _{i}z^{i}X^{0}+\mathfrak{O}(\frac{1}{%
\mu ^{2}}) \\ 
\frac{i}{2\mu }X^{0}\left( \eta _{i}\eta _{i}z^{i}-\partial _{i}\phi \right)
+\mathfrak{O}(\frac{1}{\mu ^{2}})%
\end{array}%
\right)  \label{yt}
\end{equation}%
Then, by performing the particular Kahler transformation (\ref{k1}-\ref{k3}%
), we obtain%
\begin{equation}
\tilde{\Omega}^{\prime M}=\left( 
\begin{array}{c}
X^{0} \\ 
\frac{1}{\mathrm{\mu }}X^{I} \\ 
-\frac{i}{2}X^{0} \\ 
\frac{i}{2\mu }\left( \eta _{I}\eta _{J}X^{J}-\frac{\partial \Phi }{\partial
X^{I}}\right)%
\end{array}%
\right) +\mathfrak{O}(\frac{1}{\mathrm{\mu }^{2}})
\end{equation}%
which can be splitted like 
\begin{equation}
\tilde{\Omega}^{\prime M}=\tilde{\Omega}_{\left( 0\right) }^{\prime M}+\frac{%
1}{\mathrm{\mu }}\tilde{\Omega}_{\left( 1\right) }^{\prime M}+\mathfrak{O}(%
\frac{1}{\mathrm{\mu }^{2}})
\end{equation}%
with%
\begin{equation}
\tilde{\Omega}_{\left( 0\right) }^{\prime M}=X^{0}\left( 
\begin{array}{c}
1 \\ 
0 \\ 
-\frac{i}{2} \\ 
0%
\end{array}%
\right) \qquad ,\qquad \tilde{\Omega}_{\left( 1\right) }^{\prime M}=\left( 
\begin{array}{c}
0 \\ 
X^{I} \\ 
0 \\ 
\frac{i}{2}\left( \eta _{I}\eta _{J}X^{J}-\frac{\partial \Phi }{\partial
X^{I}}\right)%
\end{array}%
\right)
\end{equation}%
where $\tilde{\Omega}_{\left( 0\right) }^{\prime M}$ is the canonical
holomorphic section of pure supergravity \textrm{\cite{Ortin:2015hya}}; and
where $\tilde{\Omega}_{\left( 1\right) }^{\prime M}$ is nothing but the
rigid $\mathring{\Omega}^{M}$ of eq(\ref{rl}) in the special coordinate
frame $X^{0}=1$. Notice that $\tilde{V}^{\prime M}$ can be also expanded
like 
\begin{equation}
\tilde{V}^{\prime M}=\tilde{V}_{grav}^{\prime M}+\frac{1}{\mathrm{\mu }}%
\tilde{V}_{\left( 1\right) }^{\prime M}+\mathcal{O}(\frac{1}{\mathrm{\mu }%
^{2}})
\end{equation}%
but, because of the form of $\mathcal{K}^{\prime }$ which is given by $-%
\frac{1}{\mathrm{\mu }^{2}}\mathcal{\mathring{K}}+\mathfrak{O}(\frac{1}{%
\mathrm{\mu }^{3}})$ as in (\ref{ke}), the $\frac{1}{\mathrm{\mu }}$-
expansion of $\tilde{V}^{\prime M}=e^{\frac{\mathcal{K}^{\prime }}{2}}\tilde{%
\Omega}^{\prime M}$ expands as follows%
\begin{equation}
e^{\frac{\mathcal{K}^{\prime }}{2}}\tilde{\Omega}^{\prime M}=\left( \tilde{%
\Omega}_{\left( 0\right) }^{\prime M}+\frac{1}{\mathrm{\mu }}\tilde{\Omega}%
_{\left( 1\right) }^{\prime M}\right) \left[ 1+\mathcal{O}\left( \frac{1}{%
\mathrm{\mu }^{2}}\right) \right]  \label{vrig}
\end{equation}%
which, up to $\frac{1}{\mathrm{\mu }^{2}}$ terms$,$ coincides exactly with
the expansion of $\tilde{\Omega}^{\prime M}$; that is 
\begin{equation}
\tilde{V}_{grav}^{\prime M}=\tilde{\Omega}_{\left( 0\right) }^{\prime
M}\qquad ,\qquad \tilde{V}_{\left( 1\right) }^{\prime M}=\mathring{\Omega}%
^{M}.
\end{equation}%
Notice moreover that, under the symplectic transformation, the special
Kahler potential (\ref{kp}) remains invariant and the prepotential $\mathcal{%
F}\left( X\right) $ (\ref{10}) gets expressed in terms of the new variables
like 
\begin{equation}
\mathcal{\tilde{F}}\left( \tilde{X}\right) =-i\left( \tilde{X}^{0}\right)
^{2}\boldsymbol{\tilde{F}}\left( \tilde{X}\right)
\end{equation}%
If substituting the $\tilde{X}^{\Lambda }$ components by their expressions $%
\tilde{X}(X)$ in terms of the old $X^{\Lambda }$'s as in (\ref{yt}), the
prepotential $\mathcal{\tilde{F}}\left( \tilde{X}\right) $ takes a fat form
like%
\begin{equation}
\mathcal{\tilde{F}}\left( \tilde{X}\right) =-i\left( X^{0}+\frac{1}{\mathrm{%
\mu }}\eta _{I}X^{I}\right) ^{2}\boldsymbol{\tilde{F}}\left[ \tilde{X}(X)%
\right]  \label{fx}
\end{equation}%
The prepotential remains invariant ($\mathcal{\tilde{F}}\left( \tilde{X}%
\right) =\mathcal{F}\left( X\right) $) since due to homogeneity and the
particular form of $\mathbb{S}_{N}^{M}$ where $\mathbb{B=C}=0$ and $\mathbb{D%
}^{T}\mathbb{A}=\mathbb{I}_{id}$, we have%
\begin{equation}
\mathcal{\tilde{F}}\left( \tilde{X}\right) =\frac{1}{2}F_{\Lambda }\mathbb{D}%
^{T}\mathbb{A}X^{\Lambda }=\frac{1}{2}X^{\Lambda }F_{\Lambda }=\mathcal{F}%
\left( X\right)
\end{equation}%
For an explicit study of the rigid limits of the coupling matrices $\mathcal{%
N}_{\Lambda \Sigma }$ and $\mathcal{U}^{MN}$, see \textrm{appendix C}.

\section{Isometries of quaternionic Kahler manifold}

\label{sec4}

In this section, we focus on the gauging quaternionic isometries in $%
\mathcal{N}=2$ gauged supergravity and develop two things: First, we build a
new family of real $4r$-dimensional quaternionic Kahler manifolds $%
\boldsymbol{M}_{QK}^{\left( ADE\right) }$ classified by rank $r$ finite
dimensional $ADE$ Lie algebras where the integer $r$ stands for the number
of hypermultiplets. Second, we give a D- brane realisation of gauged
quaternionic isometries and an interpretation of the embedding tensor $%
\vartheta _{M}^{u}$ in terms of type IIA/IIB mirror symmetry.\newline
Recall that the gauging of two abelian quaternionic isometries offers a
manner to break $\mathcal{N}=2$ supersymmetry partially \textrm{\cite%
{Ferrara:1995gu, Hansen:2013dda}}; so, we first consider the example of the
quaternionic geometry $SO\left( 1,4\right) /SO\left( 4\right) $ associated
with one hypermultiplet $\left( n_{H}=1\right) $. This matter supermultiplet
has four real scalar fields $Q^{u}$ ($u=1,2,3,4$) hosted by two SU$\left(
2\right) _{R}$ representations namely a real isosinglet $\varphi $ and a
real isotriplet $\vec{\phi}=\left( \phi ^{a}\right) $ with $a=1,2,3$.%
\begin{equation}
Q^{u}\equiv \varphi \oplus \vec{\phi}  \label{41}
\end{equation}%
Then, we develop the study of a new $\boldsymbol{M}_{QK}^{\left( ADE\right)
} $ manifolds, generalising $SO\left( 1,4\right) /SO\left( 4\right) $, and
associated with a matter sector having $n_{H}=r$ hypermultiplets where r
stands for the ranks of ADE. After that, we turn to the study of the gauging
of abelian quaternionic isometries of this family of quaternionic manifolds
by focussing the simplest $\boldsymbol{M}_{QK}^{\left( A_{1}\right) }$ with $%
A_{1}$ the Lie algebra of SU$\left( 2\right) $.

\subsection{Quaternionic Kahler manifold $\boldsymbol{M}_{QK}^{\left(
n_{H}\right) }:$ case $n_{H}=1$}

In $\mathcal{N}=2$ supergravity theory with one hypermultiplet $\boldsymbol{H%
}_{\mathcal{N}=2}$, whose four real scalars $Q^{u}$ parameterised in terms
of SU$\left( 2\right) _{R}$ singlet $\varphi $ and triplet $\phi ^{a}$ as in
(\ref{41}), the real 4- dimensional matter manifold $\boldsymbol{M}_{QK}$
has an $SU\left( 2\right) \times SP\left( 2,\mathbb{R}\right) $ holonomy
group and is given by the coset space%
\begin{equation}
\boldsymbol{M}_{QK}^{\left( n_{H}\right) }=\frac{SO\left( 1,4\right) }{%
SO\left( 4\right) }\qquad ,\qquad n_{H}=1
\end{equation}%
In what follows, we describe some relevant properties of this manifold which
are useful for the study of: $\left( i\right) $ global isometries of the
hyperKahler metric and hyperKahler 2-form as well as gaugings. $\left(
ii\right) $ the ADE- type generalisation of $SO\left( 1,4\right) /SO\left(
4\right) $; and $\left( iii\right) $ the partial supersymmetry breaking.

\subsubsection{HyperKahler metric of $\boldsymbol{M}_{QK}^{\left(
n_{H}\right) }$ and global isometries}

Using the group homomorphism $SP\left( 2,\mathbb{R}\right) \sim SU\left(
2\right) ^{\prime }$, the holonomy group of the $\boldsymbol{M}_{QK}^{\left(
n_{H}=1\right) }$ manifold becomes $SU\left( 2\right) \times SU\left(
2\right) ^{\prime }$ and is homomorphic to $SO\left( 4,\mathbb{R}\right) $.
By identifying the two SU$\left( 2\right) $ factors making this orthogonal
group, we are left with one SU$\left( 2\right) $ symmetry which is nothing
but the SU$\left( 2\right) $ R- symmetry of $\mathcal{N}=2$ supersymmetry
algebra. Under this identification, the four real scalar fields $Q^{u}$ of
the hypermultiplet are hosted by the two real irreducible representations of
SU$\left( 2\right) $ mentioned above namely the isosinglet $\varphi $ and
the isotriplet $\phi ^{a}$ like%
\begin{equation}
\varphi =Q^{u}\delta _{u}^{0}\qquad ,\qquad \phi ^{a}=Q^{u}\delta _{u}^{a}
\end{equation}%
A typical form of the quaternionic Kahler metric $ds^{2}=h_{uv}dQ^{u}dQ^{v}$
of the 4- dimensional space $SO\left( 1,4\right) /SO\left( 4\right) $,
parameterised with these real coordinate variables, is given by \textrm{\cite%
{Andrianopoli:2015rpa}} 
\begin{equation}
ds^{2}=\frac{1}{2}d\varphi ^{2}+\frac{e^{2\varphi }}{2}\left( d\phi
^{a}\delta _{ab}d\phi ^{b}\right)  \label{t}
\end{equation}%
From this relation, we read the expression of the local metric%
\begin{equation}
h_{uv}=\frac{1}{2}\left( 
\begin{array}{cc}
1 & 0 \\ 
0 & e^{2\varphi }\delta _{ab}%
\end{array}%
\right) \qquad ,\qquad h^{uv}=2\left( 
\begin{array}{cc}
1 & 0 \\ 
0 & e^{-2\varphi }\delta ^{ab}%
\end{array}%
\right)  \label{mt}
\end{equation}%
it is diagonal and depends only on the isosinglet component $\varphi $, but
not on the isotriplet $\phi ^{a}$. Moreover, the dependence on $\varphi $ is
very special in the sense that it is a function of the remarkable form $%
e^{2\varphi }$. The metric $ds^{2}$ can be also expressed like%
\begin{equation}
ds^{2}=\frac{1}{2}\left( d\varphi ^{2}+\vec{\omega}_{1}.\vec{\omega}%
_{1}\right)  \label{wa}
\end{equation}%
where the real 1-form isovector $\vec{\omega}_{1}=\left( \omega
_{1}^{a}\right) $ is given by $\omega _{1}^{a}=e^{\varphi }d\phi ^{a}$. The
normal quadratic expression (\ref{wa}) reflects the property that the metric
can be factorised in terms of vielbeins as in eq(\ref{fm}) given below. The
scalar product $\vec{\omega}_{1}.\vec{\omega}_{1}$ in above relation can be
also thought of as following from the trace $tr_{SU\left( 2\right) }\left(
\omega _{1}\otimes \omega _{1}\right) $ over quadratic product of Pauli
matrices $\tau _{a}$ involved in the matrix expansion $\omega _{1}=\tau
_{a}\omega _{1}^{a}$; and $ds^{2}=\frac{1}{4}tr_{SU\left( 2\right) }\left(
\varpi _{1}\otimes \varpi _{1}\right) $ with%
\begin{equation}
\omega _{1}=e^{\varphi }\tau _{a}d\phi ^{a}\qquad ,\qquad \varpi
_{1}=d\varphi \tau _{0}+e^{\varphi }\tau _{a}d\phi ^{a}
\end{equation}%
where $\tau _{0}$ stands for the $2\times 2$ matrix identity.

\  \  \ 

$\bullet $ \emph{Global isometries}\newline
The metric $ds^{2}$ has two special subsets of global isometries that we
describe in what follows; some of these isometries will be used later on
when we study the gauging of field translations.

$\left( i\right) $ $SO\left( 3\right) $\emph{\ symmetry group}\newline
The metric (\ref{wa}) is manifestly preserved by \emph{global rotations} of
the three component fields of the isotriplet $\phi ^{a}$ of hypermultiplet;
but living invariant the isosinglet $\varphi $. Explicitly, 
\begin{equation}
\begin{tabular}{lll}
$\phi _{a}$ & $\rightarrow $ & $\phi _{a}^{\prime }=R_{a}^{b}\phi _{b}$ \\ 
$\varphi $ & $\rightarrow $ & $\varphi ^{\prime }=\varphi $%
\end{tabular}%
\qquad ,\qquad dR_{a}^{b}=\partial _{\mu }R_{a}^{b}=0
\end{equation}%
where $R_{a}^{b}=e^{w^{a}J_{a}}$ is a 3$\times $3 orthogonal matrix with
global parameters $w^{a}$. The underlying three dimensional Lie algebra of
this $SO\left( 3\right) $ invariance is generated by three generators $J_{a}$%
; it is a non abelian symmetry isomorphic to the Lie algebra of the SU$%
\left( 2\right) $ R- symmetry with the usual Lie bracket%
\begin{equation}
\left[ J_{a},J_{a}\right] =i\varepsilon _{abc}J_{c}
\end{equation}

$\left( ii\right) $ \emph{Translation symmetry group} $\mathcal{T}_{3}$ 
\newline
The metric $ds^{2}$ is as well manifestly invariant under \emph{global
translations} of the three $\phi ^{a}$ field variables as follows%
\begin{equation}
\begin{tabular}{lll}
$\phi _{a}$ & $\rightarrow $ & $\phi _{a}^{\prime }=\phi _{a}+c_{a}$ \\ 
$\varphi $ & $\rightarrow $ & $\varphi ^{\prime }=\varphi $%
\end{tabular}%
\qquad ,\qquad dc_{a}=\partial _{\mu }c_{a}=0  \label{ga}
\end{equation}%
The three dimensional Lie algebra of $\mathcal{T}_{3}$ is abelian; it is
generated by three commuting translation generators $T_{a}=\frac{\partial }{%
\partial \phi ^{a}}.$

\  \  \  \  \  \ 

$\bullet $ \emph{Beyond}\ $SO\left( 3\right) \ltimes \mathcal{T}_{3}$\emph{\
isometry}\newline
The combination of the two above global symmetries of (\ref{wa}) is
therefore given by 
\begin{equation}
\begin{tabular}{lll}
$\phi _{a}$ & $\rightarrow $ & $\phi _{a}^{\prime }=R_{a}^{b}\phi _{b}+c_{a}$
\\ 
$\varphi $ & $\rightarrow $ & $\varphi ^{\prime }=\varphi $%
\end{tabular}%
\end{equation}%
with $\left( R_{a}^{b},c_{a}\right) $ belonging to the non abelian Euclidean
group $SO\left( 3\right) \ltimes \mathcal{T}_{3}$ symmetry with non abelian
factor $SO\left( 3\right) \simeq SU\left( 2\right) _{R}$. This is a 6-dim
Lie group with three non commuting rotation generators $J_{a}$ and three
commuting translations $T_{a}=\frac{\partial }{\partial \phi ^{a}}$; the $%
T_{a}$'s transform like a 3-vector under rotations.\newline
Notice that under global shift of the isosinglet like $\varphi \rightarrow
\varphi +c_{0}$ with $dc_{0}=0$, the invariance of the above hyperkahler
metric $h_{uv}$ requires a real rescaling of the isotriplet like%
\begin{equation}
\begin{tabular}{lll}
$\varphi ^{\prime }$ & $=$ & $\varphi +c_{0}$ \\ 
$\phi _{a}^{\prime }$ & $=$ & $e^{-c_{0}}\phi _{a}+c_{a}$%
\end{tabular}%
\qquad ,\qquad dc_{0}=0  \label{cc}
\end{equation}%
But this invariance leads to a non abelian Lie algebra and, like the $%
SO\left( 3\right) $ rotations, it is not relevant in the study of abelian
gauging of the quaternionic isometries. Non commutativity of the scale
transformations can be viewed by figuring out the generators of
infinitesimal transformations of the coordinate change (\ref{cc}) namely 
\begin{equation}
\mathrm{\delta }_{G}\varphi =c_{0}\qquad ,\qquad \mathrm{\delta }_{G}\phi
_{a}=c_{a}-c_{0}\phi _{a}
\end{equation}%
The four $T_{u}$ generators of this infinitesimal coordinate change are
given by 
\begin{equation}
T_{a}=\frac{\partial }{\partial \phi ^{a}}\qquad ,\qquad T_{0}=\frac{%
\partial }{\partial \varphi }+\phi ^{a}\frac{\partial }{\partial \phi ^{a}}
\end{equation}%
obeying the non abelian commutation relations algebra 
\begin{equation}
\left[ T_{a},T_{0}\right] =T_{a}\qquad ,\qquad \left[ T_{a},T_{b}\right] =0
\end{equation}

\subsubsection{HyperKahler 2-forms of $\boldsymbol{M}_{QK}^{\left(
n_{H}\right) }:$ case $n_{H}=1$}

The hyperKahler metric $h_{uv}$ of the 4- dimensional quaternionic Kahler
manifold $SO\left( 1,4\right) /SO\left( 4\right) $ can be also expressed in
different form by using vielbein formalism. In this approach, vielbein
1-forms $\mathcal{E}_{\dot{A}}^{A}$ carry $(\frac{1}{2},\frac{1}{2})$
quantum numbers under SO$\left( 4\right) $ and expand in the local real $%
Q^{u}$- coordinates of the manifold as follows 
\begin{equation}
\mathcal{E}_{\dot{A}}^{A}=\mathcal{E}_{\dot{A}|u}^{A}dQ^{u}  \label{14}
\end{equation}%
with $\mathcal{E}_{\dot{A}|u}^{A}$ the components along the $dQ^{u}$
directions. Using this method, the metric $h_{uv}$ and the antisymmetric
components $K_{uv}^{a}$ of the isotriplet of Kahler 2-forms $%
K_{2}^{a}=K_{uv}^{a}dQ^{u}\wedge dQ^{v}$ (hyperKahler 2-form) of the
quaternionic Kahler $SO\left( 1,4\right) /SO\left( 4\right) $ read as
follows 
\begin{equation}
\begin{tabular}{lll}
$h_{uv}$ & $=$ & $\mathcal{E}_{\dot{A}|u}^{A}\mathcal{E}_{\dot{B}%
|v}^{B}\varepsilon _{AB}\varepsilon ^{\dot{A}\dot{B}}$ \\ 
$K_{uv}^{a}$ & $=$ & $\left( \tau ^{a}\varepsilon \right) _{AB}\mathcal{E}_{%
\dot{A}|u}^{A}\mathcal{E}_{\dot{B}|v}^{B}\varepsilon ^{\dot{A}\dot{B}}$%
\end{tabular}
\label{fm}
\end{equation}%
The non vanishing components $K_{0b}^{a}$ and $K_{bc}^{a}$ of the
isotriplets of the hyperKahler 2-form are respectively proportional to $%
\delta _{b}^{a}$ and $\varepsilon ^{abc}$. The explicit expressions of
1-form vielbein quartet (\ref{14}), that splits in terms of the isosinglet $%
d\varphi $ and isotriplet $d\phi ^{a}$ like%
\begin{equation}
\mathcal{E}_{\dot{A}}^{A}=\mathcal{E}_{\dot{A}|0}^{A}d\varphi +\mathcal{E}_{%
\dot{A}|a}^{A}d\phi ^{a}
\end{equation}%
are obtained by using eq(\ref{mt}) and solving the following relation 
\begin{equation}
h_{uv}=\mathcal{E}_{\dot{A}|u}^{A}\mathcal{E}_{\dot{B}|v}^{B}\varepsilon
_{AB}\varepsilon ^{\dot{A}\dot{B}}
\end{equation}%
We find 
\begin{equation}
\begin{tabular}{lll}
$\mathcal{E}_{\dot{A}|0}^{A}$ & $=$ & $\frac{1}{2}\varepsilon ^{AB}\delta _{B%
\dot{A}}$ \\ 
$\mathcal{E}_{\dot{A}|a}^{A}$ & $=$ & $-\frac{i}{2}e^{\varphi }\delta
_{B}^{A}\left( \tau _{a}\right) _{\dot{A}}^{B}$%
\end{tabular}
\label{mf}
\end{equation}%
where one recognises the term $\mathcal{E}_{\dot{A}|a}^{A}d\phi ^{a}$ as
just $\frac{-i}{2}\delta _{B}^{A}\left( \omega _{1}\right) _{\dot{A}}^{B}$
with 1-form matrix $\left( \omega _{1}\right) _{\dot{A}}^{B}=\left( \tau
_{a}\right) _{\dot{A}}^{B}\omega _{1}^{a}$ precisely as in eq(\ref{wa}). By
using these expressions, we learn that we can express $\mathcal{E}_{\dot{A}%
}^{A}$ like $\delta _{\dot{A}}^{B}\mathcal{E}_{B}^{A}$ with%
\begin{equation}
\begin{tabular}{lll}
$\mathcal{E}_{B}^{A}$ & $=$ & $\delta _{B}^{A}d\varphi -ie^{\varphi }\left(
\tau _{a}\right) _{B}^{A}d\phi ^{a}$ \\ 
& $=$ & $\delta _{B}^{A}d\varphi -i\omega _{1}^{a}\left( \tau _{a}\right)
_{B}^{A}$%
\end{tabular}
\label{u}
\end{equation}%
Furthermore, thinking of the isotriplet of 1-forms $\omega _{1}^{a}$ as an SU%
$\left( 2\right) $ gauge connection $\omega _{1}^{a}=\omega _{u}^{a}dQ^{u}$
on the quaternionic manifold $SO\left( 1,4\right) /SO\left( 4\right) $ and
of the hyperKahler 2-forms $K_{2}^{a}=K_{uv}^{a}dQ^{u}\wedge dQ^{v}$ as
proportional to the SU$\left( 2\right) $ gauge curvature 
\begin{equation}
\Omega _{2}^{a}=d\omega _{1}^{a}+\frac{1}{2}\varepsilon ^{abc}\omega
_{1}^{b}\wedge \omega _{1}^{c}
\end{equation}%
of the connection $\omega _{1}^{a},$ we can write down a relation between
the $\Omega _{us}^{a}$ components of the curvature and the metric $h_{uv}$
namely%
\begin{equation}
h^{st}\Omega _{us}^{a}\Omega _{tv}^{b}=-\delta ^{ab}h_{uv}-\varepsilon
^{abc}\Omega _{uv}^{c}  \label{mw}
\end{equation}%
This relationship follows from properties of the three complex structures $%
\mathcal{J}^{a}$ living on quaternionic Kahler manifold obeying%
\begin{equation}
\left( \mathcal{J}^{a}\right) _{t}^{r}\left( \mathcal{J}^{b}\right)
_{s}^{t}=-\delta ^{ab}\delta _{s}^{r}+\varepsilon ^{abc}\left( \mathcal{J}%
^{c}\right) _{s}^{r}
\end{equation}%
By taking the trace over the isotriplet indices in (\ref{mw}), we can
express the quaternionic Kahler metric like $h_{uv}=-\frac{1}{3}h^{st}\Omega
_{us}^{a}\delta _{ab}\Omega _{tv}^{b}$. By using eq(\ref{t}) we learn as
well that the 1-form $\omega _{1}^{a}$ is indeed given by $e^{\varphi }d\phi
^{a}$. This property can be checked by substituting $\omega _{1}^{a}$ in the
curvature 2-forms, we obtain 
\begin{equation}
\Omega _{2}^{a}=e^{\varphi }d\varphi \wedge d\phi ^{a}+\frac{e^{2\varphi }}{2%
}\varepsilon ^{abc}d\phi ^{b}\wedge d\phi ^{c}
\end{equation}%
from which we read the curvature components 
\begin{equation}
\Omega _{0u}^{a}=\frac{1}{2}\delta _{u}^{a}e^{\varphi }\qquad ,\qquad \Omega
_{bc}^{a}=\frac{e^{2\varphi }}{2}\varepsilon ^{abc}.
\end{equation}

\subsection{Building metric for\emph{\ }$\boldsymbol{M}_{QK}^{\left(
n_{H}\right) }:$ case $n_{H}>1$}

\label{ss42}

In this subsection, we use known results on 2d- integrable models and the
geometric engineering method of 4d QFTs to reach two things: First show that
there is a correspondence between the 4- dimensional quaternionic geometry
of $SO\left( 1,4\right) /SO\left( 4\right) $ manifold (n$_{H}=1$) and the
rank $r=1$ of a hidden SU$\left( 2\right) $ symmetry. Second, use this
correspondence to study a class of 4r- dimensional quaternionic geometries
describing a matter sector of the $\mathcal{N}=2$ supergravity with $n_{H}=r$
hypermultiplets. Concretely, we give an extension of the analysis done
above, for the $SO\left( 1,4\right) /SO\left( 4\right) $ geometry of the $%
\mathcal{N}=2$ supergravity with $n_{H}=1,$ to the case of a 4r quaternionic
geometry involving several hypermultiplets; say $n_{H}=r>1$. \textrm{This
extension has been motivated by a formal similarity with two well
established approaches successfully used in QFT literature namely: }$\left(
i\right) $\textrm{\ integrable 2d- Toda theories generalising Liouville
theory and classified by ADE Lie algebras; and }$\left( ii\right) $\textrm{\
the geometric engineering of }$\mathcal{N}=2$\textrm{\ supersymmetric QFT in
4d describing the Coulomb branch of type II string compactified on local
Calabi-Yau threefolds (CY3) classified as well by ADE Lie algebras. To make
a general idea about the method, let us first give a very brief comment on
the two above mentioned approaches }and turn after to present our extension. 
\newline
For the \textrm{2d- integrable Toda models, which are classified by Lie
algebras and whose solvability is known to be due to existence of rich
symmetries, the integrability of the non linear 2d- Toda field equations of
motion having the form, }%
\begin{equation*}
\frac{\partial ^{2}u_{i}}{\partial z^{+}\partial z^{-}}-\kappa \exp
\sum_{j=1}^{r}\left( K_{ij}u_{j}\right) =0\qquad ,\qquad r\geq 1
\end{equation*}%
\textrm{\ may be also understood in terms of existence of a way to linearise
these equations like }$\partial _{+}A_{-}-\partial _{-}A_{+}+\left[
A_{+},A_{-}\right] =0$\textrm{. In these relations, the }$u_{i}=u_{i}\left(
\tau ,\sigma \right) $\textrm{\ are the 2d- Toda fields, }$K_{ij}$ is the
Cartan matrix of the underlying ADE Lie algebra with rank r and $A_{\pm
}=A_{\pm }\left[ u\right] $ related to the $u_{i}$- fields as in footnote%
\textrm{\footnote{%
\ The Lax pair leading to the Liouville equation are given by $A_{+}=\frac{%
\partial u}{\partial z}h+\kappa E_{+}$ and $A_{-}=e^{u}E_{-}$ where $%
h,E_{\pm }$ generate an su$\left( 2\right) $ Lie algebra with commutators $%
\left[ h,E_{\pm }\right] =\pm 2E_{\pm }$ and $\left[ E_{+},E_{-}\right] =h$.
For 2d Toda theories with ADE type, the corresponding Lax pair leading the
2d Toda field equations are given by $A_{+}=\sum_{i=1}^{r}h_{i}\frac{%
\partial u_{i}}{\partial z}+\kappa \sum_{i=1}^{r}E_{+i}$ and $%
A_{-}=\sum_{i=1}^{r}E_{-i}\exp \left( \sum_{j=1}^{r}K_{ij}u_{j}\right) $
where $h_{i},E_{\pm i}$ obey the ADE commutation relations $\left[
h_{i},E_{\pm j}\right] =\pm K_{ij}E_{\pm j}$ and $\left[ E_{+i},E_{-j}\right]
=\delta _{ij}h_{j}.$}}. \textrm{This linearised equation is remarkably
interpreted as the flatness condition of the curvature }$\mathrm{F}_{+-}$%
\textrm{\ of two component gauge fields given by }$A_{\pm }$\textrm{. In
this gauge language with a Lax pair }$\left( A_{+},A_{-}\right) $\textrm{\ }%
valued in the Lie algebra of ADE\textrm{, the curvature }$\mathrm{F}_{+-}$%
\textrm{\ is given\ by the commutator }$\left[ \partial _{+}+A_{+},\partial
_{-}+A_{-}\right] $\textrm{\ and the flatness condition }$\mathrm{F}_{+-}=0$
leads precisely the above Toda field equations \cite%
{Flaschka1974,Saidi:1994mk} recovering the Liouville theory as just the
leading $r=1$ case.\newline
\textrm{For the case of the geometric engineering of }$\mathcal{N}=2$\textrm{%
\ supersymmetric QFT}$_{\mathrm{4}}$'s, there is also a classification based
on Lie algebras. Though the apparent picture looks different from the Toda
theory as here one works with algebraic geometry techniques, the key idea is
formally similar in the sense it uses as well particular properties of the
ADE Lie algebras. In this method, a class of $\mathcal{N}=2$\textrm{\
supersymmetric QFT}$_{\mathrm{4}}$ are engineered as low energy effective
field theories resulting from type IIB string compactified on local
Calabi-Yau threefolds preserving eight supersymmetric charges \cite%
{Katz:1996th,S1,S2}.\textrm{\ The local CY3s are generally speaking realised
as ALE surfaces with an ADE singularity fibered on complex projective line.
The resulting }$\mathcal{N}=2$\ supersymmetric\textrm{\ field theory has a
gauge symmetry dictated by the type of the singularity. For the example of
the particular} local complex surface $\mathcal{S}$ defined by $%
z_{1}z_{2}=z_{3}^{r+1}$ where $z_{1},$ $z_{2},$ $z_{3}$ are three complex
variables; there is \textrm{an }$A_{r}$ singularity at the origin of $%
\mathcal{S}$ and then an \textrm{SU}$\left( r+1\right) $ gauge symmetry at
the level of the 4d field theory. The maximal deformation of this
singularity is given by 
\begin{equation*}
z_{1}z_{2}=z_{3}^{r+1}+\sum_{i=1}^{r}a_{i}z^{i}\qquad ,\qquad r\geq 1
\end{equation*}%
where the $a_{i}$'s are r complex moduli capturing information on the
resulting $\mathcal{N}=2$\textrm{\ supersymmetric QFT}$_{\mathrm{4}}$. Under
this complex deformation, the singularity is lifted and the \textrm{SU}$%
\left( r+1\right) $ gauge symmetry of \textrm{the }$\mathcal{N}=2$\textrm{\ }%
super QFT$_{\mathrm{4}}$\ gets broken down to the abelian U$\left( 1\right)
^{r}$\ gauge symmetry. The same picture holds for the case of generic ADE
geometries; all of them contain the SU$\left( 2\right) $ theory as
corresponding to the leading term of the family; i.e: $r=1$ case. \newline
In our way of dealing with building quaternionic Kahler geometries
associated to a number of hypermultiplets $n_{H}>1$, we borrow the idea
behind the construction of Toda theories and ADE geometries by thinking of
the $SO\left( 1,4\right) /SO\left( 4\right) $ geometry as corresponding to
the leading $r=1$ case of a family of spaces $\boldsymbol{M}_{QK}^{\left(
ADE\right) }$ indexed by ADE rank with $r\geq 1$. In this view, the number $%
n_{H}$ of hypermultiplets is therefore linked with the rank r of finite
dimensional ADE Lie algebras%
\begin{equation}
n_{H}=rank\left( ADE\right)
\end{equation}%
As such, one obtains a family of 4$n_{H}$- dimensional geometries
characterised by the rank of ADE all of them containing $SO\left( 1,4\right)
/SO\left( 4\right) $ as the leading $n_{H}=1$ which in turns is associated
with the $su\left( 2\right) $ Lie algebra having $r=1$.

\subsubsection{Quaternionic Kahler ADE type models}

A generalisation of the $SO\left( 1,4\right) /SO\left( 4\right) $ analysis
given above for $n_{H}=1$ to a higher number of hypermultiplets may be
achieved by mimicking the construction of 2d- Toda field theories \textrm{%
\cite{Aoki:1992qu,Evans:1990qq,Nohara:1990yt}} which, in turns, is obtained
by extending the well known 2d- Liouville theory \textrm{\cite%
{Ivanov:1983wp,Ginsparg:1993is,Saidi:1994mk}. Recall that 2d Liouville
theory is }associated with a hidden $su\left( 2\right) \equiv A_{1}$ behind
its integrability; the generalisation of this particular integrable 2d-
theory\ to higher finite dimensional dimensional ADE Lie algebras is given
by the integrable 2d- Toda theories having hidden ADE symmetries. Our main
motivation for the formal correspondence between the $SO\left( 1,4\right)
/SO\left( 4\right) $ metric and 2d- Liouville\textrm{\footnote{%
\ Notice that for the particular case where the u- field has no $\sigma $-
dependence; that is $u=u\left( \tau \right) $, the field equation $\frac{%
\partial ^{2}u}{\partial \bar{z}\partial z}-\kappa e^{2u}=0$ reduces to the
1- dim $\frac{\partial ^{2}u}{\partial \tau ^{2}}-\kappa e^{2u}=0$\ and
follows from the variation of the lagrangian $L=\frac{1}{2}\left( \frac{%
\partial u}{\partial \tau }\right) ^{2}+\kappa e^{2u}$. }} equation $\frac{%
\partial ^{2}u}{\partial \bar{z}\partial z}-\kappa e^{2u}=0$, with a real
2d- field $u=u\left( z,\bar{z}\right) ,$ is of the factor $e^{2\varphi }$ in
the quaternionic metric\textrm{\footnote{%
\ Notice also that if thinking of the 4d space time fields $\varphi \left( t,%
\vec{x}\right) $ and $\phi ^{a}\left( t,\vec{x}\right) $ as having no $\vec{x%
}$ space dependence and moreover related to the Liouville field $u\left(
t\right) $ like $\varphi =u\left( t\right) $ and $\phi ^{a}=tM^{a}\sqrt{2}$
with some constants $M^{a}$; then eq(\ref{qm}) reads as $ds^{2}=2Ld\tau ^{2}$
where $L$ is precisely the lagrangian of footnote 3 with $\kappa
=M^{a}M^{a}. $}} 
\begin{equation}
ds^{2}=\frac{1}{2}\left( d\varphi ^{2}+e^{2\varphi }d\phi ^{a}d\phi
^{a}\right)  \label{qm}
\end{equation}%
From the view of 2d- Liouville theory, $e^{2u}$ has an interpretation in
terms of underlying $su\left( 2\right) $ symmetry that captures the
solvability of Liouville equation formulated in terms of a vanishing
curvature of an $su\left( 2\right) $ valued gauge connection. Extending this
observation to the $SO\left( 1,4\right) /SO\left( 4\right) $ metric, the
presence of the factor $e^{2\varphi }$ in (\ref{qm}) is then very
suggestive; it can be imagined as 
\begin{equation}
e^{2\varphi }=e^{K_{11}\varphi _{1}}\qquad ,\qquad \varphi _{1}\equiv \varphi
\label{mq}
\end{equation}%
with $K_{11}=2$ referring to the usual " $1\times 1$" Cartan matrix of $%
su\left( 2\right) =A_{1}$ Lie algebra which we write as\ 
\begin{equation}
K_{A_{1}}=2  \label{11}
\end{equation}%
In other words, the metric (\ref{qm}) may be formally put in 1 to 1 with
rank of the $su\left( 2\right) $ Lie algebra and therefore we have the
following correspondence%
\begin{equation}
n_{H}=1\qquad \longleftrightarrow \qquad rank\left( su_{2}\right) =1
\end{equation}%
This correspondence valid for $n_{H}=1$ allows to ask whether this
construction can be extended to higher $n_{H}$'s and higher ranks of finite
dimensional ADE Lie algebras. An example of such extension is given by the $%
su\left( 1+n_{H}\right) =A_{r}$ Lie algebra with rank $r=n_{H}$. Below, we
construct the particular extension associated with $su\left( 1+n_{H}\right) $
Lie algebra; a similar construction can be done for generic ADE Lie
algebras. For the case of $su\left( 1+n_{H}\right) $, the Cartan matrix $%
K_{A_{r}}$ generalising $K_{A_{r}}$ (\ref{11}) is given by%
\begin{equation}
K_{A_{r}}=\left( 
\begin{array}{ccccc}
2 & -1 &  &  &  \\ 
-1 & 2 &  &  &  \\ 
&  & \ddots &  &  \\ 
&  &  & 2 & -1 \\ 
&  &  & -1 & 2%
\end{array}%
\right)  \label{mm}
\end{equation}%
The simplest example coming after (\ref{qm}-\ref{mq}) is given by the 8-
dimensional manifold based on the $su\left( 3\right) =A_{2}$ Lie algebras
with Cartan matrix as%
\begin{equation}
K_{A_{2}}=\left( 
\begin{array}{cc}
2 & -1 \\ 
-1 & 2%
\end{array}%
\right)
\end{equation}%
The geometry of this real 8d manifold involves two hypermultiplets $n_{H}=2$
with local coordinate fields described by $4+4$ real scalars $Q_{1}^{u}$ and 
$Q_{2}^{u}$ containing two $SU\left( 2\right) _{R}$ isosinglets denoted as $%
\varphi _{i}=\left( \varphi _{1},\varphi _{2}\right) ;$\ and two $SU\left(
2\right) _{R}$ isotriplets $\phi _{i}^{a}=\left( \phi _{1}^{a},\phi
_{2}^{a}\right) ,$%
\begin{equation}
\begin{tabular}{lll}
$\varphi _{i}$ & $=$ & $Q_{i}^{u}\delta _{u}^{0}$ \\ 
$\phi _{i}^{a}$ & $=$ & $Q_{i}^{u}\delta _{u}^{a}$%
\end{tabular}%
\end{equation}%
In the generic $su\left( 1+n_{H}\right) =A_{n_{H}}$ model, the
generalisation of the four scalars $\left( \varphi ,\phi ^{a}\right) $ for
one multiplet to the case of $n_{H}$ hypermultiplets is given by the
coordinate system $\left( \varphi ^{r},\phi ^{ar}\right) $ with index $%
r=1,...,n_{H}$. The quadratic $d\varphi d\varphi $ and $d\phi ^{a}d\phi ^{a}$
in the metric (\ref{qm}) can be extended like 
\begin{equation}
\begin{tabular}{l|l}
$\  \  \  \ $su$\left( 2\right) $ & $\  \  \  \ su\left( 1+n_{H}\right) $ \\ \hline
$\  \  \  \ d\varphi d\varphi $ & $\  \  \  \  \frac{1}{2}\dsum%
\limits_{r,s=1}^{n_{H}}K_{rs}d\varphi ^{r}d\varphi ^{s}$ \\ 
$\  \  \  \ d\phi ^{a}d\phi ^{a}$ \  \  \  \  \  & $\  \  \  \  \frac{1}{2}\dsum
\limits_{r,s=1}^{n_{H}}K_{rs}d\phi ^{ar}d\phi ^{as}$ \\ \hline
\end{tabular}
\label{fg}
\end{equation}%
The new real coordinates $\left( \varphi ^{r},\phi ^{ar}\right) $ carry a
vector index r transforming in the vector representation of $SO\left( n_{H},%
\mathbb{R}\right) $ isotropy group. The appearance of Cartan matrix $K_{rs}$
can be exhibited explicitly by using simple roots $\vec{\alpha}_{r}$ of the $%
su\left( 1+n_{H}\right) $ Lie algebra related to the Cartan matrix as
follows: 
\begin{equation}
\vec{\alpha}_{r}.\vec{\alpha}_{s}=K_{rs}
\end{equation}%
This feature allows to think about the local coordinates field variables of
the quaternionic Kahler manifold $\boldsymbol{M}_{QK}^{\left( n_{H}\right) }$
as follows 
\begin{equation}
\vec{\varphi}=\vec{\alpha}_{r}\varphi ^{r}\qquad ,\qquad \vec{\phi}^{a}=\vec{%
\alpha}_{r}\phi ^{ar}  \label{ff}
\end{equation}%
and generally for the matter supermultiplets 
\begin{equation}
\boldsymbol{\vec{H}}_{\mathcal{N}=2}=\vec{\alpha}_{r}\boldsymbol{H}_{%
\mathcal{N}=2}^{r}
\end{equation}%
So similar expressions to (\ref{ff}) can be also written down for the
fermionic partners $\xi _{\hat{\alpha}}^{Ar}$, the hyperini (\ref{tb1}); we
have $\vec{\xi}_{\hat{\alpha}}^{A}=\vec{\alpha}_{r}\xi _{\hat{\alpha}}^{Ar}$%
. By using the $n_{H}$ fundamental weights $\vec{\lambda}_{r}$ of the $%
su\left( 1+n_{H}\right) $ Lie algebra satisfying the property $\vec{\lambda}%
_{r}.\vec{\alpha}_{s}=\delta _{rs}$, we have%
\begin{equation}
\varphi _{r}=\vec{\lambda}_{r}.\vec{\varphi}\qquad ,\qquad \phi _{r}^{a}=%
\vec{\lambda}_{r}.\vec{\phi}^{a}\qquad ,\qquad \xi _{r}^{\hat{\alpha}A}=\vec{%
\lambda}_{r}.\vec{\xi}^{\hat{\alpha}A}
\end{equation}%
and $\boldsymbol{H}_{r}^{\mathcal{N}=2}=\vec{\alpha}_{r}.\boldsymbol{\vec{H}}%
^{\mathcal{N}=2}$ for hypermultiplets.

\subsubsection{Metrics and hyperKahler 2-forms}

Clearly, the generalisation (\ref{fg}) is a particular extension of $%
SO\left( 1,4\right) /SO\left( 4\right) $ manifold in the sense it depends on
the real $n_{H}$- dimensional vectors $\vec{\alpha}_{r}$ whose intersection
matrix $\vec{\alpha}_{r}.\vec{\alpha}_{s}$ is the $n_{H}\times n_{H}$
symmetric matrix (\ref{mm}). To build an extension of the $SO\left(
1,4\right) /SO\left( 4\right) $ metric (\ref{qm}) to geometries with $n_{H}$
hypermultiplets classified by the intersection matrix $K_{rs}$ of the $%
su\left( 1+n_{H}\right) $ Lie algebra, we extend the $e^{2\varphi }$ factor
in (\ref{qm}) like 
\begin{equation}
e^{2\varphi }\equiv e^{\vec{\alpha}.\vec{\varphi}}\qquad \rightarrow \qquad
\sum_{i=1}^{n_{H}}e^{\vec{\alpha}_{i}.\vec{\varphi}}  \label{ex}
\end{equation}%
For the case $n_{H}=1$, we have one simple root $\vec{\alpha}_{1}\equiv \vec{%
\alpha}$ and (\ref{ff}) reduces to $\vec{\varphi}=\vec{\alpha}\varphi $ and $%
\vec{\phi}^{a}=\vec{\alpha}\phi ^{a}$. Moreover, because of the property $%
\vec{\alpha}.\vec{\alpha}=2$, we have $\vec{\alpha}.\vec{\varphi}=2\varphi $
and then $e^{2\varphi }=e^{\vec{\alpha}.\vec{\varphi}}$. Therefore, a
generalisation of (\ref{qm}) to the case of several $n_{H}$ hypermultiplets
is obtained by using eqs(\ref{fg}-\ref{ex}); it reads as follows%
\begin{equation}
ds_{n_{H}}^{2}=\frac{1}{4}\left( d\vec{\varphi}.d\vec{\varphi}+\mathrm{f}%
\left( \varphi \right) d\vec{\phi}^{a}.d\vec{\phi}^{a}\right)  \label{te}
\end{equation}%
with $\mathrm{f}\left( \varphi \right) ,$ depending on the number $n_{H}$,
given by 
\begin{equation}
\mathrm{f}\left( \varphi \right) =\sum_{i=1}^{n_{H}}e^{\vec{\alpha}_{i}.\vec{%
\varphi}}
\end{equation}%
The metric $ds_{n_{H}}^{2}$ has global isometries generalising those of $%
SO\left( 1,4\right) /SO\left( 4\right) $; in particular the following
abelian ones 
\begin{equation}
\begin{tabular}{lll}
$\phi _{a}^{r}$ & $\rightarrow $ & $\phi _{a}^{r\prime }=\phi
_{a}^{r}+c_{a}^{r}$ \\ 
$\varphi ^{r}$ & $\rightarrow $ & $\varphi ^{r\prime }=\varphi ^{r}$%
\end{tabular}%
\qquad ,\qquad \partial _{\mu }c_{a}^{r}=0  \label{cz}
\end{equation}%
extending eq(\ref{ga}). Moreover, using positivity of $e^{\vec{\alpha}_{i}.%
\vec{\varphi}}$, one can rewrite the above generalised metric in the
following way 
\begin{equation}
ds^{2}=\frac{1}{4}\left( d\vec{\varphi}.d\vec{\varphi}+\vec{\omega}_{1}^{a}.%
\vec{\omega}_{1}^{a}\right)
\end{equation}%
with 1-form $\vec{\omega}_{1}^{a}=\sum_{r=1}^{n_{H}}\vec{\alpha}_{r}\omega
_{1}^{ar}$ like%
\begin{equation}
\vec{\omega}_{1}^{a}=d\vec{\phi}^{a}\sqrt{\mathrm{f}\left( \varphi \right) }%
\qquad ,\qquad \omega _{1}^{ar}=d\phi ^{ar}\sqrt{\mathrm{f}\left( \varphi
\right) }
\end{equation}%
Such construction done for $su\left( n_{H}+1\right) $ Cartan matrices $%
K_{ij}^{(A_{n_{H}})}$ given by the particular family (\ref{mm}) extends
straightforwardly to Cartan matrices $K_{ij}^{(ADE)}$ of all finite
dimensional ADE Lie algebras. Using the 1-form vielbein formalism on $%
\boldsymbol{M}_{QK}^{\left( n_{H}\right) }$, 
\begin{equation}
\mathcal{E}^{A\dot{A}r}=\mathcal{E}_{us}^{A\dot{A}r}dQ^{us}\qquad ,\qquad
u=1,2,3,4\qquad ,\qquad s=1,...,n_{H}
\end{equation}%
with%
\begin{equation}
\begin{tabular}{lll}
$\mathcal{E}_{\dot{A}|0s}^{Ar}$ & $=$ & $\frac{1}{2}\alpha
_{s}^{r}\varepsilon ^{AB}\delta _{B\dot{A}}$ \\ 
$\mathcal{E}_{\dot{A}|as}^{Ar}$ & $=$ & $-\frac{i}{2}\alpha _{s}^{r}\mathrm{f%
}\left( \varphi \right) \delta _{B}^{A}\left( \tau _{a}\right) _{\dot{A}%
}^{B} $%
\end{tabular}%
\end{equation}%
with the $\alpha _{i}^{s}$ the components of the simple root vectors $\vec{%
\alpha}_{i}$, we can also write down the metric of $\boldsymbol{M}%
_{QK}^{\left( n_{H}\right) }$,%
\begin{equation}
ds_{\left( n_{H}\right) }^{2}=h_{urvs}^{\left( n_{H}\right) }dQ^{ur}dQ^{vs}
\end{equation}%
and the components $\left( K^{\left( n_{H}\right) }\right) _{urvs}^{a}$ of
the hyperKahler 2-forms 
\begin{equation}
K_{2}^{\left( n_{H}\right) }=\left( K^{\left( n_{H}\right) }\right)
_{urvs}^{a}dQ^{ur}\wedge dQ^{vs}
\end{equation}%
For the metrics $h_{urvs}^{\left( n_{H}\right) }$, we have the following
factorisation 
\begin{equation}
h_{urvs}^{\left( n_{H}\right) }=\mathcal{E}_{\dot{A}|ur}^{Ar^{\prime
}}\delta _{r^{\prime }s^{\prime }}\mathcal{E}_{\dot{B}|vs}^{Bs^{\prime
}}\varepsilon _{AB}\varepsilon ^{\dot{A}\dot{B}}
\end{equation}%
and for $\left( K^{\left( n_{H}\right) }\right) _{urvs}^{a}$, we have%
\begin{equation}
\left( K^{\left( n_{H}\right) }\right) _{urvs}^{a}=\left( \tau
^{a}\varepsilon \right) _{AB}\mathcal{E}_{\dot{A}|ur}^{Ar^{\prime }}\delta
_{r^{\prime }s^{\prime }}\mathcal{E}_{\dot{B}|vs}^{Bs^{\prime }}\varepsilon
^{\dot{A}\dot{B}}
\end{equation}%
More comments regarding these generalized manifolds and the ADE
classification will be given in the section devoted to the conclusion and
discussions.

\subsection{Gauging quaternionic isometries}

\label{ssec43}

Following \textrm{\cite{Ferrara:1995gu}}, partial breaking of $\mathcal{N}=2$
supersymmetry in supergravity theory can be nicely realised by gauging two
abelian quaternionic isometries of the scalar manifold. This scenario
requires therefore a hypermatter sector that couples to gauge and gravity
branches. \newline
In this subsection, we study the gauging of quaternionic isometries of $%
\mathcal{N}=2$ supergravity with $n_{V}$ vector supermultiplets ($n_{V}=n$)
and $n_{H}$ hypermultiplets. We also give a realisation of this system in
terms of D- branes wrapping cycles in type II strings compactified on
Calabi-Yau threefolds.

\subsubsection{Gauging translations of $\boldsymbol{M}_{QK}^{\left(
n_{H}\right) }:$ case $n_{H}=1$}

We begin by considering the four real scalar fields $Q^{u}=\left( \varphi
,\phi ^{a}\right) $ parameterising the real 4-dim quaternionic Kahler
manifold $SO\left( 1,4\right) /SO\left( 4\right) $ with metric (\ref{t})$,$
and study the gauging of the three global translations $\mathrm{\delta }%
_{G}\phi ^{a}=c^{a}$ under which the metric $ds^{2}=h_{uv}dQ^{u}dQ^{v}$ and
the corresponding hyperKahler 2-form remain invariant. This gauging of
abelian quaternionic isometries is accompanied with three abelian gauge
fields $\mathcal{C}_{\mu }^{a}=\left( \mathcal{C}_{\mu }^{x},\mathcal{C}%
_{\mu }^{y},\mathcal{C}_{\mu }^{z}\right) $ related to the gauge fields $%
\mathcal{A}_{\mu }^{M}$ of gravity and Coulomb sectors by the so called
embedding tensor $\vartheta _{M}^{a}$ introduced in \textrm{\autoref{sec2}}
and which will described with details later on. But before going into the
gauging process, let recall some useful features for present analysis:
First, the three global translations (\ref{ga}) preserve (\ref{t}) and are
generated by three commuting operators 
\begin{equation}
T_{a}=\frac{\partial }{\partial \phi ^{a}}\qquad ,\qquad T_{a}=\delta
_{a}^{u}T_{u}\qquad ,\qquad T_{u}=\frac{\partial }{\partial Q^{u}}
\end{equation}%
Second, the differential of the field shifts $\mathrm{\delta }%
_{G}Q^{u}=c^{a}\delta _{a}^{u}$ obey the natural property%
\begin{equation}
d\left( \mathrm{\delta }_{G}Q^{u}\right) =0  \label{dg}
\end{equation}%
due to%
\begin{equation}
dc^{a}=dx^{\mu }\partial _{\mu }c^{a}=0
\end{equation}%
The real constants $c^{a}$ are just shifts of the origin of the $Q^{u}$-
coordinate frame; and alike the $Q^{u}$'s which, in type II string on CY3,
have an interpretation in terms of the geometric and stringy moduli, the
gauging of two of these three $c^{a}$'s can be given as well an
interpretation in type II strings on CY3 with a homological basis of
symplectic 3-cycles given by 
\begin{equation}
\Pi _{3}^{M}=\left( \mathfrak{A}_{3}^{\Lambda },\mathfrak{B}_{3\Lambda
}\right)
\end{equation}%
The $c^{a}$'s can be interpreted in terms of turning on fluxes $\pi ^{M}$ of
an external field strength 3-forms, say $\boldsymbol{H}_{3}=d\boldsymbol{B}%
_{2}$; the fluxes $\pi ^{M}$ are through the 3-cycles $\Pi _{3}^{M}$ and are
related to $c^{a}$ as follows%
\begin{equation}
c^{a}=\vartheta _{M}^{a}\pi ^{M}  \label{tc}
\end{equation}%
with%
\begin{equation}
\pi ^{M}=\dint \nolimits_{\Pi _{3}^{M}}\boldsymbol{H}_{3}\qquad ,\qquad \pi
^{M}=\left( p^{\Lambda },q_{\Lambda }\right)
\end{equation}%
In (\ref{tc}), the real quantity $\vartheta _{M}^{a}$ is the embedding
tensor carrying two indices $\left( ^{a},_{M}\right) $; it transforms in the
bi-fundamental of $SO\left( 3\right) \times SP\left( 2n+2\right) $; other
comments on this remarkable tensor will be given below; and more general
properties can be found in \textrm{\cite{deWit:2005ub}}. \newline
By gauging the scalar field translations, the above three parameters $c^{a}$
are no longer constants; they are coordinate dependent and, to avoid
confusion, they will be denoted here below as follows 
\begin{equation}
c^{a}\qquad \rightarrow \qquad -\xi ^{a}\left( x\right)  \label{ksi}
\end{equation}%
So the local field translations in the quaternionic Kahler manifold $%
SO\left( 1,4\right) /SO\left( 4\right) $ read as 
\begin{equation}
\mathrm{\delta }_{G}Q^{u}=-\xi ^{a}\delta _{a}^{u}\qquad ,\qquad \partial
_{\mu }\xi ^{a}=0  \label{ks}
\end{equation}%
By exhibiting the local gauge parameters $\xi ^{a}$, the variation operator $%
\mathrm{\delta }_{G}$ can be expressed in terms of the Killing generators $%
\mathrm{T}_{a}$ like $\mathrm{\delta }_{G}=\xi ^{a}\mathrm{T}_{a}$; and the
action of these Killing operators on the field coordinates $Q^{u}$ is given
by $\mathrm{T}_{a}Q^{u}=\kappa _{a}^{u}$ with Killing vectors $\kappa
_{a}^{u}$ as follows 
\begin{equation}
\kappa _{a}^{u}=-\delta _{a}^{u}  \label{kv}
\end{equation}%
Notice that under gauging of the translations, the differential the
coordinate field variations $d\left( \mathrm{\delta }_{G}Q^{u}\right) $ is
no longer vanishing; by using (\ref{ks}), we get%
\begin{equation}
d\left( \mathrm{\delta }_{G}Q^{u}\right) =-d\xi ^{a}\delta _{a}^{u}
\label{vk}
\end{equation}%
To have the property (\ref{dg}) valid even for local field translations, we
need to introduce a triplet $\mathcal{C}_{1}^{a}=\mathcal{C}_{1}^{u}\delta
_{u}^{a}$ of 1-form connections living on $SO\left( 1,4\right) /SO\left(
4\right) $ with abelian gauge transformation given by the opposite of (\ref%
{vk}) namely 
\begin{equation}
\mathrm{\delta }_{G}\mathcal{C}_{1}^{u}=d\xi ^{a}\delta _{a}^{u}\qquad
,\qquad \mathcal{C}_{1}^{a}=\mathcal{C}_{\mu }^{a}dx^{\mu }
\end{equation}%
By help of these three gauge connections, the previous global constraint $%
d\left( \mathrm{\delta }_{G}Q^{u}\right) =0$ gets now mapped to $\mathrm{%
\delta }_{G}\left( \mathcal{D}Q^{u}\right) =0$ with gauge covariant
differential operator $\mathcal{D}$ as%
\begin{equation}
\mathcal{D}Q^{u}=\left( d+\mathcal{C}_{1}\right) Q^{u}
\end{equation}%
In the above expression, we have used the convention notation $\mathcal{C}%
_{1}=\mathcal{C}_{1}^{a}\mathrm{T}_{a}$ describing a 1-form gauge potential
valued in the abelian Lie algebra of the translation group $\mathcal{T}_{3}$
generated by the three $\mathrm{T}_{a}$'s. In terms of the space time
coordinates, we have $\mathcal{C}_{1}^{a}=\mathcal{C}_{\mu }^{a}dx^{\mu }$
with gauge transformations as $\mathrm{\delta }_{G}\mathcal{C}_{\mu
}^{a}=\partial _{\mu }\xi ^{a}$. The space time gauge covariant derivative $%
\mathcal{D}_{\mu }Q^{u}$ is then defined as 
\begin{equation}
\mathcal{D}_{\mu }Q^{u}=\left( \partial _{\mu }+\mathcal{C}_{\mu }^{a}%
\mathrm{T}_{a}\right) Q^{u}  \label{pc}
\end{equation}%
It is manifestly invariant under local translations $\mathrm{\delta }%
_{G}Q^{u}=-\xi ^{a}\delta _{a}^{u}$. Notice that by substituting $\mathrm{T}%
_{a}Q^{u}=\delta _{a}^{u}$, the covariant derivative $\mathcal{D}_{\mu
}Q^{u} $ can be also defined like 
\begin{equation}
\mathcal{D}_{\mu }Q^{u}=\partial _{\mu }Q^{u}+\mathcal{C}_{\mu }^{a}\delta
_{a}^{u}  \label{cp}
\end{equation}

* \emph{Implementing symplectic structure}\newline
The abelian 1-form gauge connections $\mathcal{C}_{1}^{a}=\mathcal{C}_{\mu
}^{a}dx^{\mu }$ introduced above form a real isotriplet of $SO\left(
3\right) \simeq SU\left( 2\right) _{R}$ as follows 
\begin{equation}
\mathcal{C}_{\mu }^{a}=\left( 
\begin{array}{c}
\mathcal{C}_{\mu }^{x} \\ 
\mathcal{C}_{\mu }^{y} \\ 
\mathcal{C}_{\mu }^{z}%
\end{array}%
\right)  \label{rs}
\end{equation}%
To interpret these $\mathcal{C}_{\mu }^{a}$ as gauge fields of the $\mathcal{%
N}=2$ supergravity theory, we have to think about them as given by some
linear combinations of the graviphoton $A_{\mu }^{0}$ and the gauge fields $%
A_{\mu }^{i}$ of the Coulomb branch as well as their magnetic counterparts $%
\tilde{A}_{\mu }^{0}$ and $\tilde{A}_{\mu }^{i}$. The relation between $%
\mathcal{C}_{\mu }^{a}$ and $\mathcal{A}_{\mu }^{M}$ is captured by the
embedding tensor introduced previously; it reads as 
\begin{equation}
\mathcal{C}_{\mu }^{a}=\vartheta _{M}^{a}\mathcal{A}_{\mu }^{M}  \label{cac}
\end{equation}%
Using the electric and magnetic components of the embedding tensor $%
\vartheta _{M}^{a}=\left( \theta _{\Lambda }^{a},\tilde{\theta}^{\Lambda
a}\right) $\ and those of the gauge potential fields $\mathcal{A}_{\mu
}^{M}=\left( A_{\mu }^{\Lambda },\tilde{A}_{\mu \Lambda }\right) $, the
relation (\ref{cac}) expands as $\mathcal{C}_{\mu }^{a}=\theta _{\Lambda
}^{a}A_{\mu }^{\Lambda }+\tilde{\theta}^{a\Lambda }\tilde{A}_{\mu \Lambda }$
and reads in terms of the graviphoton and Coulomb branch directions like 
\begin{equation}
\mathcal{C}_{\mu }^{a}=\left( \theta _{0}^{a}A_{\mu }^{0}+\tilde{\theta}^{a0}%
\tilde{A}_{\mu 0}\right) +\left( \theta _{i}^{a}A_{\mu }^{i}+\tilde{\theta}%
^{ai}\tilde{A}_{\mu i}\right)
\end{equation}%
However, to break local supersymmetry partially, we need two massive gauge
fields as shown on the following decomposition of the $\mathcal{N}=2$
supergravity multiplets in terms of the $\mathcal{N}=1$ ones \textrm{\cite%
{Ferrara:1983gn,Ferrara:1995gu,Fre:1996js}} 
\begin{equation}
\begin{tabular}{l|l}
$\  \  \  \  \  \  \  \  \  \  \  \  \  \  \mathcal{N}=2$ {\small massless repres } & $\  \
\  \  \  \  \  \  \  \  \mathcal{N}=1$ {\small repres} \\ \hline
${\small (2,}\frac{3}{2}^{{\small 2}}{\small ,1)\oplus (1,}\frac{1}{2}^{%
{\small 2}}{\small ,0}^{{\small 2}}{\small )\oplus (}\frac{1}{2}^{{\small 2}}%
{\small ,0}^{{\small 4}}{\small )}$ \  \  \  \  \  \  \  \  & $\  \  \  \ {\small (2,}%
\frac{3}{2}{\small )\oplus (}\frac{3}{2}{\small ,1,1,}\frac{1}{2}{\small %
)\oplus (}\frac{1}{2}{\small ,0}^{{\small 2}}{\small )}^{{\small 2}}\left. 
\begin{array}{c}
\text{ } \\ 
\text{ } \\ 
\text{ }%
\end{array}%
\right. $ \\ \hline
\end{tabular}
\label{grv}
\end{equation}%
where ${\small (}\frac{3}{2}{\small ,1,1,}\frac{1}{2}{\small )}$ is a
massive $\mathcal{N}=1$ supersymmetric gravitino multiplet having two
massive gauge fields. This property is implemented by setting to zero one of
the three components in eq(\ref{rs}); for instance by taking $\mathcal{C}%
_{\mu }^{z}=0$; so we have%
\begin{equation}
\mathcal{C}_{\mu }^{a}=\left( 
\begin{array}{c}
\mathcal{C}_{\mu }^{x} \\ 
\mathcal{C}_{\mu }^{y} \\ 
0%
\end{array}%
\right) \qquad \rightarrow \qquad \mathcal{C}_{\mu }^{m}=\left( 
\begin{array}{c}
C_{\mu } \\ 
\tilde{C}_{\mu }%
\end{array}%
\right)  \label{sr}
\end{equation}%
Notice that by requiring $\mathcal{C}_{\mu }^{z}=0$ in eq(\ref{sr}), the $%
SO\left( 3\right) \simeq SU\left( 2\right) _{R}$ rotation symmetry of $%
\mathcal{C}_{\mu }^{a}$ breaks down to $SO\left( 2\right) \simeq U\left(
1\right) _{R}$ rotating the two components of $\mathcal{C}_{\mu }^{m}$. This
is an important feature of partial breaking of $\mathcal{N}=2$ local
supersymmetry since the symmetry reduction%
\begin{equation}
SO\left( 3\right) \rightarrow SO\left( 2\right) \qquad \sim \qquad SU\left(
2\right) _{R}\rightarrow U\left( 1\right) _{R}
\end{equation}%
is a necessary condition for partial supersymmetry breaking in $\mathcal{N}%
=2 $ supersymmetric theory.

\  \  \  \ 

* \emph{Brane interpretation}\newline
The 1-form gauge potential doublet $\mathcal{C}_{1}^{m}=\mathcal{C}_{\mu
}^{m}dx^{\mu }$ can be given a nice interpretation in the class of $\mathcal{%
N}=2$ theories embedded into type IIA string on CY3$_{IIA}$. There, these $%
\mathcal{C}_{1}^{m}$'s descend from the gauge potential $\mathcal{C}_{3}$ of
a D2- brane wrapping 2-cycles $\mathfrak{A}_{2}^{m}$ in CY3$_{IIA}$ as
follows%
\begin{equation}
\mathcal{C}_{1}^{m}=\dint \nolimits_{\mathfrak{A}_{2}^{m}}\mathcal{C}_{3}
\label{ccm}
\end{equation}%
Observe that $SO\left( 2\right) $ rotation symmetry of $\mathcal{C}_{\mu
}^{m}$ requires a CY3$_{IIA}$ with at least two 2-cycles $\mathfrak{A}%
_{2}^{m}$. Observe as well that to implement the $SP\left( 2n_{H}\right) $
holonomy symmetry group of the quaternionic Kahler sector, we have to add
two other gauge potentials $\mathcal{\tilde{C}}_{1m}$ namely the magnetic
duals of the $\mathcal{C}_{1}^{m}$ ones given by eq(\ref{ccm}). These $%
\mathcal{\tilde{C}}_{1m}$'s descend from the 4-form potential $\mathcal{%
\tilde{C}}_{5}$ of a D4- brane wrapping two 4- cycles $\mathfrak{V}_{4}^{m}$
of the CY3$_{IIA}$ 
\begin{equation}
\mathcal{\tilde{C}}_{1m}=\dint \nolimits_{\mathfrak{V}_{4}^{m}}\mathcal{%
\tilde{C}}_{5}
\end{equation}%
Therefore, the implementation of the $SP\left( 2n_{H}\right) $ symplectic
symmetry in the hypermatter sector requires a gauge field $\mathcal{C}_{\mu
}^{m\underline{\alpha }}$ carrying two indices: the index $m=1,2$ for the $%
SO\left( 2\right) \sim U\left( 1\right) _{R}$ symmetry and the index 
\underline{$\alpha $}$=1,2$ for $SP\left( 2n_{H}\right) $ with one
hypermultiplet%
\begin{equation}
\mathbb{C}_{\mu }^{m\underline{\alpha }}=\left( 
\begin{array}{c}
\mathcal{C}_{1}^{m} \\ 
\mathcal{\tilde{C}}_{1}^{m}%
\end{array}%
\right)  \label{ccom}
\end{equation}%
By taking into account the magnetic sector, the embedding relation (\ref{cac}%
) gets extended like%
\begin{equation}
\mathbb{C}_{\mu }^{m\underline{\alpha }}=\Theta _{M}^{m\underline{\alpha }}%
\mathcal{A}_{\mu }^{M}
\end{equation}%
with generalised embedding tensor $\Theta _{M}^{m\underline{\alpha }}$ given
by%
\begin{equation}
\Theta _{M}^{m\underline{\alpha }}=\left( 
\begin{array}{c}
\vartheta _{M}^{m} \\ 
\tilde{\vartheta}_{M}^{m}%
\end{array}%
\right)
\end{equation}%
Notice finally the in the above D- brane interpretation of the gauging
process, we have used type IIA strings on CY3$_{IIA}$ picture; however the
special Kahler construction of \textrm{\autoref{sec3}} is embedded in type
IIB string on CY3$_{IIB}$. To overcome this apparent difficulty, we use type
II strings duality as described below.

\  \  \  \ 

* \emph{Mirror symmetry}\newline
By using mirror symmetry between type IIA string on CY3$_{IIA}$ and type IIB
string on CY3$_{IIB}$, the two gauge field potentials $\mathcal{C}_{\mu
}^{m} $ can be also interpreted in terms of linear combinations of the gauge
potentials $A_{\mu }^{\Lambda }$ and their magnetic duals $\tilde{A}_{\mu
\Lambda }$ living in the gravity and Coulomb branches; a similar thing can
be done for $\mathcal{\tilde{C}}_{\mu m}$; it will be understood here below.
Recall that in type IIB string on CY3$_{IIB}$, the gauge potentials $A_{\mu
}^{\Lambda }$ and $\tilde{A}_{\mu \Lambda }$ constitute together a $SP\left(
2n+2\right) $ symplectic vector denoted as%
\begin{equation}
\mathcal{A}_{\mu }^{M}=\left( 
\begin{array}{c}
A_{\mu }^{\Lambda } \\ 
\tilde{A}_{\mu \Lambda }%
\end{array}%
\right) \qquad ,\qquad \Lambda =0,1,...,n  \label{al}
\end{equation}%
The gauge potentials $A_{1}^{\Lambda }=A_{\mu }^{\Lambda }dx^{\mu }$ and $%
\tilde{A}_{1\Lambda }=\tilde{A}_{\mu \Lambda }dx^{\mu }$ descend from the 4-
form gauge potential $\mathcal{C}_{4}$ of a D3- brane wrapping 3- cycles $%
\mathfrak{A}_{3}^{\Lambda }$ and $\mathfrak{B}_{3\Lambda }$ in the CY3$%
_{IIB} $ as follows%
\begin{equation}
\begin{tabular}{lll}
$A_{1}^{\Lambda }$ & $=$ & $\dint \nolimits_{\mathfrak{A}_{3}^{\Lambda }}%
\mathcal{C}_{4}$ \\ 
$\tilde{A}_{1\Lambda }$ & $=$ & $\dint \nolimits_{\mathfrak{B}_{3\Lambda }}%
\mathcal{C}_{4}$%
\end{tabular}
\label{aam}
\end{equation}%
The relationship between the $\mathcal{C}_{\mu }^{m}$'s and $\mathcal{A}%
_{\mu }^{M}$ can be written as in (\ref{cac}) where $\vartheta _{M}^{m}$ is
the embedding tensor; the last quantity is in the bi-fundamental of 
\begin{equation}
SO\left( 2\right) \times SP\left( 2n+2\right)
\end{equation}%
where $n=n_{_{V}}$ the number of vector multiplets; it has two index legs;
one leg m in the sector type IIA string on CY3$_{IIA}$ and the other leg M
in the sector of type string IIB on CY3$_{IIB}$. The tensor $\vartheta
_{M}^{m}$ can be imagined as an object realising mirror symmetry which
exchanges Kahler and complex structures of CY3$_{IIA}$ and CY3$_{IIB}$.

\  \  \  \ 

* \emph{a comment on extension to n}$_{\emph{H}}$\emph{\ hypermultiplets}%
\newline
Here we briefly describe the extension of the analysis done for $n_{H}=1$ to
the general case of several hypermultiplets within the ADE geometry picture;
this generalisation will be used in the discussion section - see eqs(\ref%
{vkk}-\ref{vee}) - to motivate the structure of the rigid limits of the
scalar potential and the anomaly for ADE geometries. Of particular interest
for us is the form of the generalised embedding tensor and the corresponding
moment maps. To that purpose, notice that for the case of a matter sector
having n$_{H}$ hypermultiplets with $4n_{H}$ scalars described by the r
quartets $Q^{ur}$, eqs(\ref{cp},\ref{rs},\ref{cac}) concerning the gauge
fields $\mathcal{C}_{\mu }^{a}$ and the embedding tensor $\vartheta _{M}^{a}$
extend like 
\begin{equation}
\begin{tabular}{lll}
$\mathcal{D}_{\mu }Q^{ur}$ & $=$ & $\partial _{\mu }Q^{ur}+\delta _{a}^{u}%
\mathcal{C}_{\mu }^{ar}$ \\ 
$\mathcal{C}_{\mu }^{ar}$ & $=$ & $\vartheta _{M}^{ar}\mathcal{A}_{\mu }^{M}$%
\end{tabular}
\label{ar}
\end{equation}%
where the generalised gauge fields $\mathcal{C}_{\mu }^{ar}$ carry an extra
index r running from 1 to $n_{H},$ 
\begin{equation}
\mathcal{C}_{\mu }^{ar}=\left( 
\begin{array}{c}
\mathcal{C}_{\mu }^{xr} \\ 
\mathcal{C}_{\mu }^{yr} \\ 
\mathcal{C}_{\mu }^{zr}%
\end{array}%
\right) \qquad ,\qquad r=1,...,n_{H}
\end{equation}%
and where the new embedding tensor $\vartheta _{M}^{ar}$ carry as well the $%
SO\left( n_{H}\right) $ quantum number r as follows 
\begin{equation}
\vartheta _{M}^{ar}=\left( 
\begin{array}{c}
\theta _{\Lambda }^{ar} \\ 
\tilde{\theta}^{a\Lambda r}%
\end{array}%
\right) \qquad ,\qquad r=1,...,n_{H}  \label{at}
\end{equation}%
The electric $\theta _{\Lambda }^{ar}$ an magnetic $\tilde{\theta}^{a\Lambda
r}$ components given by%
\begin{equation}
\theta _{\Lambda }^{ar}=\left( 
\begin{array}{c}
\theta _{0}^{ar} \\ 
\theta _{i}^{ar}%
\end{array}%
\right) \qquad ,\qquad \tilde{\theta}^{a\Lambda r}=\left( 
\begin{array}{c}
\tilde{\theta}^{a0r} \\ 
\tilde{\theta}^{air}%
\end{array}%
\right)  \label{tat}
\end{equation}%
with $\theta _{0}^{ar}$, $\tilde{\theta}^{a0r}$ are the components along the
graviphoton direction and $\theta _{i}^{ar}$, $\tilde{\theta}^{air}$
associated with the Coulomb branch dimensions.

\subsubsection{More on the embedding tensor $\protect \vartheta _{M}^{m}$:
case $n_{H}=1$}

To make contact between the gauge potentials $\mathcal{C}_{\mu }^{m}$ of eq(%
\ref{ccom}) and the $\mathcal{A}_{\mu }^{M}$'s of eq(\ref{al}), we use the
embedding tensor $\vartheta _{M}^{m}$ of (\ref{tc}) to relate them as in (%
\ref{cac}). The $\vartheta _{M}^{m}$ carries two indices: the $SO\left(
2\right) $ orthogonal $m=1,2$ and the $SP\left( 2n+2\right) $ symplectic $%
M=0,...,2n+2$; it splits into electric and magnetic components like%
\begin{equation}
\vartheta _{M}^{m}=\left( 
\begin{array}{c}
\theta _{\Lambda }^{m} \\ 
\tilde{\theta}^{\Lambda m}%
\end{array}%
\right) \qquad ,\qquad \vartheta ^{Mm}=\left( 
\begin{array}{c}
-\tilde{\theta}^{\Lambda m} \\ 
\theta _{\Lambda }^{m}%
\end{array}%
\right)  \label{77}
\end{equation}%
By using the components of $\mathcal{C}_{\mu }^{m}$ eq(\ref{ccom}), we have 
\begin{equation}
\mathcal{C}_{\mu }^{m}=\theta _{\Lambda }^{m}A_{\mu }^{\Lambda }+\tilde{%
\theta}^{\Lambda m}\tilde{A}_{\mu \Lambda }  \label{aac}
\end{equation}
Recall that in a generic $\mathcal{N}=2$ abelian gauge theory with
electric/magnetic duality, one has $2n+2$ abelian gauge fields: $\left(
n+1\right) $ gauge fields of electric type%
\begin{equation}
A_{\mu }^{\Lambda }=\left( 
\begin{array}{c}
A_{\mu }^{0} \\ 
A_{\mu }^{i}%
\end{array}%
\right)
\end{equation}%
and $\left( n+1\right) $ magnetic duals%
\begin{equation}
\tilde{A}_{\mu \Lambda }=\left( 
\begin{array}{c}
\tilde{A}_{\mu 0} \\ 
\tilde{A}_{\mu i}%
\end{array}%
\right)
\end{equation}%
These $\left( n+1\right) +\left( n+1\right) $ gauge fields combine together
into the SP$\left( 2n+2\right) $ symplectic vector like in (\ref{al}).%
\newline
Substituting $\mathcal{C}_{\mu }^{m}$ in terms of $\mathcal{A}_{\mu }^{M}$
back into the gauge covariant (\ref{pc}), we have%
\begin{equation}
\mathcal{D}_{\mu }Q^{u}=\left( \partial _{\mu }+\mathcal{A}_{\mu }^{M}%
\mathbf{\kappa }_{M}\right) Q^{u}  \label{ccd}
\end{equation}%
with generators given by 
\begin{equation}
\mathbf{\kappa }_{M}=\vartheta _{M}^{m}\mathrm{T}_{m}\qquad ,\qquad
Y^{\Lambda }=\theta ^{\Lambda m}\mathrm{T}_{m}\qquad ,\qquad \tilde{Y}%
_{\Lambda }=\tilde{\theta}_{\Lambda }^{m}\mathrm{T}_{m}  \label{kx}
\end{equation}%
and where the embedding tensor $\vartheta _{M}^{m}$ appears as the gauge
coupling constants of the $\mathcal{A}_{\mu }^{M}$'s to the $Q^{u}$'s. The
embedding tensor $\vartheta _{M}^{m}$ satisfies a set of constraint
relations \textrm{\cite{deWit:2005ub, Andrianopoli:2015rpa}}; one of them is
the following one which is required to keep the charges mutually local, 
\begin{equation}
\vartheta _{M}^{[m}\vartheta _{N}^{n]}=0  \label{13}
\end{equation}%
To solve this constraint, it is interesting to express it into a manifestly
symplectic invariant manner like $\vartheta _{M}^{m}\mathcal{C}%
^{MN}\vartheta _{N}^{n}=0$; by expanding the symplectic trace $\vartheta
_{M}^{m}\vartheta ^{nM}=0$, we can rewrite (\ref{13}) like%
\begin{equation}
\theta _{\Lambda }^{m}\tilde{\theta}^{\Lambda n}-\tilde{\theta}^{\Lambda
m}\theta _{\Lambda }^{n}=0\qquad ,\qquad m,n=1,2  \label{ctt}
\end{equation}%
with%
\begin{equation}
\theta _{\Lambda }^{m}\tilde{\theta}^{\Lambda n}=\theta _{0}^{m}\tilde{\theta%
}^{0n}+\theta _{i}^{m}\tilde{\theta}^{in}
\end{equation}%
A class of solutions of the constraint eq(\ref{ctt}) is obtained by
constraining each of the two terms to vanish, that is $\theta _{\Lambda }^{m}%
\tilde{\theta}^{\Lambda n}=\theta _{\Lambda }^{n}\tilde{\theta}^{\Lambda
m}=0 $. This constraint can be equivalently stated by splitting $\theta
_{\Lambda }^{m}=\left( \theta _{0}^{m},\theta _{i}^{m}\right) $; this leads
to 
\begin{equation}
\theta _{0}^{m}\tilde{\theta}^{0n}+\theta _{i}^{m}\tilde{\theta}%
^{in}=0\qquad ,\qquad m,n=1,2
\end{equation}%
The above constraint relation can be solved in two basic manners as follows:
either by constraining each one of the two terms of the above sum to vanish
identically as 
\begin{equation}
\theta _{0}^{m}\tilde{\theta}^{0n}=\theta _{i}^{m}\tilde{\theta}^{in}=0
\label{041}
\end{equation}%
or by compensation of the two terms like 
\begin{equation}
\theta _{0}^{m}\tilde{\theta}^{0n}=-\theta _{i}^{m}\tilde{\theta}^{in}\neq 0
\label{042}
\end{equation}%
Let us describe briefly the solutions of these two sets of conditions:

$\bullet $ \emph{Solution of eq(\ref{041})}\newline
Eq(\ref{041}) can be solved in four manners as follows:%
\begin{equation}
\begin{tabular}{llllll}
&  & $\theta _{0}^{m}$ \  \  \  & $\tilde{\theta}^{0m}$ \  \  \  & $\theta
_{i}^{m}$ \  \  \  & $\tilde{\theta}^{im}$ \\ \hline
$\left( i\right) $ & : & $e^{m}$ & $0$ & $g^{m}\zeta _{i}$ \  \  \  & $0$ \\ 
$\left( ii\right) $ & : & $e^{m}$ & $0$ & $0$ & $\tilde{g}^{m}\tilde{\zeta}%
_{{}}^{i}$ \\ 
$\left( iii\right) $ & : & $0$ & $\tilde{e}^{m}$ & $g^{m}\zeta _{i}$ & $0$
\\ 
$\left( iv\right) $ & : & $0$ & $\tilde{e}^{m}$ & $0$ & $\tilde{g}^{m}\tilde{%
\zeta}_{{}}^{i}$ \\ \hline
\end{tabular}
\label{s41}
\end{equation}%
Under the symplectic transformation (\ref{spf}), we have $\vartheta
_{M}^{m}\rightarrow \vartheta _{N}^{m}\left( \mathbb{S}^{-1}\right)
_{M}^{N}. $ So these solution are mapped into the following ones still
preserving the constraint relation (\ref{13})%
\begin{equation}
\begin{tabular}{llllll}
&  & $\theta _{0}^{\prime m}$ \  \  \  & $\tilde{\theta}^{\prime 0m}$ \  \  \  & 
$\theta _{i}^{\prime m}$ \  \  \  & $\tilde{\theta}^{\prime im}$ \\ \hline
$\left( i\right) $ & : & $e^{m}$ & $0$ & $\mathrm{\mu }g^{m}\zeta _{i}-\eta
_{i}e^{m}$ \  \  \  & $0$ \\ 
$\left( ii\right) $ & : & $e^{m}$ & $\frac{1}{\mathrm{\mu }}\eta _{i}\tilde{%
\zeta}^{i}\tilde{g}^{m}$ & $-\eta _{i}e^{m}$ & $\frac{1}{\mathrm{\mu }}%
\tilde{g}^{m}\tilde{\zeta}^{i}$ \\ 
$\left( iii\right) $ & : & $0$ & $\tilde{e}^{m}$ & $\mathrm{\mu }g^{m}\zeta
_{i}$ & $0$ \\ 
$\left( iv\right) $ & : & $0$ & $\tilde{e}^{m}+\frac{1}{\mathrm{\mu }}\eta
_{i}\tilde{g}^{m}\tilde{\zeta}^{i}$ & $0$ & $\frac{1}{\mathrm{\mu }}\tilde{g}%
^{m}\tilde{\zeta}^{i}$ \\ \hline
\end{tabular}
\label{sett}
\end{equation}%
Notice that the above transformed solutions satisfy eq(\ref{13}); the
solutions $\left( i\right) ,$ $\left( iii\right) $ and $\left( iv\right) $
are somehow particular; they carry either electric or magnetic charges in
each of the graviphoton and Coulomb branch sectors. However, the solution $%
\left( ii\right) $ is dyonic; and the constraint (\ref{13}) is ensured by
compensation like $\tilde{\theta}_{0}^{m}\theta ^{n0}=-\tilde{\theta}%
_{i}^{m}\theta ^{ni}.$\newline
For the particular case of one vector multiplet in the Coulomb branch, the
embedding tensor $\vartheta _{M}^{m}$ takes the following form in terms of
the electric and magnetic coupling constants 
\begin{equation}
\vartheta _{M}^{m}=\left( 
\begin{array}{cc}
\theta _{0}^{1} & \theta _{0}^{2} \\ 
\theta _{1}^{1} & \theta _{1}^{2} \\ 
\tilde{\theta}^{01} & \tilde{\theta}^{02} \\ 
\tilde{\theta}^{11} & \tilde{\theta}^{12}%
\end{array}%
\right) =\left( 
\begin{array}{cc}
e_{1} & e_{2} \\ 
g_{1} & g_{2} \\ 
\tilde{e}_{1} & \tilde{e}_{2} \\ 
\tilde{g}_{1} & \tilde{g}_{2}%
\end{array}%
\right)
\end{equation}%
Two particular examples are respectively given by the embedding tensors $%
\left( \vartheta _{M}^{m}\right) _{elect}$ and $\left( \vartheta
_{M}^{m}\right) _{dyonic}$ of the electric and dyonic gauging studied in 
\textrm{\cite{Ferrara:1995xi,Andrianopoli:2015rpa}} 
\begin{equation}
\left( \vartheta _{M}^{m}\right) _{elect}=\left( 
\begin{array}{cc}
e_{1} & e_{2} \\ 
g_{1} & 0 \\ 
0 & 0 \\ 
0 & 0%
\end{array}%
\right) \qquad ,\qquad \left( \vartheta _{M}^{m}\right) _{dyonic}=\left( 
\begin{array}{cc}
e_{1} & e_{2} \\ 
0 & 0 \\ 
0 & 0 \\ 
\tilde{g}_{1} & 0%
\end{array}%
\right)
\end{equation}%
Putting (\ref{sett}) back into (\ref{aac}), we have the following relations
between $C_{\mu }^{m}$ and the gauge fields associated with the special
Kahler geometry 
\begin{equation}
\begin{tabular}{lllll}
$\left( i\right) $ & : & $C_{\mu }^{m}$ & $=$ & $e^{m}A_{\mu
}^{0}+g^{m}\zeta _{i}A_{\mu }^{i}$ \\ 
$\left( ii\right) $ & : & $C_{\mu }^{m}$ & $=$ & $e^{m}A_{\mu }^{0}+\tilde{g}%
^{m}\tilde{\zeta}^{i}\tilde{A}_{\mu i}$ \\ 
$\left( iii\right) $ & : & $C_{\mu }^{m}$ & $=$ & $\tilde{e}^{m}\tilde{A}%
_{\mu 0}+g^{m}\zeta _{i}A_{\mu }^{i}$ \\ 
$\left( iv\right) $ & : & $C_{\mu }^{m}$ & $=$ & $\tilde{e}^{m}\tilde{A}%
_{\mu 0}+\tilde{g}^{m}\tilde{\zeta}^{i}\tilde{A}_{\mu i}$%
\end{tabular}%
\end{equation}%
where the first (fourth) combination is purely electric (magnetic) while the
second and the third are dyonic. By setting 
\begin{equation}
\tilde{A}_{\mu }=\tilde{\zeta}^{i}\tilde{A}_{\mu i}\qquad ,\qquad A_{\mu
}=\zeta _{i}A_{\mu }^{i}  \label{ch}
\end{equation}%
the above relations read as%
\begin{equation}
\begin{tabular}{lllll}
$\left( i\right) $ & : & $\mathcal{C}_{\mu }^{m}$ & $=$ & $e^{m}A_{\mu
}^{0}+g^{m}A_{\mu }$ \\ 
$\left( ii\right) $ & : & $\mathcal{C}_{\mu }^{m}$ & $=$ & $e^{m}A_{\mu
}^{0}+\tilde{g}^{m}\tilde{A}_{\mu }$ \\ 
$\left( iii\right) $ & : & $\mathcal{C}_{\mu }^{m}$ & $=$ & $\tilde{e}^{m}%
\tilde{A}_{\mu 0}+g^{m}A_{\mu }$ \\ 
$\left( iv\right) $ & : & $\mathcal{C}_{\mu }^{m}$ & $=$ & $\tilde{e}^{m}%
\tilde{A}_{\mu 0}+\tilde{g}^{m}\tilde{A}_{\mu }$%
\end{tabular}%
\end{equation}%
$\bullet $ \emph{Solution of eq(\ref{042})}\newline
By using the factorisation $\theta _{i}^{m}=g^{m}\zeta _{i}$ and $\tilde{%
\theta}^{im}=\tilde{g}^{m}\tilde{\zeta}^{i}$ and by setting $\zeta _{i}%
\tilde{\zeta}^{i}=-\epsilon \left \vert \zeta \right \vert ^{2}$ with $%
\epsilon =\pm 1$, eq(\ref{042}) can be put into the following form%
\begin{equation}
\theta _{0}^{m}\tilde{\theta}^{0n}=\epsilon \left \vert \zeta \right \vert
^{2}g^{m}\tilde{g}^{n}
\end{equation}%
whose a solution is given by%
\begin{equation}
\theta _{0}^{m}=\varepsilon _{1}\left \vert \zeta \right \vert g^{m}\qquad
,\qquad \tilde{\theta}^{0n}=\varepsilon _{2}\left \vert \zeta \right \vert 
\tilde{g}^{n}  \label{gm}
\end{equation}%
with $\varepsilon _{1}\varepsilon _{2}=\epsilon $ solved like%
\begin{equation}
\begin{tabular}{|llllll|}
\hline
$\  \  \  \varepsilon _{1}$ & $=$ \  \  \  & $+$ \  \  \  & $+$ \  \  \  & $-$ \  \  \ 
& $-$ \  \  \  \\ 
$\  \  \  \varepsilon _{2}$ & $=$ & $+$ & $-$ & $+$ & $-$ \\ 
$\  \  \  \epsilon $ & $=$ & $+$ & $-$ & $-$ & $+$ \\ \hline
\end{tabular}%
\end{equation}%
By substituting (\ref{gm}) back into $\mathcal{C}_{\mu }^{m}=\theta
_{\Lambda }^{m}A_{\mu }^{\Lambda }+\tilde{\theta}^{\Lambda m}\tilde{A}_{\mu
\Lambda }$ and using the change (\ref{ch}), we obtain%
\begin{equation}
\mathcal{C}_{\mu }^{m}=g^{m}\left( \varepsilon _{1}\left \vert \zeta \right
\vert A_{\mu }^{0}+A_{\mu }\right) +\tilde{g}^{m}\left( \varepsilon
_{2}\left \vert \zeta \right \vert \tilde{A}_{\mu 0}+\tilde{A}_{\mu }\right)
\end{equation}%
which can be put into the form%
\begin{equation}
\mathcal{C}_{\mu }^{m}=g^{m}G_{\mu }+\tilde{g}^{m}\tilde{G}_{\mu }
\end{equation}%
with $\tilde{G}_{\mu }$ and $G_{\mu }$ given by%
\begin{equation}
G_{\mu }=A_{\mu }+\varepsilon _{1}\left \vert \zeta \right \vert A_{\mu
}^{0}\qquad ,\qquad \tilde{G}_{\mu }=\tilde{A}_{\mu }+\varepsilon _{2}\left
\vert \zeta \right \vert \tilde{A}_{\mu 0}
\end{equation}

\subsubsection{Symplectic moment maps\emph{\ }$\mathcal{P}_{M}^{a}$}

Because of the factorisation $\boldsymbol{M}_{SK}\times \boldsymbol{M}_{QK}$
of the scalar manifold in $\mathcal{N}=2$ supergravity, the symplectic
moment maps on the target space are of two types: $\left( i\right) $
isosinglet moment maps $\mathcal{P}_{M}^{0}$ transforming as a symplectic
vector associated with the gauging of isometries of the special Kahler
manifold $\boldsymbol{M}_{SK}$; and $\left( ii\right) $ isotriplets $%
\mathcal{P}_{M}^{a}$ transforming as well as a symplectic vector; but
associated with the quaternionic Kahler $\boldsymbol{M}_{QK}$. Here, we are
interested into those $\mathcal{P}_{M}^{a}$'s induced by the gauging of the%
\textrm{\ translations\footnote{%
\ for convenience we shall the keep manifest the SU$\left( 2\right) _{R}$
symmetry by making an analysis as if we three gauged translations; later on
we impose the condition $\xi _{3}=0.$}} on the 4- dimensional quaternionic
manifold namely 
\begin{equation}
\delta _{G}Q^{u}=-\xi ^{a}\delta _{a}^{u}
\end{equation}%
These $\mathcal{P}_{M}^{a}$'s have a quite similar structure as the
embedding tensor $\vartheta _{M}^{a}$ in the sense they also carry two
vector indices of different kinds; the index $M$ for the $SP\left(
2n+2\right) $ symplectic group in $\boldsymbol{M}_{SK}$ and the index $%
a=1,2,3$ for SO$\left( 3\right) \simeq SU\left( 2\right) _{R}$. Our interest
into these $\mathcal{P}_{M}^{a}$'s is because they play an important role in
the study of the scalar potential in $\mathcal{N}=2$ supergravity namely 
\begin{equation}
\mathcal{V}_{\text{sugra}}^{\mathcal{N}=2}=\mathcal{G}_{i\bar{j}}W^{ai}\bar{W%
}^{a\bar{j}}+2N^{a}\bar{N}^{a}-12S^{a}\bar{S}^{a}
\end{equation}%
For example, we have $W^{ai}\sim \mathcal{P}_{M}^{a}\mathcal{G}^{i\bar{j}}%
\bar{U}_{\bar{j}}^{M}$. They are also important for the study of the central
anomaly term 
\begin{equation}
\mathcal{C}_{A}^{B}=\mathcal{C}_{a}(\tau ^{a})_{B}^{A}
\end{equation}%
appearing in the $\mathcal{N}=2$ supercurrent algebra with typical
anticommutation relations as follows \textrm{\cite%
{Hughes:1986dn,Ferrara:1995xi,Partouche:1996yp}}%
\begin{equation}
\left \{ \mathcal{J}^{0A}\left( x\right) ,\mathcal{\bar{J}}_{B}^{0}\left(
y\right) \right \} =\delta _{3}\left( x-y\right) \mathcal{H}_{A}^{B}
\label{as}
\end{equation}%
with 2$\times $2 matrix operator $\mathcal{H}_{A}^{B}$ as%
\begin{equation}
\mathcal{H}_{A}^{B}=\delta _{B}^{A}\sigma _{\mu }\mathcal{T}^{\mu 0}+%
\mathcal{C}_{a}(\tau ^{a})_{B}^{A}  \label{sa}
\end{equation}%
where $\mathcal{J}_{\alpha A}^{0}\left( x\right) ,$ $\mathcal{\bar{J}}_{\dot{%
\alpha}A}^{0}\left( x\right) $ and $\mathcal{T}_{\mu }^{0}\left( x\right) $
are the time components of the supersymmetric current densities $\mathcal{J}%
_{\alpha A}^{\nu },$ $\mathcal{\bar{J}}_{\dot{\alpha}A}^{\nu }$ and $%
\mathcal{T}_{\mu }^{\nu }$. The time component densities in the current
superalgebra (\ref{sa}) are related to the usual $Q_{\alpha }^{A}$, $\bar{Q}%
_{\dot{\alpha}B}$ and $P_{\mu }$ charges of the $\mathcal{N}=2$
supersymmetric QFT$_{4}$ as follows 
\begin{equation}
Q_{\alpha A}=\int d^{3}x\mathcal{J}_{\alpha A}^{0}\qquad ,\qquad \bar{Q}_{%
\dot{\alpha}A}=\int d^{3}x\mathcal{\bar{J}}_{\dot{\alpha}A}^{0}\qquad
,\qquad P_{\mu }=\int d^{3}x\mathcal{T}_{\mu }^{0}
\end{equation}%
To obtain the explicit expression of the $\mathcal{P}_{M}^{a}$'s, we start
form the Killing vector fields 
\begin{equation}
\mathrm{T}_{a}=T_{a}^{u}\frac{\partial }{\partial Q^{u}}\qquad ,\qquad
T_{a}^{u}=-\delta _{a}^{u}  \label{ta}
\end{equation}%
and use the embedding tensor $\vartheta _{M}^{a}$ to map them into a
symplectic Killing vector fields $\kappa _{M}$ as follows 
\begin{equation}
\mathbf{\kappa }_{_{M}}=\vartheta _{M}^{a}\mathrm{T}_{a}\qquad ,\qquad 
\mathbf{\kappa }_{_{M}}^{a}=-\vartheta _{M}^{a}  \label{115}
\end{equation}%
By using the vector field language $\mathcal{L}_{\kappa _{M}}Q^{u}\equiv 
\mathcal{L}_{M}Q^{u}$ with Lie derivative $\mathcal{L}_{M}$, reading, in
terms of exterior differential $d$ and contraction $i_{M}$ operators like $%
\mathcal{L}_{M}=d\circ i_{M}+i_{M}\circ d$, the $\mathcal{P}_{M}^{a}$'s can
be defined up to compensator shifts $W_{M}^{a}$ as%
\begin{equation}
\mathcal{P}_{M}^{a}=-i_{M}\left( \omega _{1}^{a}\right) =-\vartheta
_{M}^{u}\omega _{u}^{a}
\end{equation}%
with $\omega _{1}^{a}=\omega _{u}^{a}dQ^{u}$. The differential form is the
1-form SU$\left( 2\right) $ connection of the quaternionic Kahler geometry
given by eqs(\ref{wa},\ref{u}); by substituting $\omega _{1}^{a}=e^{\varphi
}d\phi ^{a}$, we obtain the following relation between the moment maps and
the embedding tensor%
\begin{equation}
\mathcal{P}_{M}^{a}=e^{\varphi }\vartheta _{M}^{a}  \label{mp}
\end{equation}%
Notice that in the case of several hypermultiplets, the moment maps carry as
well an SO$\left( n_{H}\right) $ index $r=1,...,n_{H}$, and the above
relation extends like $\mathcal{P}_{M}^{ar}=\sqrt{\mathrm{f}\left( \varphi
\right) }\vartheta _{M}^{ar}$.

\section{Rescalings and rigid limit of $\mathcal{N}=2$ supergravity}

\label{sec5}

In this section, we first use supersymmetry and the structure of the 4d-
TAUB-NUT hyperKahler metric to propose another way to rescale fields of $%
\mathcal{N}=2$ gauged supergravity. Then, we use the results of sections 3
and 4 to study the rigid limit of the scalar potential of gauged
supergravity, the rigid limit of Ward identities underlying the gauging and
the matrix anomaly of the $\mathcal{N}=2$ supercurrent algebra.\newline
To that purpose, we shall proceed progressively as follows: $\left( i\right) 
$ we use rescaling properties of the fields of the $\mathcal{N}=2$
supergravity in order to determine the leading terms in $\frac{1}{\mathrm{%
\mu }}$- expansions. $\left( ii\right) $ we compute the value of observables
in the rigid limit $\frac{1}{\mathrm{\mu }}\rightarrow 0$; the parameter $%
\mathrm{\mu }$ of the development is given by the ratio $\frac{M_{pl}}{%
{\Large \Lambda }}$ with $\Lambda $ standing for the scale of partial
supersymmetry breaking; say the $m_{3/2}$ mass of the $\mathcal{N}=1$
massive gravitino multiplet ${\small (}\frac{3}{2}{\small ,1,1,}\frac{1}{2}%
{\small )}$ eq(\ref{grv}).

\subsection{Rescaling of fields in $\mathcal{N}=2$ gauged supergravity}

To determine the rescaling dimensions of the component fields involved in $%
\mathcal{N}=2$ supergravity, we will use supersymmetric representations to
think about rescalings as a general property shared by all those field
components belonging to the same $\mathcal{N}=2$ supermultiplet. In other
words, fields in the same $\mathcal{N}=2$ supersymmetric representation
share the same rescaling behaviour with respect to the scales $\Lambda $ and 
$M_{pl}$. To that end, let us use the language of "$\mathcal{N}=2$
superfields" $\Phi (x^{\mu },\theta _{\alpha }^{A},\bar{\theta}_{A}^{\dot{%
\alpha}})$ to think of these supersymmetric representations; this allows to
deduce directly the rescaling properties of the various component fields of $%
\Phi ^{\mathcal{N}=2}$ knowing those of the supercoordinates $x^{\mu
},\theta _{\alpha }^{A},\bar{\theta}_{A}^{\dot{\alpha}}$. So the first thing
to begin with is the rescaling of the local supercoordinate variables $%
\mathcal{Z}^{\mathcal{N}=2}=\left( x^{\mu },\theta _{\alpha }^{A},\bar{\theta%
}_{A}^{\dot{\alpha}}\right) $ of the $\mathcal{N}=2$ superspace. These
rescalings are given by the canonical dimensions expressed in terms of
Planck scale $M_{pl}$ as follows%
\begin{equation}
\left( x^{\mu },\theta _{\alpha }^{A},\bar{\theta}_{A}^{\dot{\alpha}}\right)
\qquad \rightarrow \qquad \left( x^{\mu }M_{pl},\theta _{\alpha }^{A}\sqrt{%
M_{pl}},\bar{\theta}_{A}^{\dot{\alpha}}\sqrt{M_{pl}}\right)
\end{equation}%
The two supersymmetric parameters $\epsilon _{\alpha }^{A}$ are then
rescaled in same manner as $\theta _{\alpha }^{A}$ as they are just shifts
of the Grassman variables: $\epsilon _{\alpha }^{A}\rightarrow \epsilon
_{\alpha }^{A}\sqrt{M_{pl}}$. Similar expressions can be written down for
the superspace derivatives.\newline
Regarding the superfields $\Phi ^{\mathcal{N}=2}$, their rescaling property
depend on the type of the $\mathcal{N}=2$ supermultiplet we are interested
in namely the gravity multiplet $\boldsymbol{G}_{\mathcal{N}=2}\sim {\small %
(2,}\frac{3}{2}^{2}{\small ,1)}$, the vector supermultiplet $\boldsymbol{V}_{%
\mathcal{N}=2}\sim {\small (1,}\frac{1}{2}^{2}{\small ,0}^{2}{\small )}$ and
the hypermultiplet $\boldsymbol{H}_{\mathcal{N}=2}\sim {\small (}\frac{1}{2}%
^{2}{\small ,0}^{4}{\small )}$. The rescaling properties of these
supersymmetric representations are described in what follows.

\subsubsection{Case of $\mathcal{N}=2$ gravity and vector multiplets}

As there is no simple off shell description of $\mathcal{N}=2$
supersymmetric gauge theory preserving manifestly the SU$\left( 2\right) $
R-symmetry, we shall consider only those field components contributing to
the study of the gauging isometries of the scalar manifold, to Ward
identities and to the induced scalar potential.

\  \  \  \ 

$\bullet $ $\mathcal{N}=2$ \emph{gravity} \emph{multiplet}\newline
The component fields of the gravity multiplet carrying physical degrees of
freedom are the graviton vierbein $e_{\mu }^{m}$, the graviphoton $A_{\mu
}^{0}$ and the two gravitini $\psi _{\mu }^{\alpha A}$; they are
respectively rescaled as follows%
\begin{equation}
\begin{tabular}{l|l}
dimensionless fields \  \  \  \  \  & \  \  \  \  \  \ rescaled fields \  \  \  \  \  \  \
\  \  \  \\ \hline
$\  \  \  \  \  \  \  \  \  \  \  \  \  \  \  \ e_{\mu }^{m}$ & $\  \  \  \  \  \  \  \  \  \  \  \  \
\  \  \  \frac{1}{M_{pl}}e_{\mu }^{m}$ \\ 
$\  \  \  \  \  \  \  \  \  \  \  \  \  \  \  \ A_{\mu }^{0}$ & $\  \  \  \  \  \  \  \  \  \  \  \  \
\  \  \  \frac{1}{M_{pl}}A_{\mu }^{0}$ \\ 
$\  \  \  \  \  \  \  \  \  \  \  \  \  \  \  \  \psi _{\mu }^{\alpha A}$ & $\  \  \  \  \  \  \  \
\  \  \  \  \  \  \  \  \frac{1}{M_{pl}^{3/2}}\psi _{\mu }^{\alpha A}$ \\ \hline
\end{tabular}
\label{gr}
\end{equation}%
These rescalings involve only the Planck scale $M_{pl}$ but no scale $%
\Lambda $; in the rigid limit $\frac{1}{M_{pl}}\rightarrow 0$, this
supermultiplet decouples.\newline
\  \  \  \ 

$\bullet $ $\mathcal{N}=2$ \emph{vector multiplet}\newline
The basic component fields making $\mathcal{N}=2$ vector multiplets $%
\boldsymbol{V}_{\mathcal{N}=2}^{i}$ the in Wess-Zumino gauge are as follows 
\begin{equation}
\boldsymbol{V}_{\mathcal{N}=2}^{i}\equiv z^{i},\text{ }A_{\mu }^{i},\text{ }%
\lambda _{\hat{\alpha}}^{Ai};\text{ }\left( D^{i}\right) _{B}^{A}  \label{ve}
\end{equation}%
with $z^{i}$ standing for the complex scalar fields and the Majorana spinors 
$\lambda _{\hat{\alpha}}^{Ai}=\left( \lambda _{\alpha }^{Ai},\bar{\lambda}_{%
\dot{\alpha}}^{Ai}\right) $ referring to the two gauginos. The extra 2$%
\times $2 extra matrices $\left( D^{i}\right) _{B}^{A}$ expands in terms of
isotriplets $D_{a}^{i}$ like 
\begin{equation}
\left( D^{i}\right) _{B}^{A}=\sum_{a=1}^{3}D_{a}^{i}\left( \tau ^{a}\right)
_{B}^{A}
\end{equation}%
with $\tau ^{a}$ the usual Pauli matrices. The $D_{a}^{i}$'s refer to the
auxiliary fields which scale as mass$^{2}$; they contribute to the scalar
potential $\mathcal{V}_{scal}^{\mathcal{N}=2}$ scaling as mass$^{4}$. The
rescaling properties of the fields of (\ref{ve}) depend on the
supersymmetric breaking scale $\Lambda $; they are given by%
\begin{equation}
\begin{tabular}{l|l}
\  \  \  \  \ dimensionless fields \  \  \  \  \  \  & \  \  \  \  \  \  \  \  \ rescaled
fields \  \  \  \  \  \  \  \  \  \  \\ \hline
$\  \  \  \  \  \  \  \  \  \  \  \  \  \  \  \ z^{i}$ & $\  \  \  \  \  \  \  \  \  \  \  \  \  \  \  \ 
\frac{1}{\Lambda }z^{i}$ \\ 
$\  \  \  \  \  \  \  \  \  \  \  \  \  \  \  \ A_{\mu }^{i}$ & $\  \  \  \  \  \  \  \  \  \  \  \  \
\  \  \  \frac{1}{\Lambda }A_{\mu }^{i}$ \\ 
$\  \  \  \  \  \  \  \  \  \  \  \  \  \  \  \  \lambda _{\alpha }^{iA}$ & $\  \  \  \  \  \  \  \
\  \  \  \  \  \  \  \  \frac{1}{\Lambda \sqrt{M_{pl}}}\lambda _{\alpha }^{iA}$ \\ 
$\  \  \  \  \  \  \  \  \  \  \  \  \  \  \  \  \left( D^{i}\right) _{B}^{A}$ & $\  \  \  \  \
\  \  \  \  \  \  \  \  \  \  \  \frac{1}{\Lambda M_{pl}}\left( D^{i}\right) _{B}^{A}$
\\ \hline
\end{tabular}
\label{fe}
\end{equation}%
These rescalings are different from those used in \textrm{\cite%
{Andrianopoli:2015rpa}} where $z^{i}$ was rescaled like $\frac{1}{M_{pl}}%
z^{i}$. In our way of doing, all fields of the vector supermultiplets $%
\boldsymbol{V}_{\mathcal{N}=2}^{i}$ have the same rescaling with respect to $%
\Lambda $; that is 
\begin{equation}
\boldsymbol{V}_{\mathcal{N}=2}^{i}\rightarrow \frac{\Lambda }{M_{pl}}%
\boldsymbol{V}_{\mathcal{N}=2}^{i}
\end{equation}%
Notice that in the language of $\mathcal{N}=1$ superfields, the $D_{a}^{i}$
isotriplets of auxiliary fields can be imagined as the complex $\mathrm{F}%
^{i}$ auxiliary fields of $\mathcal{N}=1$ chiral multiplets and the real $%
\mathrm{D}^{i}$ auxiliary fields of $\mathcal{N}=1$ vector ones 
\begin{equation}
D_{a}^{i}\text{ \  \ }\sim \text{ \  \ }\mathrm{F}^{i},\text{ \ }\mathrm{\bar{F%
}}^{i},\text{ \ }\mathrm{D}^{i}
\end{equation}%
The rescaling property of these auxiliary fields is useful for determining
the rescalings of the Fayet-Iliopoulos coupling constants $\nu
_{i}^{a}=\left( \nu _{i}^{+},\nu _{i}^{-},\nu _{i}^{0}\right) $ contributing
to the $\mathcal{N}=1$ supersymmetric scalar potential $\mathcal{V}_{scal}^{%
\mathcal{N}=1}$. Notice also that by using the reduced scale parameter $%
\mathrm{\mu }=\frac{M_{pl}}{{\Large \Lambda }}$, the field rescalings (\ref%
{fe}) read as follows 
\begin{equation}
\begin{tabular}{lll}
$z^{i}\rightarrow \frac{\mathrm{\mu }}{M_{pl}}z^{i}$ & $\qquad ,\qquad $ & $%
\lambda _{\alpha }^{iA}\rightarrow \frac{\mathrm{\mu }}{M_{pl}^{3/2}}\lambda
_{\alpha }^{iA}$ \\ 
$A_{\mu }^{i}\rightarrow \frac{\mathrm{\mu }}{M_{pl}}A_{\mu }^{i}$ & $\qquad
,\qquad $ & $D_{a}^{i}\rightarrow \frac{\mathrm{\mu }}{M_{pl}^{2}}D_{a}^{i}$%
\end{tabular}
\label{za}
\end{equation}%
From these rescalings of the fields in the Coulomb branch sector, we deduce
other rescaling behaviours; for example the metric $\mathcal{G}_{i\bar{j}}=%
\frac{\partial ^{2}\mathcal{K}}{\partial z^{i}\partial z^{\bar{j}}}$ of the
special Kahler manifold which is rescaled like 
\begin{equation}
\mathcal{G}_{i\bar{j}}\rightarrow \frac{M_{pl}^{2}}{\mathrm{\mu }^{2}}%
\mathcal{G}_{i\bar{j}}  \label{resg}
\end{equation}%
This is \ because the derivatives of the Kahler potential $\mathcal{K}$ are
not affected by the rescalings since $\mathcal{K}$ is given by the logarithm
of a volume, i.e: $\mathcal{K}\sim \ln \left( V_{vol}\right) $; the
rescaling shifts then $\mathcal{K}$ by a constant. Moreover, using the space
time property $\partial _{\mu }\rightarrow \frac{1}{M_{pl}}\partial _{\mu }$%
, it follows that gauge covariant derivative $\mathcal{D}_{\mu }=\partial
_{\mu }+\mathcal{A}_{\mu }^{M}\mathbf{\kappa }_{M}$ (\ref{ccd}) having the
explicit form 
\begin{equation}
\mathcal{D}_{\mu }=\partial _{\mu }+\left( A_{\mu }^{0}Y_{0}+\tilde{A}_{\mu
0}\tilde{Y}^{0}\right) +\left( A_{\mu }^{i}Y_{i}+\tilde{A}_{\mu i}\tilde{Y}%
^{i}\right)
\end{equation}%
is rescaled in same manner like $\partial _{\mu }$, thus 
\begin{equation}
\mathcal{D}_{\mu }\rightarrow \frac{1}{M_{pl}}\mathcal{D}_{\mu }
\end{equation}%
By help of eqs(\ref{gr}-\ref{za}), we learn that the $\tilde{Y}_{0}$ and $%
\tilde{Y}_{i}$ generators should transform differently like 
\begin{equation}
\tilde{Y}_{0}\rightarrow \tilde{Y}_{0}\qquad ,\qquad \tilde{Y}%
_{i}\rightarrow \frac{1}{\mathrm{\mu }}\tilde{Y}_{i}
\end{equation}%
Similar relations may be derived as well for the isotriplets of
Fayet-Iliopoulos coupling constants $\nu _{i}^{a}$; by using the fact that
the scalar potential $\mathcal{V}_{scal}^{\mathcal{N}=2}$ of the theory is
related to variations of the fermions $\delta _{B}\lambda _{\alpha }^{iA}$
induced by the gauged translations\textrm{\ like} 
\begin{equation}
\mathcal{V}_{scal}^{\mathcal{N}=2}\sim \mathcal{G}_{i\bar{j}}\left( \delta
_{A}\lambda ^{iB}\right) \left( \delta ^{B}\lambda _{A}^{\bar{j}}\right) +...
\end{equation}%
we have 
\begin{equation}
\mathcal{V}_{scal}^{\mathcal{N}=2}\qquad \rightarrow \qquad \frac{1}{%
M_{pl}^{4}}\mathcal{V}_{scal}^{\mathcal{N}=2}  \label{Vprl}
\end{equation}%
Moreover, knowing that generic scalar potentials $\mathcal{V}_{scal}^{%
\mathcal{N}=2}$ in $\mathcal{N}=2$ supersymmetric theory involve FI terms of
type $\nu _{i}^{a}D_{a}^{i}$ as well as quadratic terms type $\mathcal{G}_{i%
\bar{j}}D_{a}^{i}\delta ^{ab}D_{b}^{j}$; and using the rescaling property $%
D_{a}^{i}\rightarrow \frac{\mathrm{\mu }}{M_{pl}^{2}}D_{a}^{i}$, we conclude
that FI coupling constants $\nu _{i}^{a}$ rescale like 
\begin{equation}
\nu _{i}^{a}\rightarrow \frac{1}{\mathrm{\mu }M_{pl}^{2}}\nu _{i}^{a}
\label{nu}
\end{equation}

\subsubsection{Case of hypermultiplets}

The rescaling properties of the component fields of the hypermatter
superfield $\boldsymbol{H}_{\mathcal{N}=2}$ may be obtained in similar
manner as the gravity multiplet; except that now the field variables
parameterising the metric $ds^{2}=\frac{1}{2}\left( d\varphi
^{2}+e^{2\varphi }d\phi ^{a}d\phi ^{a}\right) $ of the quaternionic Kahler
manifold (\ref{qm}) have a different scaling mass with respect to the
canonical one. The point is that because the factor $e^{2\varphi }$ is
dimensionless, the isosinglet variable $\varphi $ should be a dimensionless
field variable; the same thing should hold for the isotriplet $\phi ^{a}$
due to consistency since $d\phi ^{a}d\phi ^{a}$ should have same scaling as $%
d\varphi ^{2}$. Let us give some details on this issue; in particular on how 
$\left( \varphi ,\phi ^{a}\right) $ can be rescaled.

\  \  \  \ 

$\bullet $ \emph{Rescaling the coordinate fields }$\left( \varphi ,\phi
^{a}\right) $\newline
If denoting by $\left( h^{0},h^{a}\right) $ the real four scalars the
hypermultiplet having the right canonical dimensions ($h^{0},$ $h^{a}\sim
mass$), and by $\left( \zeta _{\alpha }^{A},\bar{\zeta}_{\dot{\alpha}%
A}\right) $ the hyperini partners (with dimension $mass^{3/2}$); then one
can write down a relationship between the supermultiplets $\left( \varphi
,\phi ^{a};\xi _{\alpha }^{A},\bar{\xi}_{\dot{\alpha}}^{A}\right) $ of eq(%
\ref{tb1}), used in the metric building, and the $\left( h^{0},h^{a};\zeta
_{\alpha }^{A},\bar{\zeta}_{\dot{\alpha}}^{A}\right) $ by using a massive
parameter $M$. Thinking of this parameter $M$ as given by the Planck mass $%
M_{pl}$, we then have 
\begin{equation}
\begin{tabular}{lllllll}
$\varphi $ & $=$ & $\frac{h^{0}}{M_{pl}}$ & $\qquad ,\qquad $ & $\phi ^{a}$
& $=$ & $\frac{h^{a}}{M_{pl}}$ \\ 
$\xi _{\alpha }^{A}$ & $=$ & $\frac{\zeta _{\alpha }^{A}}{M_{pl}}$ & $\qquad
,\qquad $ & $\bar{\xi}_{\dot{\alpha}}^{A}$ & $=$ & $\frac{\bar{\zeta}_{\dot{%
\alpha}}^{A}}{M_{pl}}$%
\end{tabular}
\label{fh}
\end{equation}%
However, this is one way to think about the rescaling as the real four
fields $\left( h^{0},h^{a}\right) $ are not the unique way to describe the
scalars of the hypermultiplet. In fact the scalars of the hypermultiplet can
be also described in terms of a complex field doublet $f^{A}$ and its
complex conjugate $\bar{f}_{A}$; this is an interesting option especially
that the partial breaking of $\mathcal{N}=2$ supersymmetry requires complex
scalar fields; this option is also interesting because it allows to recover
the right scaling dimensions of the Fayet-Iliopoulos coupling constants.
Therefore, we propose to think about the real four variables $Q^{u}=\left(
\varphi ,\phi ^{a}\right) $ and their fermionic partners $\left( \xi
_{\alpha }^{A},\bar{\xi}_{\dot{\alpha}}^{A}\right) $ as composite fields
respecting the SU$\left( 2\right) $ representation group properties 
\begin{equation}
\mathbf{2}\otimes \mathbf{\bar{2}}=\mathbf{1}\oplus \mathbf{3}\qquad ,\qquad 
\mathbf{2}\otimes \mathbf{1}=\mathbf{2}
\end{equation}%
Focussing on scalars and using $2\otimes \bar{2}=1\oplus 3$, the real four
field variables $\varphi $ and $\phi ^{a}$ are given by composites of a
complex scalar isodoublets $f^{A}$ and $\bar{f}_{A}$ as follows%
\begin{equation}
\varphi \equiv f\bar{f}\sim \mathbf{1}\qquad ,\qquad \phi ^{a}\equiv f\tau
^{a}\bar{f}\sim \mathbf{3}  \label{gf}
\end{equation}%
where here $f^{A}$ is assumed a \emph{dimensionless} complex field doublet ($%
f^{A}\sim \mathbf{2}$); the right dimension will be restored later by
performing\ a rescaling by using Planck mass ($f^{A}=\frac{1}{M_{pl}}%
f^{\prime A}$ with $f^{\prime A}$ scaling as mass). Similar relations to the
scalars (\ref{gf}) can be written down for the fermionic fields $\xi
_{\alpha }^{A}$ and $\bar{\xi}_{\dot{\alpha}}^{A}$; they are also given by
composites involving the scalar field doublets $f^{A}$ and $\bar{f}_{A}$ and
a Dirac spinor $\left( \psi _{\alpha },\bar{\chi}_{\dot{\alpha}}\right) $
transforming as an isosinglet of SU$\left( 2\right) $ as in eq(\ref{tb1}).
We have 
\begin{equation}
\xi _{\alpha }^{A}\equiv f^{A}\psi _{\alpha }+\varepsilon ^{AB}\bar{f}%
_{B}\chi _{\alpha }\qquad ,\qquad \bar{\xi}_{\dot{\alpha}}^{A}\equiv \bar{f}%
_{A}\bar{\psi}_{\dot{\alpha}}-\varepsilon _{AB}f^{B}\bar{\chi}_{\dot{\alpha}}
\label{hf}
\end{equation}%
By exhibiting explicitly the SU$\left( 2\right) $ index in eq(\ref{gf}), we
have%
\begin{equation}
\begin{tabular}{lll}
$\varphi $ & $\equiv $ & $f^{A}\bar{f}_{A}$ \\ 
$\phi ^{a}$ & $\equiv $ & $f^{A}\left( \tau ^{a}\right) _{A}^{B}\bar{f}_{B}$%
\end{tabular}%
\end{equation}%
Rescaling the complex field doublet $f^{A}$ by their canonical mass
dimension like $f^{A}\rightarrow \frac{1}{M_{pl}}f^{A}$, the component
fields $\left( \varphi ,\phi ^{a}\right) $ of the hypermultiplet get
rescaled by a factor $\frac{1}{M_{pl}^{2}}$ as follows%
\begin{equation}
\begin{tabular}{l|l}
\ {\small dimensionless fields \  \  \  \  \ } & \  \ {\small fields with
dimensions \  \  \  \ } \\ \hline
$\  \  \  \  \  \  \  \  \  \  \  \varphi $ & $\  \  \  \  \  \  \  \  \  \  \  \frac{1}{M_{pl}^{2}%
}\varphi $ \\ 
$\  \  \  \  \  \  \  \  \  \  \  \phi ^{a}$ & $\  \  \  \  \  \  \  \  \  \  \  \frac{1}{%
M_{pl}^{2}}\phi ^{a}$ \\ \hline
\end{tabular}
\label{hyre}
\end{equation}%
\begin{equation*}
\end{equation*}%
Typical hyperKahler metrics with a parameter\ $\mathrm{\lambda }=\frac{1}{%
M_{pl}^{2}}$ scaling as mass$^{-2}$ have been constructed in hyperKahler
manifolds literature in terms of self interacting hypermultiplets by using
the $\mathcal{N}=2$ harmonic superspace method \textrm{\cite%
{Galperin:1985de,Saidi:2008au,ElHassouni:1987nr}}. A simple example of such
metrics is given by the $\mathcal{N}=2$ 4d- TAUB-NUT model whose bosonic
Lagrangian density reads as follows 
\begin{equation}
\mathcal{L}_{TB}\left( f,\bar{f}\right) =h_{A}^{B}\partial _{\mu
}f^{A}\partial ^{\mu }\bar{f}_{B}+\bar{g}_{AB}\partial _{\mu }f^{A}\partial
^{\mu }f^{B}+g^{AB}\partial _{\mu }\bar{f}_{A}\partial ^{\mu }\bar{f}_{B}
\label{tb}
\end{equation}%
with metric components $h_{A}^{B},$ $g^{AB},$ $\bar{g}_{AB}$ given by 
\begin{equation}
\begin{tabular}{lll}
$h_{A}^{B}$ & $=$ & $\delta _{A}^{B}\left( 1+\lambda f\bar{f}\right) -\frac{%
2+\lambda f\bar{f}}{2\left( 1+\lambda f\bar{f}\right) }\left( \lambda f^{A}%
\bar{f}_{B}\right) $ \\ 
$g^{AB}$ & $=$ & $\frac{2+\lambda f\bar{f}}{2\left( 1+\lambda f\bar{f}%
\right) }\left( \lambda f^{A}f^{B}\right) $ \\ 
$\bar{g}_{AB}$ & $=$ & $\frac{2+\lambda f\bar{f}}{2\left( 1+\lambda f\bar{f}%
\right) }\left( \lambda \bar{f}_{A}\bar{f}_{B}\right) $%
\end{tabular}
\label{bt}
\end{equation}%
where $M=\frac{1}{2\sqrt{\lambda }}$ is the mass of the Taub- NUT black hole
with horizon at $r=M$. In terms of the scalar fields, this horizon is
associated with the limit $f\bar{f}=0$ and corresponds to the vanishing of
the relation $f\bar{f}=2M\left( r-M\right) >0$. For other explicit metrics
generalising (\ref{tb}) to higher 4r- dimensional hyperKahler manifolds; see 
\textrm{\cite{Saidi:2008au}}. Our interst in giving (\ref{tb}) is just to
motivate the rescaling (\ref{hyre}) by thinking of the dimensionless fields $%
\left( \varphi ,\phi ^{a}\right) $ like 
\begin{equation}
\varphi =\lambda f\bar{f}\qquad ,\qquad \phi ^{a}=\lambda f\tau ^{a}\bar{f}
\end{equation}%
In what follows, we shall use the rescaling picture (\ref{hyre}) to deal
with the real scalar fields $Q^{u}=\left( \varphi ,\phi ^{a}\right) $ of
hypermultiplets; this way of doing constitutes another difference with the
approach of \textrm{\cite{Andrianopoli:2015rpa}}.

\  \ 

$\bullet $ \emph{Rescaling of embedding tensor }$\vartheta _{M}^{m}$\newline
Because of the structure of the covariant derivative $D_{\mu }=\partial
_{\mu }+\mathcal{C}_{\mu }^{m}T_{m}$, following from eq(\ref{pc}) with
rescaling property $D_{\mu }\rightarrow \frac{1}{M_{pl}}D_{\mu }$, and due
to the relation $T_{m}\sim \frac{\partial }{\partial \phi ^{m}}$ which
rescales as the inverse of the field variable $\phi ^{m}$, that is $%
T_{m}\rightarrow M_{pl}^{2}T_{m}$, then the gauge fields $\mathcal{C}_{\mu
}^{m}$ should be rescaled like 
\begin{equation}
\mathcal{C}_{\mu }^{a}\rightarrow \frac{1}{M_{pl}^{3}}\mathcal{C}_{\mu }^{a}
\end{equation}%
Moreover, using the relationship between $\mathcal{C}_{\mu }^{m}$ and the
gauge fields $\mathcal{A}_{\mu }^{M}$ namely $\mathcal{C}_{\mu
}^{m}=\vartheta _{M}^{m}\mathcal{A}_{\mu }^{M}$ we can determine the
rescaling of the components of the embedding tensor $\vartheta _{M}^{m}$.
Since fields in pure gravity sector and Coulomb branch have different
scalings as shown by (\ref{gr}) and (\ref{fe}), it it interesting to split $%
\vartheta _{M}^{m}\mathcal{A}_{\mu }^{M}$ as follows%
\begin{equation}
\begin{tabular}{lll}
$\mathcal{C}_{\mu }^{m}$ & $=$ & $\theta _{\Lambda }^{m}A_{\mu }^{\Lambda }+%
\tilde{\theta}^{\Lambda m}\tilde{A}_{\mu \Lambda }$ \\ 
& $=$ & $\left( \theta _{0}^{m}A_{\mu }^{0}+\tilde{\theta}^{0m}\tilde{A}%
_{\mu 0}\right) +\left( \theta _{i}^{m}A_{\mu }^{i}+\tilde{\theta}^{im}%
\tilde{A}_{\mu i}\right) $%
\end{tabular}%
\end{equation}%
From this expansion, we deduce the following rescaling properties of the
components of the embedding tensor%
\begin{equation}
\begin{tabular}{lllllll}
$\tilde{\theta}_{0}^{m}$ & $\rightarrow $ & $\frac{1}{M_{pl}^{2}}\tilde{%
\theta}_{0}^{m}$ & $\qquad ,\qquad $ & $\theta ^{m0}$ & $\rightarrow $ & $%
\frac{1}{M_{pl}^{2}}\theta ^{m0}$ \\ 
$\tilde{\theta}_{i}^{m}$ & $\rightarrow $ & $\frac{1}{\mathrm{\mu }M_{pl}^{2}%
}\tilde{\theta}_{i}^{m}$ & $\qquad ,\qquad $ & $\theta ^{mi}$ & $\rightarrow 
$ & $\frac{1}{\mathrm{\mu }M_{pl}^{2}}\theta ^{mi}$%
\end{tabular}
\label{tet}
\end{equation}%
The different behaviours of the rescalings of the 0- components $\vartheta _{%
\underline{\tau }}^{m}=\left( \theta _{0}^{m},\tilde{\theta}^{m0}\right) $
along the graviphoton dimension and the \b{M}-th components $\vartheta _{%
\text{\b{M}}}^{m}=\left( \theta _{i}^{m},\tilde{\theta}^{mi}\right) $ along
the Coulomb branch directions show that the rescaled embedding tensor $\hat{%
\vartheta}_{M}^{m}$ can be split in terms of the $\frac{1}{\mathrm{\mu }}$-
parameter like 
\begin{equation}
\vartheta _{M}^{m}\qquad \rightarrow \qquad \hat{\vartheta}_{M}^{m}=\frac{1}{%
M_{pl}^{2}}\left( \delta _{M}^{\underline{\tau }}\vartheta _{\underline{\tau 
}}^{m}+\frac{1}{\mathrm{\mu }}\delta _{M}^{\underline{M}}\vartheta _{%
\underline{M}}^{m}\right)  \label{st}
\end{equation}%
where the pure gravity contribution $\vartheta _{\underline{\tau }}^{m}$
appears as the leading term and the Coulomb branch one $\vartheta _{\text{\b{%
M}}}^{m}$ as the next term. In matrix notation, we have 
\begin{equation}
\hat{\vartheta}_{M}^{m}=\underline{\mathbf{\vartheta }}_{M}^{m}+\frac{1}{%
\mathrm{\mu }}\mathring{\vartheta}_{M}^{m}  \label{ts}
\end{equation}%
with%
\begin{equation}
\underline{\mathbf{\vartheta }}_{M}^{m}=\frac{1}{M_{pl}^{2}}\left( 
\begin{array}{c}
\theta _{0}^{m} \\ 
0 \\ 
\tilde{\theta}^{m0} \\ 
0%
\end{array}%
\right) \qquad ,\qquad \mathring{\vartheta}_{M}^{m}=\frac{1}{M_{pl}^{2}}%
\left( 
\begin{array}{c}
0 \\ 
\theta _{i}^{m} \\ 
0 \\ 
\tilde{\theta}^{mi}%
\end{array}%
\right)  \label{ss}
\end{equation}%
Notice that the rescaling of $\tilde{\theta}^{mi}$ and $\theta _{i}^{m}$ may
be compared with the rescalings of the FI coupling constants $\nu _{i}^{a}$
given by (\ref{nu}). Furthermore, using the expression of the Killing vector
field $\mathbf{\kappa }_{M}=\vartheta _{M}^{m}\mathrm{T}_{m}$ with
components 
\begin{equation}
\mathbf{\kappa }_{M}=\left( 
\begin{array}{c}
Y_{\Lambda } \\ 
\tilde{Y}^{\Lambda }%
\end{array}%
\right) \qquad ,\qquad Y_{\Lambda }=\left( 
\begin{array}{c}
Y_{0} \\ 
Y_{i}%
\end{array}%
\right) \qquad ,\qquad \tilde{Y}^{\Lambda }=\left( 
\begin{array}{c}
\tilde{Y}^{0} \\ 
\tilde{Y}^{i}%
\end{array}%
\right)
\end{equation}%
and%
\begin{equation}
\begin{tabular}{lllllll}
$Y_{0}$ & $=$ & $\theta _{0}^{m}\mathrm{T}_{m}$ & $\qquad ,\qquad $ & $%
\tilde{Y}^{0}$ & $=$ & $\tilde{\theta}^{0m}\mathrm{T}_{m}$ \\ 
$Y_{i}$ & $=$ & $\theta _{i}^{m}\mathrm{T}_{m}$ & $\qquad ,\qquad $ & $%
\tilde{Y}^{i}$ & $=$ & $\tilde{\theta}^{im}\mathrm{T}_{m}$%
\end{tabular}%
\end{equation}%
we deduce from (\ref{tet}) the following rescalings%
\begin{equation}
\begin{tabular}{lllllll}
$Y_{0}$ & $\rightarrow $ & $Y_{0}$ & $\qquad ,\qquad $ & $\tilde{Y}^{0}$ & $%
\rightarrow $ & $\tilde{Y}^{0}$ \\ 
$Y_{i}$ & $\rightarrow $ & $\frac{1}{\mathrm{\mu }}Y_{i}$ & $\qquad ,\qquad $
& $\tilde{Y}^{i}$ & $\rightarrow $ & $\frac{1}{\mathrm{\mu }}\tilde{Y}^{i}$%
\end{tabular}%
\end{equation}%
Before proceeding, let us collect in the two following tables the rescaling
behaviours of useful objects; they are needed later on%
\begin{equation}
\begin{tabular}{|l|l|l|l|l|l|l|}
\hline
{\small components fields } & $z^{i}$ & $\left( A_{\mu }^{i},\tilde{A}_{\mu
i}\right) $ \  & $\left( A_{\mu }^{0},\tilde{A}_{\mu 0}\right) $ \  & $\  \
Q^{u}$ \  & $\  \  \mathcal{C}_{\mu }^{a}$ & $\  \  \mathcal{G}_{i\bar{j}}$ \\ 
\hline
{\small rescaling factors} & $\frac{\mathrm{\mu }}{M_{pl}}$ & $\  \  \frac{%
\mathrm{\mu }}{M_{pl}}$ & $\  \  \  \frac{1}{M_{pl}}$ & $\  \frac{1}{M_{pl}^{2}}$
& $\  \  \frac{1}{M_{pl}^{3}}$ \  & $\  \  \frac{M_{pl}^{2}}{\mathrm{\mu }^{2}}$
\  \\ \hline
\end{tabular}
\label{ress}
\end{equation}%
and%
\begin{equation}
\begin{tabular}{|l|l|l|l|l|l|}
\hline
{\small embedding tensor} & $\left( \theta ^{mi},\tilde{\theta}%
_{i}^{m}\right) $ & $\left( \theta ^{m0},\tilde{\theta}_{0}^{m}\right) $ & $%
\left( Y^{i},\tilde{Y}_{i}\right) $ \  & $\left( Y^{0},\tilde{Y}_{0}\right) $
\  & $\  \  \mathrm{T}_{m}$ \  \\ \hline
{\small rescaling factors} & $\  \  \  \  \frac{1}{\mathrm{\mu }M_{pl}^{2}}$ & $%
\  \  \  \  \frac{1}{M_{pl}^{2}}$ & $\  \  \  \frac{1}{\mathrm{\mu }}$ & $\  \  \  \ 1$
& $\ M_{pl}^{2}$ \\ \hline
\end{tabular}
\label{tett}
\end{equation}%
\begin{equation*}
\text{ \  \ }
\end{equation*}%
Using the vielbein $\mathcal{E}^{A\dot{A}}=\mathcal{E}_{u}^{A\dot{A}}dQ^{u}$
on the quaternionic Kahler manifold (\ref{14}), we determine the rescaling
of $\mathcal{E}_{u}^{A\dot{A}}$; it is given by $\mathcal{E}^{A\dot{A}%
}\rightarrow M_{pl}^{2}\mathcal{E}_{u}^{A\dot{A}}.$

\subsection{Rigid limit of Ward identities}

\label{ssec52}

In this subsection we study the Ward identities of gauging quaternionic
isometries of the scalar manifold of $\mathcal{N}=2$ supergravity, their
rigid limit and the induced scalar potential $\mathcal{\mathring{V}}_{\text{%
kah}}^{\mathcal{N}=2}$. These limits will be used in next subsection for the
derivation of partial supersymmetry breaking conditions.

\subsubsection{Ward identities and induced scalar potential}

\label{rsp}

$\mathcal{N}=2$ local supersymmetry can be broken partially\textrm{\ by
gauging two abelian quaternionic isometries of the scalar manifold\ }of $%
\mathcal{N}=2$ supergravity theory. To tier $\mathcal{N}=2$ supersymmetry
and the quaternionic gauging isometries, we have to modify the usual
supersymmetric transformations of the fermions of the $\mathcal{N}=2$
supergravity theory by adding extra terms proportional to embedding tensor $%
{\small \vartheta }_{M}^{m}$ as in eq(\ref{et}) reported below. \newline
Following \textrm{\cite{Andrianopoli:2015rpa, D'Auria:1990fj, Louis:2012ux,
Dall'Agata:2006vd, Andrianopoli:1996cm}}, the gauging of abelian
quaternionic isometries affects the underlying supersymmetric properties of $%
\mathcal{N}=2$ supergravity; it requires modifying the usual supersymmetric
transformations $\mathrm{\delta }_{\epsilon }^{{\small (0)}}\chi _{\alpha }$
of the fermions $\chi _{\hat{\alpha}}$\ of the $\mathcal{N}=2$ local theory,
namely the gauginos $\lambda _{\hat{\alpha}}^{Ai}$, the hyperini $\zeta _{%
\hat{\alpha}}^{A}$ and the gravitini $\psi _{\hat{\alpha}\mu }^{A}$, like%
\begin{equation}
\begin{tabular}{lll}
$\mathrm{\delta }_{\epsilon }\lambda ^{Ai}$ & $=$ & $\mathrm{\delta }%
_{\epsilon }^{{\small (0)}}\lambda ^{Ai}+\mathrm{\delta }_{\epsilon }^{%
{\small (\vartheta )}}\lambda ^{Ai}$ \\ 
$\mathrm{\delta }_{\epsilon }\zeta ^{A}$ & $=$ & $\mathrm{\delta }_{\epsilon
}^{{\small (0)}}\zeta ^{A}+\mathrm{\delta }_{\epsilon }^{{\small (\vartheta )%
}}\zeta ^{A}$ \\ 
$\mathrm{\delta }_{\epsilon }\psi _{\mu }^{A}$ & $=$ & $\mathrm{\delta }%
_{\epsilon }^{{\small (0)}}\psi _{\mu }^{A}+\mathrm{\delta }_{\epsilon }^{%
{\small (\vartheta )}}\psi _{\mu }^{A}$%
\end{tabular}
\label{et}
\end{equation}%
where spinor indices $\hat{\alpha}=\left( \alpha ,\dot{\alpha}\right) $ have
has been dropped out for simplicity. The extra $\mathrm{\delta }_{\epsilon
}^{{\small (\vartheta )}}\chi $'s in above relations depend on the embedding
tensor $\vartheta _{M}^{a}$ of (\ref{77}) and encode data on the induced
scalar potential $\mathcal{V}_{\text{sugra}}^{\mathcal{N}=2}$ of the $%
\mathcal{N}=2$ supergravity theory whose structure will be described in a
moment; see \textrm{eqs(\ref{wid1}-\ref{wb3})}. The extra terms $\mathrm{%
\delta }_{\epsilon }^{{\small (\vartheta )}}\chi $ in (\ref{et}) depend on
the Killing vectors $\kappa _{M}^{u}=-\vartheta _{M}^{a}\delta _{a}^{u}$ (%
\ref{115}) and moment maps $\mathcal{P}_{M}^{a}$ (\ref{mp}) of the
quaternionic Kahler manifold $\boldsymbol{M}_{QK}$ and have the typical form
(\ref{stt}).\newline
Substituting $\kappa _{M}^{u}=-\vartheta _{M}^{a}\delta _{a}^{u}$ and using
the vielbein expression $\mathcal{E}_{uB}^{A}=\frac{1}{\mathrm{2}}\delta
_{B}^{A}\delta _{u}^{0}-\frac{i}{\mathrm{2}}e^{\varphi }\delta
_{u}^{a}\left( \tau _{a}\right) _{B}^{A}$ as well as $\mathcal{P}%
_{M}^{a}=e^{\varphi }\vartheta _{M}^{a}$ (\ref{mp}); then expanding the
traceless matrices $\left( W^{i}\right) _{B}^{A},$ $N_{B}^{A}$ and $%
S_{B}^{A} $ as 
\begin{equation}
\begin{tabular}{lll}
$\left( W^{i}\right) _{B}^{A}$ & $=$ & $W^{ia}\left( \tau _{a}\right) ^{AB}$
\\ 
$N_{B}^{A}$ & $=$ & $N^{a}\left( \tau _{a}\right) _{B}^{A}$ \\ 
$S_{B}^{A}$ & $=$ & $S^{a}\left( \tau _{a}\right) _{B}^{A}$%
\end{tabular}%
\end{equation}%
eqs(\textrm{\ref{wb1}}) can be brought to the form of complex isotriplet
equations%
\begin{equation}
\begin{tabular}{lll}
$W^{ia}$ & $=$ & $-i\mathcal{P}_{M}^{a}\mathcal{G}^{i\bar{j}}\bar{U}_{\bar{j}%
}^{M}$ \\ 
$N^{a}$ & $=$ & $i\mathcal{P}_{M}^{a}\bar{V}^{M}$ \\ 
$S^{a}$ & $=$ & $\frac{i}{2}\mathcal{P}_{M}^{a}V^{M}$%
\end{tabular}
\label{wb2}
\end{equation}%
Putting these relations back into (\ref{wid1}), we obtain the explicit
expression of the induced potential $\mathcal{V}_{\text{sugra}}^{\mathcal{N}%
=2}$ in terms of moment maps and geometric objects of the scalar manifold of
the $\mathcal{N}=2$ supergravity. By taking the trace over SU$\left(
2\right) $ R-symmetry indices, we obtain 
\begin{equation}
\mathcal{V}_{\text{sugra}}^{\mathcal{N}=2}=\frac{1}{2}\text{\textrm{Tr}}%
\left( -12\bar{S}_{C}^{A}S_{B}^{C}+2\bar{N}_{\dot{C}}^{A}N_{B}^{\dot{C}%
}+\sum_{i,j=1}^{n_{v}}\mathcal{G}_{i\bar{j}}\left( W^{i}\right)
_{C}^{A}\left( \bar{W}^{\bar{j}}\right) _{B}^{C}\right)
\end{equation}%
which reads also as%
\begin{equation}
\mathcal{V}_{\text{sugra}}^{\mathcal{N}=2}=-12\bar{S}^{a}S^{a}+2\bar{N}%
^{a}N^{a}+\sum_{i,j=1}^{n_{v}}\mathcal{G}_{i\bar{j}}W^{ia}\bar{W}^{\bar{j}a}
\label{wb3}
\end{equation}%
By using eqs(\ref{wb2}), we can bring the above expression of the scalar
potential first to the form%
\begin{equation}
\mathcal{V}_{\text{sugra}}^{\mathcal{N}=2}=\left[ \mathcal{G}^{i\bar{k}}\bar{%
U}_{\bar{k}}^{M}U_{k}^{N}\right] \mathcal{P}_{N}^{a}\mathcal{P}_{M}^{a}-%
\mathcal{P}_{N}^{a}\mathcal{P}_{M}^{a}V^{M}\bar{V}^{N}
\end{equation}%
and then to 
\begin{equation}
\mathcal{V}_{\text{sugra}}^{\mathcal{N}=2}=-\frac{1}{2}\mathcal{M}^{MN}%
\mathcal{P}_{N}^{a}\mathcal{P}_{M}^{a}-2V^{M}\bar{V}^{N}\mathcal{P}_{N}^{a}%
\mathcal{P}_{M}^{a}  \label{sps}
\end{equation}%
with $\mathcal{U}^{MN}=U_{i}^{M}\mathcal{G}^{i\bar{j}}\bar{U}_{\bar{j}}^{N}$
and the symmetric $\mathcal{M}^{MN}$ as in eq(\ref{tmn}).

\subsubsection{Rigid limit of Ward identities}

\label{sss522}

Here we study the rigid limits of the Ward identities, the scalar potential
and the matrix anomaly of the $\mathcal{N}=2$ supercurrent algebra. These
limits are obtained by working out the $\mathrm{\mu }$- expansions of the
fermion shifts $W^{ia},$ $N^{a}$ and $S^{a}$, \textrm{given in the \autoref%
{appendixE}}, and then of (\textrm{\ref{wid1}}) and (\ref{wb3}). Before
going into details let us first restore the scaling dimensions of the
various quantities involved in the computations. By implementing the
canonical dimension of $\mathcal{V}_{\text{sugra}}^{\mathcal{N}=2}$ using
Planck mass, the left hand of eq(\ref{wb3}) can be rewritten as follows 
\begin{equation}
\frac{1}{M_{pl}^{4}}\mathcal{V}_{\text{sugra}}^{\mathcal{N}=2}=-12\bar{S}%
^{a}S^{a}+2\bar{N}^{a}N^{a}+\sum_{i,j=1}^{n_{v}}\mathcal{G}_{i\bar{j}}W^{ia}%
\bar{W}^{\bar{j}a}
\end{equation}%
where now $\mathcal{V}_{\text{sugra}}^{\mathcal{N}=2}$ scales as mass$^{4}$.
Multiplying both sides of this equation by $M_{pl}^{4}$, the above relation
can be then brought to the form 
\begin{equation}
\mathcal{V}_{\text{sugra}}^{\mathcal{N}=2}=-12\bar{S}^{\prime a}S^{\prime
a}+2\bar{N}^{\prime a}N^{\prime a}+\mathcal{G}_{i\bar{j}}W^{\prime ai}\bar{W}%
^{^{\prime }a\bar{j}}  \label{prf}
\end{equation}%
where the primed $W^{\prime ia},$ $N^{\prime a}$ and $S^{\prime a}$ scale as
mass$^{2}$; they are related to the old dimensionless $W^{ia},$ $N^{a}$ and $%
S^{a}$ like%
\begin{equation}
\begin{tabular}{lll}
$W^{\prime ai}$ & $=$ & $M_{pl}^{2}$ $W^{ai}$ \\ 
$N^{\prime a}$ & $=$ & $M_{pl}^{2}$ $N^{a}$ \\ 
$S^{\prime a}$ & $=$ & $M_{pl}^{2}$ $S^{a}$%
\end{tabular}%
\end{equation}%
Below, we shall drop out the prime indices in (\ref{prf}) and think about $%
W^{ia},$ $N^{a}$ and $S^{a}$ as well as on the $\mathcal{V}_{\text{sugra}}^{%
\mathcal{N}=2}$ and the anomaly matrix $C_{A}^{B}$ as dimensionful objects.
To deal with the rigid limit of (\ref{prf}), we use the fermionic
transformations (\ref{stt}) to think of $\mathcal{V}_{\text{sugra}}^{%
\mathcal{N}=2}$ as given by the sum of three contributions $\mathcal{V}_{%
\text{kah}}^{\mathcal{N}=2},$ $\mathcal{V}_{\text{hyper}}^{\mathcal{N}=2}$, $%
\mathcal{V}_{\text{gra}}^{\mathcal{N}=2}$ respectively associated with the
transformations $\mathrm{\delta }_{\epsilon }\lambda ^{Ai},$ $\mathrm{\delta 
}_{\epsilon }\zeta ^{A},$ $\mathrm{\delta }_{\epsilon }\psi _{\mu }^{A}$ of
fermions belonging in the three supersymmetric field representations $%
\boldsymbol{V}_{\mathcal{N}=2}^{i},$ $\boldsymbol{H}_{\mathcal{N}=2}$ and $%
\boldsymbol{G}_{\mathcal{N}=2}$ of eq(\ref{21}) involved in the $\mathcal{N}%
=2$ supergravity. So, we have, 
\begin{equation}
\mathcal{V}_{\text{sugra}}^{\mathcal{N}=2}=\mathcal{V}_{\text{kah}}^{%
\mathcal{N}=2}+\mathcal{V}_{\text{hyper}}^{\mathcal{N}=2}-\mathcal{V}_{\text{%
gra}}^{\mathcal{N}=2}
\end{equation}%
where we have set%
\begin{equation}
\begin{tabular}{lll}
$\mathcal{V}_{\text{kah}}^{\mathcal{N}=2}$ & $=$ & $\mathcal{G}_{i\bar{j}%
}W^{ai}\delta _{ab}\bar{W}^{b\bar{j}}$ \\ 
$\mathcal{V}_{\text{hyper}}^{\mathcal{N}=2}$ & $=$ & $2\bar{N}^{a}\delta
_{ab}N^{b}$ \\ 
$\mathcal{V}_{\text{gra}}^{\mathcal{N}=2}$ & $=$ & $12\bar{S}^{a}\delta
_{ab}S^{b}$%
\end{tabular}%
\end{equation}%
with $W^{ai},N^{a},$ $S^{a}$ scaling as mass$^{2}$; but functions of the
dimensionless expansion parameter $\mathrm{\mu }$ as described in what
follows.

$b)$ \emph{Rigid limit of Ward identities}\newline
To determine the rigid limit of Ward identities (\ref{wid1}), we use the $%
\frac{1}{\mathrm{\mu }}$- expansion of the scalar potential $\mathcal{V}_{%
\text{sugra}}^{\mathcal{N}=2}$, \textrm{given in the} \textrm{\autoref%
{appendixE}, }and the development $\mathcal{V}_{\text{sugra}}^{\mathcal{N}%
=2}=\sum_{n}\mathrm{\mu }^{-n}\mathcal{V}_{(n)}^{\mathcal{N}=2}$ to
determine the rigid limit of the Ward identities that we re-express as
follows 
\begin{equation}
\mathcal{G}_{i\bar{j}}\left( W^{i}\right) _{C}^{A}\left( \bar{W}^{\bar{j}%
}\right) _{B}^{C}=\delta _{B}^{A}.\mathcal{V}_{\text{sugra}}^{\mathcal{N}%
=2}-2\bar{N}_{\dot{C}}^{A}N_{B}^{\dot{C}}+12\bar{S}_{C}^{A}S_{B}^{C}
\label{lw}
\end{equation}%
The left hand side of above relation gives the contribution coming from the
gauged supersymmetric transformations $\mathrm{\delta }^{\left( \vartheta
\right) }\lambda _{\hat{\alpha}}^{A}$ of the two gauginos (\ref{stt}); its
trace gives the contribution to the scalar potential in the Coulomb branch.

\  \  \  \ 

$\bullet $ \emph{Rigid Ward identities}\newline
These \textrm{Ward identities} are obtained by taking the rigid limit of the
supergravity ones (\ref{lw}); they are given by%
\begin{equation}
\sum_{i,j=1}^{n_{v}}\mathcal{\mathring{G}}_{i\bar{j}}(\mathring{W}%
^{i})_{C}^{A}(\overline{\mathring{W}}^{\bar{j}})_{B}^{C}=\delta _{B}^{A}%
\mathcal{\mathring{V}}_{\text{kah}}^{\mathcal{N}=2}+\boldsymbol{\mathring{C}}%
_{B}^{A}  \label{wd}
\end{equation}%
with $\mathcal{\mathring{V}}_{\text{kah}}^{\mathcal{N}=2}$ like in (\ref{vkr}%
). The extra non diagonal $\boldsymbol{\mathring{C}}_{B}^{A}$ , which
corresponds to the rigid limit of the anomalous term of the $\mathcal{N}=2$
supercurrent algebra, is an interesting term as it captures crucial data on $%
\mathcal{N}=2$ supersymmetry breaking. It scales in same manner as the
scalar potential $\mathcal{\mathring{V}}_{\text{kah}}^{\mathcal{N}=2}$ and
reads as follows%
\begin{equation}
\boldsymbol{\mathring{C}}_{B}^{A}=12\overline{\mathring{S}}_{C}^{A}\mathring{%
S}_{B}^{C}-2\overline{\mathring{N}}_{\dot{C}}^{A}\mathring{N}_{B}^{\dot{C}}
\label{cab}
\end{equation}%
where the derivation of the expressions of $\mathring{S}_{B}^{C}$ and $%
\mathring{N}_{B}^{\dot{C}}$\ is reported in \autoref{appendixE}. By
replacing $\mathring{N}_{B}^{A}$ and $\mathring{S}_{B}^{A}$ by their values (%
\ref{na},\ref{an}) in terms of the embedding tensor components, this anomaly
can be expressed like $\boldsymbol{\mathring{C}}_{B}^{A}=\sum_{c=1}^{3}%
\boldsymbol{\zeta }_{c}\left( \tau ^{c}\right) _{B}^{A}$ with isovector $%
\boldsymbol{\zeta }_{c}$ as 
\begin{equation}
\boldsymbol{\zeta }_{c}=\frac{1}{2}\varepsilon _{abc}\left( \tilde{\theta}%
^{a0}\theta _{0}^{b}-\theta _{0}^{a}\tilde{\theta}^{b0}\right)  \label{xc}
\end{equation}%
Putting back into (\ref{wd}), the explicit form of rigid limit of Ward
identities in terms of the embedding tensor read then as follows%
\begin{equation}
\mathcal{\mathring{G}}_{i\bar{j}}(\mathring{W}^{i})_{C}^{A}(\overline{%
\mathring{W}}^{\bar{j}})_{B}^{C}=\delta _{B}^{A}\left[ \frac{1}{2}\mathcal{%
\mathring{M}}^{\underline{M}\underline{N}}\mathring{\vartheta}_{\underline{N}%
}^{a}\mathring{\vartheta}_{\underline{M}}^{a}\right] +\frac{1}{2}\varepsilon
_{abc}\left( \tilde{\theta}^{a0}\theta _{0}^{b}-\theta _{0}^{a}\tilde{\theta}%
^{b0}\right) \left( \tau ^{c}\right) _{B}^{A}  \label{rwi}
\end{equation}%
Notice the three following features concerning this analysis: $\left(
i\right) $ the antisymmetric tensor $\tilde{\theta}^{a0}\theta _{0}^{b}-%
\tilde{\theta}^{b0}\theta _{0}^{a}$ involved in the isovector (\ref{xc}) is
a remarkable quantity; it can be also expressed like $\vartheta _{\underline{%
\tau }}^{a}\mathcal{C}^{\underline{\tau }\underline{\sigma }}\vartheta _{%
\underline{\sigma }}^{b}$ with $\vartheta _{\underline{\tau }}^{a}=\left(
\theta _{0}^{a},\tilde{\theta}^{a0}\right) $ standing for an SP$\left( 2,%
\mathbb{R}\right) $ vector and $\mathcal{C}^{\underline{\tau }\underline{%
\sigma }}$ the metric of this symplectic group%
\begin{equation}
\mathcal{C}^{\underline{\tau }\underline{\sigma }}=\left( 
\begin{array}{cc}
0 & -1 \\ 
1 & 0%
\end{array}%
\right)
\end{equation}%
This SP$\left( 2,\mathbb{R}\right) $ should be thought of as the subgroup of
SP$\left( 2n+2,\mathbb{R}\right) $ associated with the pure supergravity
sector.\newline
$\left( ii\right) $ the isotriplet $\boldsymbol{\zeta }_{c}$ is also
invariant under the SP$\left( 2n,\mathbb{R}\right) $ symplectic symmetry of
the rigid theory; a property which may be explicitly exhibited by using the
constraint relation $\vartheta _{M}^{a}\mathcal{C}^{MN}\vartheta _{N}^{b}=0$
which splits like $\vartheta _{\underline{\tau }}^{a}\mathcal{C}^{\underline{%
\tau }\underline{\sigma }}\vartheta _{\underline{\sigma }}^{b}=-\mathring{%
\vartheta}_{\underline{M}}^{a}\mathcal{C}^{\underline{M}\underline{N}}%
\mathring{\vartheta}_{\underline{N}}^{a}$. So the above SP$\left( 2,\mathbb{R%
}\right) $ invariant isovector $\boldsymbol{\zeta }_{c}=\frac{1}{2}%
\varepsilon _{abc}\vartheta _{\underline{\tau }}^{a}\mathcal{C}^{\underline{%
\tau }\underline{\sigma }}\vartheta _{\underline{\sigma }}^{b}$ can be as
well written like%
\begin{equation}
\boldsymbol{\zeta }_{c}=-\frac{1}{2}\varepsilon _{abc}\mathring{\vartheta}_{%
\underline{M}}^{a}\mathcal{C}^{\underline{M}\underline{N}}\mathring{\vartheta%
}_{\underline{N}}^{a}  \label{cx}
\end{equation}%
which coincides with the SP$\left( 2n,\mathbb{R}\right) $ invariant
isovector of the rigid theory of \textrm{\cite{Andrianopoli:2015wqa}. The
two ways (\ref{cab}) and (\ref{cx}) of expressing the anomaly show that }the
partial breaking of the rigid theory can be derived either by using the SP$%
\left( 2n,\mathbb{R}\right) $ symplectic structure of the vector multiplet
of the observable sector as in \textrm{\cite{Andrianopoli:2015wqa}}; or by
using the SP$\left( 2,\mathbb{R}\right) $ symplectic structure associated
with the graviphoton of the hidden sector. The two equivalent expressions
show as well that in the rigid limit, the SP$\left( 2n+2,\mathbb{R}\right) $
symplectic symmetry of the $\mathcal{N}=2$ supergravity theory gets broken
down to $SP\left( 2n,\mathbb{R}\right) \times SP\left( 2,\mathbb{R}\right) $.%
\newline
$\left( iii\right) $ Eq(\ref{cab}) is a general relation as it is directly
expressed in terms of the embedding tensor components. From this formula, we
can recover the expression of the traceless matrix $\boldsymbol{\mathring{C}}%
_{B}^{A}$ obtained in literature. By choosing the components of the
embedding tensor like%
\begin{equation}
\begin{tabular}{lllllll}
$\tilde{\theta}^{a0}$ & $=$ & $\eta _{i}m^{ia}$ & $\qquad ,\qquad $ & $%
\theta _{0}^{a}$ & $=$ & $e^{a}$ \\ 
$\tilde{\theta}^{ia}$ & $=$ & $m^{ia}$ & $\qquad ,\qquad $ & $\theta
_{i}^{a} $ & $=$ & $-\eta _{i}e^{a}$%
\end{tabular}
\label{ab}
\end{equation}%
where the parameter $m^{ia}$, $e^{a}$ and $\eta _{i}$ are as in \textrm{\cite%
{Andrianopoli:2015rpa}, we end with the }following moment maps%
\begin{equation}
\mathcal{P}_{\underline{M}}^{a}=%
\begin{pmatrix}
-e_{i}^{a} \\ 
m^{ia}%
\end{pmatrix}%
\qquad ,\qquad \mathcal{P}_{\underline{\tau }}^{a}=%
\begin{pmatrix}
e^{a} \\ 
m^{a}%
\end{pmatrix}%
\end{equation}%
with $e_{i}^{a}\equiv \eta _{i}e^{a}$ and $m^{a}\equiv \eta _{i}m^{ia}$. The
obove $\mathcal{P}_{\underline{M}}^{a}$ is precisely the moment maps
obtained in the rigid theory of \textrm{\cite{Andrianopoli:2015wqa}}. Notice
that in the old frame, that is before performing the symplectic
transformation, the choice (\ref{ab}) corresponds to an embedding tensor of
the form,%
\begin{equation}
\vartheta _{M}^{a}=%
\begin{pmatrix}
e^{a} \\ 
0 \\ 
0 \\ 
\mu m^{ia}%
\end{pmatrix}%
\end{equation}%
and should compared with the solution $(ii)$ of (\ref{s41}).

\section{Partial supersymmetry breaking}

\label{sec6}

In this section, we study the conditions for partial supersymmetry breaking
in the rigid limit of $\mathcal{N}=2$ gauged supergravity. By taking this
limit, one can distinguish two sectors: $(i)$ the observable sector
characterised by the contribution of the vector multiplet to the Ward
identities; and $(ii)$ the hidden sector with contribution to Ward
identities coming from gravity and hypermatter. To study the breaking we
will consider below these two sectors; first, we focus on the partial
breaking of $\mathcal{N}=2$ supersymmetry in the observable sector by
deriving the scalar potential and the anomaly in the rigid limit. Then, we
study of the partial breaking in the hidden sector by following the same
approach.

\subsection{Observable sector}

With the analysis on the $\frac{1}{\mathrm{\mu }}$- expansion of Ward
identities of $\mathcal{N}=2$ gauged supergravity given in \textrm{\autoref%
{sss522}}, we can now study the conditions for partial breaking of $\mathcal{%
N}=2$ supersymmetry in the rigid limit. From eq(\ref{rwi}), it follows that
the rigid limit of the Ward identities of the gauging of two abelian
isometries of quaternionic manifold $\boldsymbol{M}_{QK}$ can be put into
the the following hermitian $2\times 2$ matrix form 
\begin{equation}
\sum_{i,j=1}^{n_{v}}\mathcal{\mathring{G}}_{i\bar{j}}(\mathring{W}%
^{i})_{C}^{A}(\overline{\mathring{W}}^{\bar{j}})_{B}^{C}=\boldsymbol{H}%
_{B}^{A}  \label{hg}
\end{equation}%
with%
\begin{equation}
\boldsymbol{H}_{B}^{A}=%
\begin{pmatrix}
\mathcal{\mathring{V}}_{\text{kah}}^{\mathcal{N}=2}+\zeta _{3} & \zeta
_{1}-i\zeta _{2} \\ 
\zeta _{1}+i\zeta _{2} & \mathcal{\mathring{V}}_{\text{kah}}^{\mathcal{N}%
=2}-\zeta _{3}%
\end{pmatrix}%
\end{equation}%
By help of a unitary transformation $\boldsymbol{U}$ on the matrices $(%
\mathring{W}^{i})_{C}^{A}$, the hermitian matrix $\boldsymbol{H}$ in the
right hand side of (\ref{hg}) can be put into a diagonal form $\boldsymbol{%
\hat{H}}=\left( \boldsymbol{UHU}^{\dagger }\right) $ given by%
\begin{equation}
\boldsymbol{\hat{H}}_{B}^{A}=%
\begin{pmatrix}
\mathcal{\mathring{V}}_{\text{kah}}^{\mathcal{N}=2}+\sqrt{|\zeta |^{2}} & 0
\\ 
0 & \mathcal{\mathring{V}}_{\text{kah}}^{\mathcal{N}=2}-\sqrt{|\zeta |^{2}}%
\end{pmatrix}%
\end{equation}%
where 
\begin{equation}
|\zeta |^{2}\equiv \left( \zeta _{1}\right) ^{2}+\left( \zeta _{2}\right)
^{2}+\left( \zeta _{3}\right) ^{2}  \label{vze}
\end{equation}%
standing for the norm of the real isovector $\zeta _{a}$. Notice that the
gauging of two abelian quaternionic isometries requires an embedding tensor $%
\vartheta _{M}^{a}=\left( \vartheta _{M}^{m},\vartheta _{M}^{3}\right) $ of
the form%
\begin{equation}
\vartheta _{M}^{m}=\left( 
\begin{array}{c}
\theta _{\Lambda }^{m} \\ 
\tilde{\theta}^{m\Lambda }%
\end{array}%
\right) \qquad ,\qquad m=1,2\qquad ,\qquad \Lambda =0,i=1,...,n
\end{equation}%
and 
\begin{equation}
\vartheta _{M}^{3}=0
\end{equation}%
Therefore the isovector $\boldsymbol{\zeta }_{c}$ in (\ref{xc}) and $|\zeta
|^{2}$ should be read as 
\begin{equation}
\boldsymbol{\zeta }_{3}=\frac{1}{2}\varepsilon _{3mn}\left( \tilde{\theta}%
^{m0}\theta _{0}^{n}-\theta _{0}^{m}\tilde{\theta}^{n0}\right) \qquad
,\qquad |\zeta |^{2}=\left( \zeta _{3}\right) ^{2}  \label{z3}
\end{equation}%
This feature, required by the gauging of two abelian quaternionic
isometries, means that the isovector $\boldsymbol{\zeta }_{c}$ has in fact
one component as follows 
\begin{equation}
\boldsymbol{\zeta }_{c}=\frac{1}{2}\left( 
\begin{array}{c}
0 \\ 
0 \\ 
\tilde{\theta}^{10}\theta _{0}^{2}-\theta _{0}^{1}\tilde{\theta}^{20}%
\end{array}%
\right)
\end{equation}%
It breaks the SU$\left( 2\right) $ R- symmetry down to U$\left( 1\right)
_{R} $ in agreement with partial breaking of $\mathcal{N}=2$ supersymmetry.
So the matrix eq(\ref{hg}) may be interpreted like%
\begin{equation}
\sum_{i,j=1}^{n_{v}}\mathcal{\mathring{G}}_{i\bar{j}}(\mathring{W}^{\prime
i})_{C}^{A}(\overline{\mathring{W}}^{\prime \bar{j}})_{B}^{C}=%
\begin{pmatrix}
\mathcal{\mathring{V}}_{\text{kah}}^{\mathcal{N}=2}+|\zeta _{3}| & 0 \\ 
0 & \mathcal{\mathring{V}}_{\text{kah}}^{\mathcal{N}=2}-|\zeta _{3}|%
\end{pmatrix}%
\end{equation}%
with $(\mathring{W}^{\prime i})_{C}^{A}=\boldsymbol{U}_{B}^{A}(\mathring{W}%
^{i})_{C}^{B}$. The $\mathcal{N}=2$ supersymmetry in the rigid limit of the
gauged supergravity theory is then partially broken if the following
condition holds 
\begin{equation}
\mathcal{\mathring{V}}_{\text{kah}}^{\mathcal{N}=2}=|\zeta _{3}|\qquad
,\qquad |\zeta _{3}|\neq 0  \label{cpb}
\end{equation}%
it is recovered if $|\zeta _{3}|\rightarrow 0$. This condition agrees with
known results in this matter as recently done in \textrm{\cite%
{Andrianopoli:2015wqa}}.

\subsection{Hidden sector}

In the hidden sector of $\mathcal{N}=2$ gauged supergravity concerning the
gravity and matter supermultiplets, the contribution of gauging quaternionic
isometries to the rigid limit of the Ward identities can be expressed like%
\begin{equation}
2\overline{\mathring{N}}_{\dot{C}}^{A}\mathring{N}_{B}^{\dot{C}}-12\overline{%
\mathring{S}}_{C}^{A}\mathring{S}_{B}^{C}=\delta _{B}^{A}.\mathcal{V}_{\text{%
sugra}}^{\mathcal{N}=2}-\sum_{i,j=1}^{n_{v}}\mathcal{G}_{i\bar{j}}\left( 
\mathring{W}^{i}\right) _{C}^{A}(\overline{\mathring{W}}^{\bar{j}})_{B}^{C}
\end{equation}%
This quantity reads in terms of the moment maps and the covariantly
covariant sections along the graviphoton direction as follows%
\begin{equation}
2\overline{\mathring{N}}_{\dot{C}}^{A}\mathring{N}_{B}^{\dot{C}}-12\overline{%
\mathring{S}}_{C}^{A}\mathring{S}_{B}^{C}=-2\delta _{B}^{A}\mathcal{P}_{%
\underline{\rho }}^{a}\mathcal{P}_{\underline{\sigma }}^{a}V^{\underline{%
\rho }}\bar{V}^{\underline{\sigma }}-i\varepsilon _{abc}\left( \tau
^{c}\right) _{B}^{A}\mathcal{P}_{\underline{\rho }}^{a}\mathcal{P}_{%
\underline{\sigma }}^{b}V^{\underline{\rho }}\bar{V}^{\underline{\sigma }}
\end{equation}%
Substituting the $\mathcal{P}_{\underline{\sigma }}^{a}$'s and the $V^{%
\underline{\rho }}$'s by their expressions, we can put the above expression
into the form%
\begin{equation}
2\overline{\mathring{N}}_{\dot{C}}^{A}\mathring{N}_{B}^{\dot{C}}-12\overline{%
\mathring{S}}_{C}^{A}\mathring{S}_{B}^{C}=2\delta _{B}^{A}\Delta \mathcal{%
\mathring{V}-}\boldsymbol{\mathring{C}}_{B}^{A}
\end{equation}%
where the rigid anomaly $\boldsymbol{\mathring{C}}_{B}^{A}$ is as in eq(\ref%
{cab}) and where $\Delta \mathcal{\mathring{V}}$ is given by (\ref{vhid}).
Thinking of $2\delta _{B}^{A}\Delta \mathcal{\mathring{V}-}\boldsymbol{%
\mathring{C}}_{B}^{A}$ as a hermitian 2$\times $2 matrix $\mathbb{H}_{B}^{A}$
as%
\begin{equation}
\mathbb{H}_{B}^{A}=%
\begin{pmatrix}
2\Delta \mathcal{\mathring{V}}-\zeta _{3} & -\zeta _{1}+i\zeta _{2} \\ 
-\zeta _{1}-i\zeta _{2} & 2\Delta \mathcal{\mathring{V}}+\zeta _{3}%
\end{pmatrix}
\label{hhid}
\end{equation}%
We can diagonalise it by help of a unitary transformation to bring it to the
following form%
\begin{equation}
\widehat{\mathbb{H}}_{B}^{A}=%
\begin{pmatrix}
2\Delta \mathcal{\mathring{V}}+\sqrt{|\zeta |^{2}} & 0 \\ 
0 & 2\Delta \mathcal{\mathring{V}}-\sqrt{|\zeta |^{2}}%
\end{pmatrix}%
\end{equation}%
where $|\zeta |^{2}$ is as in (\ref{vze}). Like for the observable sector
and because of the gauging of two abelian isometries, which breaks SU$\left(
2\right) _{R}$ symmetry down to U$\left( 1\right) _{R}$, the above diagonal
matrix reduces to%
\begin{equation}
\widehat{\mathbb{H}}_{B}^{A}=%
\begin{pmatrix}
2\Delta \mathcal{\mathring{V}}+|\zeta _{3}| & 0 \\ 
0 & 2\Delta \mathcal{\mathring{V}}-|\zeta _{3}|%
\end{pmatrix}%
\end{equation}%
Thus, $\mathcal{N}=2$ supersymmetry in the hidden supergravity sector is
partially broken if the following conditions hold 
\begin{equation}
-2\Delta \mathcal{\mathring{V}}=\pm |\zeta _{3}|\qquad ,\qquad |\zeta
_{3}|\neq 0  \label{vx}
\end{equation}%
Replacing $\Delta \mathcal{\mathring{V}}$ and $|\zeta _{3}|$ by their
respective expressions as in eqs(\ref{vhid}) and\ (\ref{z3}), we obtain%
\begin{equation}
\dsum \limits_{m=1,2}\left( \tilde{\theta}^{m0}\tilde{\theta}^{m0}+4\theta
_{0}^{m}\theta _{0}^{m}\right) =\pm \dsum \limits_{m=1,2}\varepsilon
_{3mn}\left( \tilde{\theta}^{m0}\theta _{0}^{n}-\theta _{0}^{m}\tilde{\theta}%
^{n0}\right)
\end{equation}%
and leads to the following conditions on the elctric $\theta _{0}^{n}$ and
the magnetic $\tilde{\theta}^{n0}$ components of the embedding tensor%
\begin{equation}
\left( \tilde{\theta}^{10}\right) ^{2}+\left( \tilde{\theta}^{20}\right)
^{2}+4\left( \theta _{0}^{1}\right) ^{2}+4\left( \theta _{0}^{2}\right)
^{2}\mp \left( \tilde{\theta}^{10}\theta _{0}^{2}-\tilde{\theta}^{20}\theta
_{0}^{1}\right) =0  \label{xv}
\end{equation}%
Comparing these conditions with the analysis of \textrm{\cite%
{Andrianopoli:2015rpa}}, where the particular choice $\theta _{0}^{1}=\tilde{%
\theta}^{20}=0$ has been used for the values of the embedding tensor, the
above constraint relations reduce to%
\begin{equation}
\left( \theta _{0}^{2}\right) ^{2}\mp \tilde{\theta}^{10}\theta _{0}^{2}+%
\frac{1}{4}\left( \tilde{\theta}^{10}\right) ^{2}=0
\end{equation}%
These constraints factorise like $\left( \theta _{0}^{2}\mp \frac{1}{2}%
\tilde{\theta}^{10}\right) ^{2}=0$ and are solved like $\tilde{\theta}%
^{10}=\pm 2\theta _{0}^{2}$ in perfect agreement with the result of \textrm{%
\cite{Andrianopoli:2015rpa}}.

\section{Conclusion and discussions}

\label{sec7}

Using a novel manner of rescaling fields in $\mathcal{N}=2$ gauged
supergravity, we have developed in this paper the explicit derivation of the
rigid limit of Ward identities of the gauging of two abelian quaternionic
isometries agreeing with known results in literature; the two abelian
gaugings are necessary for partial supersymmetry breaking. This explicit
analysis allowed us to figure out the $\frac{1}{\mathrm{\mu }}$- expansion ($%
\frac{1}{\mathrm{\mu }}$ of order $\frac{m_{3/2}}{M_{pl}}<<1$) of the basic
quantities of this $\mathcal{N}=2$ theory such as the symplectic sections of
special Kahler geometry $\boldsymbol{M}_{SK}$; the embedding tensor $%
\vartheta _{M}^{m}$ and the moment maps $\mathcal{P}_{M}^{m}$ as well as
Ward identities. It also allowed us to determine the rigid limit of the
induced scalar potential $\mathcal{V}_{kah}^{\mathcal{N}=2}$ in the
observable and hidden sectors; and the rigid limit of the central anomaly $%
\mathcal{C}_{B}^{A}=\zeta _{a}\left( \tau ^{a}\right) _{B}^{A}$ of the $%
\mathcal{N}=2$ supercurrent algebra as well as the conditions for partial
supersymmetry breaking. Known results in literature have been recovered by
making particular choices of the embedding tensor components; and new
features have been also obtained.

\  \  \  \  \newline
To study the $\frac{1}{\mathrm{\mu }}$- expansion of Ward identities of $%
\mathcal{N}=2$ gauged supergravity and their rigid limit, we have proceeded
as follows: $\left( a\right) $ We have used the approach of \textrm{\cite%
{Andrianopoli:2015rpa}} based on working with a special holomorphic
prepotential $\mathcal{F}$ depending, in addition to the usual complex
holomorphic $X^{\Lambda }\left( z\right) $ of the Coulomb branch, on $n_{V}$
parameters $\eta _{i}$ whose geometric interpretation has been given in
present study. $\left( b\right) $ We have organised the presentation of our
analysis into five complementary sections in order to exhibit explicitly the
contributions of each one of the three sectors of the theory namely gravity,
gauge and matter. Our motivation in doing so is to take the opportunity of
this explicit analysis in order to $\left( i\right) $ complete partial
results in literature, $\left( ii\right) $ shed more light on some crucial
steps in the derivation of the rigid limit; and $\left( iii\right) $ give
some generalisations like the ADE classification developed in \textrm{%
\autoref{ss42}}. The main lines of the rigid limit study performed in
present paper may be summarised in \textrm{three principal steps }as
follows: \newline
First, we started by introducing some basic tools on the scalar manifold $%
\boldsymbol{M}_{scal}$ of $\mathcal{N}=2$ supergravity which factorises as $%
\boldsymbol{M}_{SK}\times \boldsymbol{M}_{QK}$ with $\boldsymbol{M}_{SK}$
standing for the special Kahler submanifold, associated with the Coulomb
branch, and $\boldsymbol{M}_{QK}$ referring to the submanifold associated
with the matter sector. We have also recalled useful relations on Ward
identities in $\mathcal{N}=2$ gauged supergravity and on solutions of the
fermionic shifts $\left( W^{i}\right) _{B}^{A}$, $N_{\alpha }^{A}$ and $%
S_{B}^{A}$, induced by the gauging of quaternionic isometries, in terms of
the covariantly holomorphic sections $V^{M}\left( z,\bar{z}\right) $ living
on $\boldsymbol{M}_{SK}$ as well as the vielbein $\mathcal{E}_{u\alpha }^{A}$
and the moment maps $\mathcal{P}_{M}^{m}$ of the special quaternionic $%
\boldsymbol{M}_{QK}$ given by eqs(\ref{wb1}). \newline
In the second step, we have studied the $\frac{1}{\mathrm{\mu }}$- expansion
of the various $SP\left( 2n_{V}+2\right) $ symplectic sections of $%
\boldsymbol{M}_{SK}$ with expansion parameter $\frac{1}{\mathrm{\mu }}=\frac{%
\Lambda }{M_{pl}}$ where the massive scale $\Lambda <<M_{pl}$ is of order of 
$m_{3/2},$ the mass of gravitino multiplet generated by partial
supersymmetry breaking as sketched by eq(\ref{grv}). These expansions are
needed for the determination of the rigid limit of Ward identities $%
\dsum_{t}\gamma _{t}\delta _{A}\chi _{t}^{C}\delta _{C}\chi _{t}^{B}=\delta
_{A}^{B}\mathcal{V}_{\text{sugra}}^{\mathcal{N}=2}$ as well as the
derivation of the rigid limit of the induced scalar potential (\ref{vkr})
and the rigid limit of the anomalous central charge matrix (\ref{cab}, \ref%
{xc},\ref{cx}) of the $\mathcal{N}=2$ supercurrent algebra (\ref{as}-\ref%
{sett}). After that we have studied the gauging of abelian isometries of a
family of special quaternionic manifolds $\boldsymbol{M}_{QK}^{\left(
n_{H}=r\right) }$ whose first element is given by $\boldsymbol{M}%
_{QK}^{\left( n_{H}=1\right) }=SO\left( 1,4\right) /SO\left( 4\right) $
involving one hypermultiplet. Next, we have addressed the solving of the
embedding tensor constraint equation (\ref{13}). We derived different
solutions for (\ref{13}) which include the interesting dyonic solution as
shown by eqs(\ref{s41},\ref{sett}). We have also given an interpretation of
the embedding tensor $\vartheta _{M}^{u}$ in terms of D-branes of type II
strings on Calabi-Yau threefolds and type IIA/IIB mirror symmetry. \newline
In the third step, we have first described the rescalings of the various
fields of the $\mathcal{N}=2$ gauged supergravity. In this regards, we have
proposed another way to implement the scale dimensions that leads to the
right mass dimensions of the components of the embedding tensor which can be
interpreted in terms of FI couplings scaling as mass$^{2}$ as in eq(\ref{nu}%
); see also eqs(\ref{ress}-\ref{tett}). In our approach fields belonging to
the same $\mathcal{N}=2$ representation have the same dependence on the
scale $\Lambda $ as shown on eqs(\ref{gr},\ref{fe}) and (\ref{fh},\ref{hyre}%
). Then, we have derived the rigid limit of Ward identities of gauged
quaternionic isometries of the special quaternionic manifold $\boldsymbol{M}%
_{QK}^{\left( n_{H}=1\right) }$. As a result in this direction, we have
shown that the rigid limit of the induced scalar potential is given by eqs(%
\ref{2m}-\ref{vkr}). We have also derived the explicit expression of the
rigid limit $\mathring{C}_{B}^{A}$ of the anomalous central extension matrix 
$C_{B}^{A}$ of the supercurrent algebra in $\mathcal{N}=2$ supergravity (\ref%
{as}-\ref{sa}) and determined the conditions for partial supersymmetry
breaking both in observable and hidden sectors; see eq(\ref{cpb},\ref{vx}-%
\ref{xv}). In this context, known results in literature have been rederived
by making particular choices of the embedding tensor components.

\  \  \  \newline
We end this study by making a rough discussion regarding the gauging of
abelian quaternionic isometries in $\mathcal{N}=2$ supergravity with several
hypermultiplets parameterising $\boldsymbol{M}_{QK}^{\left( ADE\right) }$; a
rigorous analysis requires the derivation of exact the solutions of Ward
identities extending eqs(\textrm{\ref{wb1}}). Here, we use general arguments
to derive the structure of the rigid limits of the induced scalar potential $%
\mathcal{\mathring{V}}_{\text{kah}}^{\left( ADE\right) }$ and the matrix
anomaly $\left( C^{ADE}\right) _{B}^{A}$ in the observable sector. \newline
We begin by making a comment on the extension and describing the motivation
behind the ADE generalisation. As a comment, notice that in our ADE model to
deal with the matter sector of $\mathcal{N}=2$ gauged supergravity, the
number $n_{H}$ of hypermultiplets is interpreted as the rank $r$ of a finite
dimensional ADE Lie algebra; the simplest model corresponds therefore to $%
su\left( 2\right) $ algebra; it concerns the $SO\left( 1,4\right) /SO\left(
4\right) $ geometry extensively studied in the present paper. The ADE
geometry for several hypermultiplets is motivated from a remarkable property
of the 4- dimensional real manifold $SO\left( 1,4\right) /SO\left( 4\right) $
of the $\mathcal{N}=2$ gauged supergravity with $n_{H}=1$. For this simple
model, the factor $e^{2\varphi }$, involved in the metric (\ref{qm}) of the
coset $SO\left( 1,4\right) /SO\left( 4\right) $, may be put in
correspondence with the $e^{2u}$ term appearing in the 2d- Liouville theory
with integrable field equation given by 
\begin{equation}
\frac{\partial ^{2}u}{\partial \bar{z}\partial z}+\kappa e^{2u}=0\qquad
,\qquad u=u\left( z,\bar{z}\right)
\end{equation}%
The integrability of this 2d- field equation is known to be captured by a
Lax pair $L_{\pm }$ valued in a hidden $su\left( 2\right) \simeq A_{1}$
algebra having a flat curvature. By using the formal correspondence between
the exponential factors appearing in the two theories ( $e^{2u}%
\leftrightarrow e^{2\varphi })$, the metric (\ref{qm}) can be conjectured to
have as well a hidden $su\left( 2\right) $ symmetry where the number $%
n_{H}=1 $ is interpreted as the rank of $su\left( 2\right) $. Within this
view, we have built\textrm{\ in \autoref{ss42}} a family of 4r- dimensional
manifolds $\boldsymbol{M}_{QK}^{\left( ADE\right) }$ with metric as in (\ref%
{te}) generalising the (\ref{qm}) of the 4- dimensional $\boldsymbol{M}%
_{QK}^{\left( A_{1}\right) }\sim SO\left( 1,4\right) /SO\left( 4\right) $.
These ADE geometries can be put in turns into correspondence with integrable
ADE field equations of 2d- Toda field theories based on finite dimensional
ADE Lie algebras%
\begin{equation}
\begin{tabular}{l|l|l}
{\small Integrable 2d- QFT} & {\small hypermultiplet geometry} & {\small %
hidden symmetry} \\ \hline
{\small \  \  \ 2d- Liouville } & $\  \  \  \  \  \  \  \  \  \  \  \  \  \boldsymbol{M}%
_{QK}^{\left( A_{1}\right) }$ & $\  \  \ su\left( 2\right) $ \\ 
{\small \  \  \  \  \  \  \  \ }$\  \  \downarrow $ & \  \  \  \  \  \  \  \  \  \  \  \ $\  \  \
\  \downarrow $ & \  \  \ $\  \  \downarrow $ \\ 
{\small \  \  \  \ 2d- Toda } & $\  \  \  \  \  \  \  \  \  \  \  \  \  \boldsymbol{M}%
_{QK}^{\left( ADE\right) }$ & \  \  \ ADE \\ \hline
\end{tabular}%
\end{equation}%
\begin{equation*}
\end{equation*}%
Within this generalisation, we can extend the analysis done for $\boldsymbol{%
M}_{QK}^{\left( A_{1}\right) }$ to $\boldsymbol{M}_{QK}^{\left( ADE\right) }$%
; in particular the question concerning the explicit derivation of the
expressions of the rigid limits of the induced scalar potential (\ref{vkr})
and the anomaly (\ref{cx}) in the observable sector. Recall that for $%
n_{H}=1 $, the rigid limit of the induced scalar potential is given by 
\begin{equation}
\mathcal{\mathring{V}}_{\text{kah}}^{\mathcal{N}=2}=\frac{1}{2}\mathring{%
\vartheta}_{\underline{M}}^{a}\mathcal{\mathring{M}}^{\underline{M}%
\underline{N}}\mathring{\vartheta}_{\underline{N}}^{a}
\end{equation}%
with SP$\left( 2n_{V}\right) $ symplectic tensor $\mathcal{\mathring{M}}^{%
\underline{M}\underline{N}}$ living on $\boldsymbol{M}_{SK}$, and so
independent on the $\boldsymbol{M}_{QK}^{\left( A_{1}\right) }$ quaternionic
geometry as shown by (\ref{rlm}). The dependence on properties of $%
\boldsymbol{M}_{QK}^{\left( A_{1}\right) }$ is captured by the embedding
tensor $\mathring{\vartheta}_{\underline{M}}^{a}$ which, according to 
\textrm{(\ref{at}-\ref{tat})}, has the generic form $\mathring{\vartheta}_{%
\underline{M}}^{as}$ with an extra index running as $s=1,...n_{H}$. By help
of these arguments, the natural candidate extending $\mathcal{\mathring{V}}_{%
\text{kah}}^{\mathcal{N}=2}$ for the case of ADE geometries reads as follows 
\begin{equation}
\mathcal{\mathring{V}}_{\text{kah}}^{\left( ADE\right) }=\frac{1}{4}K_{rs}%
\mathring{\vartheta}_{\underline{M}}^{ar}\mathcal{\mathring{M}}^{\underline{M%
}\underline{N}}\mathring{\vartheta}_{\underline{N}}^{as}  \label{vkk}
\end{equation}%
where the symmetric $K_{rs}$ is the $n_{H}\times n_{H}$ Cartan matrix of ADE
Lie algebras. For the case $n_{H}=1$, we recover the expression of $\mathcal{%
\mathring{V}}_{\text{kah}}^{\mathcal{N}=2}$. The structure of the matrix
anomaly $\left( C^{ADE}\right) _{B}^{A}$ associated with the $\boldsymbol{M}%
_{QK}^{\left( ADE\right) }$ geometries can be derived by using the same
arguments as for the induced scalar potential. Expression $\left(
C^{ADE}\right) _{B}^{A}$ in terms of the development $\boldsymbol{\zeta }%
_{c}^{\left( ADE\right) }\left( \tau ^{c}\right) _{B}^{A}$, we can write
down the generalisation of $\boldsymbol{\zeta }_{c}^{\left( A_{1}\right) }=%
\frac{1}{2}\varepsilon _{abc}\left( \tilde{\theta}^{a0}\theta
_{0}^{b}-\theta _{0}^{a}\tilde{\theta}^{b0}\right) $ of eq(\ref{xc}). It
reads as%
\begin{equation}
\boldsymbol{\zeta }_{c}^{\left( ADE\right) }=\frac{1}{4}\varepsilon
_{abc}K_{rs}\left( \tilde{\theta}^{a0s}\theta _{0}^{br}-\theta _{0}^{as}%
\tilde{\theta}^{b0r}\right)
\end{equation}%
where $\theta _{0}^{ar}$ and $\tilde{\theta}^{a0s}$ are the electric and
magnetic components of the embedding tensors $\vartheta _{\Lambda }^{mr}$
and $\tilde{\theta}^{a\Lambda s}$ along the graviphoton direction as in (\ref%
{tat}). By using the constraint eqs(\ref{13}), which generalises for the
case of ADE geometries like $K_{rs}\vartheta _{M}^{mr}\mathcal{C}%
^{MN}\vartheta _{N}^{ns}=0$, we can re- express the above isovector $%
\boldsymbol{\zeta }_{c}^{\left( ADE\right) }$ in terms of the electric $%
\theta _{i}^{ar}$ and magnetic $\tilde{\theta}^{ais}$ along the Coulomb
branch directions. We have 
\begin{equation}
\boldsymbol{\zeta }_{c}^{\left( ADE\right) }=-\frac{1}{4}\varepsilon
_{abc}K_{rs}\vartheta _{\underline{M}}^{ar}\mathcal{C}^{\underline{M}%
\underline{N}}\vartheta _{\underline{N}}^{bs}  \label{vee}
\end{equation}%
where $\vartheta _{\underline{M}}^{ar}=\delta _{\underline{M}}^{N}\vartheta
_{N}^{ar}$ and where the components of the embedding tensors $\vartheta _{%
\underline{M}}^{ar}$ are as in eq(\ref{at}). Here, the antisymmetric $%
\mathcal{C}^{\underline{M}\underline{N}}$ is the metric of the SP$\left(
2n_{V}\right) $ symplectic symmetry of the rigid theory, and the symmetric $%
n_{H}\times n_{H}$ matrix $K_{rs}$ is the Cartan matrix of finite
dimensional ADE Lie algebras. The condition for partial breaking of $%
\mathcal{N}=2$ supersymmetry is the same as in (\ref{cpb}). It would be
interesting to check explicitly the above relations for $\mathcal{\mathring{V%
}}_{\text{kah}}^{\left( ADE\right) }$ and $\boldsymbol{\zeta }_{c}^{\left(
ADE\right) }$; this needs working out the exact solutions of Ward identities
for the generalised geometries; progress in this direction will be reported
in a future occasion. 
\begin{equation*}
\end{equation*}

\begin{acknowledgement*}
M. Vall would like to thank L. Andrianopoli for discussions.
\end{acknowledgement*}

\appendix

\section*{Appendices}

In this appendix, which is organised into five sections, we give some
general results on $\mathcal{N}=2$ supergravity which are useful for our
study; but which have not been explicited in the core of the paper; we also
give some details regarding sections 3 and 5. In \autoref{appendixA}, we
recall some results on the Lagrangian of $\mathcal{N}=2$ supergravity for
both ungauged and gauged versions by focusing on the bosonic part of the
degrees of freedom of the theory described in \autoref{sec2}. We also give
the general expressions of the constraints on the embedding tensor $%
\vartheta _{M}^{u}$. In \autoref{appendixB}, we collect useful tools on the
special Kahler and the quaternionic Kahler manifolds, and also give some
explicit computations; certain of the geometrical tools are standard ones;
but for completeness of our study, we have jugged interesting to recall them
here for direct access. In \autoref{appendixC}, we study the rigid limit of
the coupling matrices $\mathcal{N}_{\Lambda \Sigma }$ and $\mathcal{U}^{MN}$%
. In \autoref{appendixD}, we describe the solutions of the Ward identities
in $\mathcal{N}=2$ gauged supergravity by focussing on abelian quaternionic
isometries and in \autoref{appendixE}, we give the rigid limit of the scalar
potential.

\section{ Appendix A: $\mathcal{N}=2$ supergravities in 4d}

\label{appendixA}

First, we describe the bosonic part of the Lagrangian density of 4d $%
\mathcal{N}=2$ ungauged supergravity in presence of $n_{V}$ abelian vector
supermultiplets, with gauge fields $A_{\mu }^{i}$, and $n_{H}$
hypermultiplets. Then, we give its gauged extension with gauging isometries
in the scalar manifold $\boldsymbol{M}_{SK}\times \boldsymbol{M}_{QK}$.

\subsection{Ungauged supergravity}

First recall that the bosonic fields of the 4d $\mathcal{N}=2$ ungauged
supergravity as appearing in the three supermultiplets (\ref{21}) are given
by: $\left( i\right) $ the space time metric $g_{\mu \nu }=e_{\mu }^{m}\eta
_{mn}e_{\nu }^{n}$ with vierbein $e_{\mu }^{m}$, $\left( ii\right) $ the
complex scalars $z^{i}$ of the $n_{V}$ vector supermultiplets; and $\left(
iii\right) $ the real 4n$_{H}$ scalars $Q^{u}$ of the hypermultiplets. The
list of the abelian gauge fields is given by $A_{\mu }^{\Lambda }=\left(
A_{\mu }^{0},A_{\mu }^{i}\right) $ with $A_{\mu }^{0}$ the graviphoton. 
\newline
Following \textrm{\cite{Gallerati:2016oyo,Trigiante:2016mnt}}, the bosonic
part of the Lagrangian density of the 4d $\mathcal{N}=2$ ungauged
supergravity is given by%
\begin{equation}
\begin{tabular}{lll}
$e^{-1}\mathcal{L}_{ungauged}^{bos}$ & $=$ & $-\frac{R}{2}+\frac{1}{2}\,%
\mathcal{G}_{i\bar{\jmath}}\partial _{\mu }z^{i}\partial ^{\mu }\bar{z}^{%
\bar{\jmath}}+\frac{1}{2}\,h_{uv}\partial _{\mu }Q^{u}\partial ^{\mu }Q^{v}$
\\ 
&  & $+\frac{1}{4}I_{\Sigma \Lambda }{\mathcal{F}}_{\mu \nu }^{\Sigma }{\ 
\mathcal{F}}^{\Lambda \mu \nu }+\frac{1}{8}R_{\Sigma \Lambda }\varepsilon
^{\mu \nu \rho \sigma }{\mathcal{F}}_{\mu \nu }^{\Sigma }{\mathcal{F}}_{\rho
\sigma }^{\Lambda }$%
\end{tabular}
\label{a1}
\end{equation}%
where $e=\det \left( e_{\mu }^{m}\right) $ and where 
\begin{equation}
{\mathcal{F}}_{\mu \nu }^{\Lambda }=\partial _{\mu }A_{\nu }^{\Lambda
}-\partial _{\nu }A_{\mu }^{\Lambda }
\end{equation}%
is the field strength of the vector fields $A_{\mu }^{\Lambda }$. The local
gauge metrics $I_{\Sigma \Lambda }=\func{Im}\mathcal{N}_{\Sigma \Lambda }$
and $R_{\Sigma \Lambda }=\func{Re}\mathcal{N}_{\Sigma \Lambda }$ generalize
respectively the usual $-1/g^{2}$ and the \textrm{theta-term} of the
Yang-Mills kinetic terms with gauge coupling $g$. Moreover, $\mathcal{G}_{i%
\bar{\jmath}}=\mathcal{G}_{i\bar{\jmath}}\left( z,\bar{z}\right) $ is the
metric of the complex n- dimensional special Kahler manifold $\boldsymbol{M}%
_{SK}$ parameterised by the complex scalars $z^{i}$ of the Coulomb branch.
The local field matrix$\,h_{uv}=h_{uv}\left( Q\right) $ is the hyperkahler
metric of the real 4n$_{H}$ quaternionic manifold $\boldsymbol{M}_{QK}$.
Notice that in eq(\ref{a1}), there is no scalar potential term; it can be
generated, without breaking of the supersymmetry, only through the gauging
isometries procedure.

\subsection{Dyonic gauged $\mathcal{N}=2$ supergravity}

Following \textrm{\cite{deWit:2005ub, Andrianopoli:2015rpa},} the dyonic
gauging of a subgroup \b{H} of the global isometry group \b{G} of the scalar
manifold of $\mathcal{N}=2$ gauged supergravity 
\begin{equation}
\boldsymbol{M}_{scal}=\boldsymbol{M}_{SK}\times \boldsymbol{M}_{QK}
\label{sc}
\end{equation}%
can be encoded in the embedding tensor $\vartheta _{M}^{u}$ having two
indices (two legs), one in the Coulomb branch and the other in the matter
one. $\vartheta _{M}^{u}$ is not an arbitrary quantity, it obeys constraint
relations. Let us first give properties of this tensor and turns after to
the Lagrangian density.

\  \  \  \ 

$\bullet $ \emph{Constraints on} $\vartheta _{M}^{u}$\newline
Denoting by $T_{\underline{\alpha }}=\left( t_{\underline{a}},t_{\underline{m%
}}\right) $ the set of generators of global isometry group \b{G} of (\ref{sc}%
) and by $X_{M}=\left( X_{\Lambda },X^{\Lambda }\right) $ a generators basis
of a subgroup \b{H} $\subset $ \b{G}, the embedding can be stated as $%
X_{M}=\vartheta _{M}^{\underline{\alpha }}T_{\underline{\alpha }}$; and
reads explicitly, by exhibiting the electric $t_{\underline{a}}$ and
magnetic $t_{\underline{m}}$, as follows%
\begin{equation}
X_{M}=\vartheta _{M}^{\underline{a}}t_{\underline{a}}+\vartheta _{M}^{ 
\underline{m}}t_{\underline{m}}
\end{equation}%
with action of the matrix representation $\left( X_{M}\right) _{N}^{P}\equiv
X_{MN}^{P}$ like%
\begin{equation}
\left( X_{M}\right) _{N}^{P}=\vartheta _{M}^{\underline{\alpha }}\left( T_{ 
\underline{\alpha }}\right) _{N}^{P}\,
\end{equation}%
Consistency of the gauging of the subgroup \b{H} is guaranteed by the
following set of linear and quadratic constraints relations on the embedding
tensor%
\begin{equation}
\begin{tabular}{lll}
$\vartheta _{M}^{\underline{a}}\vartheta _{N}^{\underline{b}}f_{\underline{a}
\underline{b}}^{\underline{c}}+\left( X_{M}\right) _{N}^{P}\vartheta _{P}^{ 
\underline{c}}$ & $=$ & $0$ \\ 
$\vartheta _{M}^{\underline{m}}\vartheta _{N}^{\underline{n}}f_{\underline{m}
\underline{n}}^{\underline{p}}+\left( X_{M}\right) _{N}^{P}\vartheta _{P}^{ 
\underline{p}}$ & $=$ & $0$%
\end{tabular}%
\end{equation}%
and%
\begin{equation}
X_{(MNP)}\equiv \mathcal{C}_{(PQ}X_{MN)}^{Q}=0  \label{et1}
\end{equation}%
as well as%
\begin{equation}
\begin{tabular}{lll}
$\vartheta _{M}^{\underline{a}}\vartheta _{N}^{\underline{b}}f_{\underline{a}
\underline{b}}^{\underline{c}}+\left( X_{M}\right) _{N}^{P}\vartheta _{P}^{ 
\underline{c}}$ & $=$ & $0$ \\ 
$\vartheta _{M}^{\underline{m}}\vartheta _{N}^{\underline{n}}f_{\underline{m}
\underline{n}}^{\underline{p}}+\left( X_{M}\right) _{N}^{P}\vartheta _{P}^{ 
\underline{p}}$ & $=$ & $0$%
\end{tabular}
\label{et2}
\end{equation}%
and%
\begin{equation}
\begin{tabular}{lll}
$\vartheta _{M}^{\underline{a}}\mathcal{C}^{MN}\vartheta _{N}^{\underline{b}%
} $ & $=$ & $0$ \\ 
$\vartheta _{M}^{\underline{a}}\mathcal{C}^{MN}\vartheta _{N}^{\underline{n}%
} $ & $=$ & $0$ \\ 
$\vartheta _{M}^{\underline{m}}\mathcal{C}^{MN}\vartheta _{N}^{\underline{n}%
} $ & $=$ & $0$%
\end{tabular}
\label{et4}
\end{equation}%
The $f_{\underline{a}\underline{b}}^{\underline{c}}$ and $f_{\underline{m} 
\underline{n}}^{\underline{p}}$\ in (\ref{et2}) are the structure constants
respectively associated with $t_{\underline{a}},t_{\underline{m}}$. The
conditions (\ref{et2}) are closure constraints, they follow from 
\begin{equation}
\lbrack X_{M},\,X_{N}]=-X_{MN}{}^{P}\,X_{P}  \label{salg}
\end{equation}%
Moreover, the first two equalities in (\ref{et4}) descend from (\ref{et1})
and (\ref{et2}) while the third one has to be imposed independently \textrm{%
\  \cite{deWit:2005ub,Andrianopoli:2015rpa}}. By using the embedding tensor,
we can define gauged quantities $\digamma _{M}$ out of ungauged counterparts 
$\digamma _{\underline{\alpha }}$; F the examples the gauged Killing vectors 
$\left( \kappa _{M}^{i},\kappa _{M}^{u}\right) $ on $\boldsymbol{M}%
_{SK}\times \boldsymbol{M}_{QK}$ and the moment maps $\left( \mathcal{P}%
_{M}^{0},\mathcal{P}_{M}^{a}\right) $, we have 
\begin{equation}
\begin{tabular}{lll}
$\kappa _{M}^{i}$ & $=$ & $\vartheta _{M}^{\underline{a}}\,k_{\underline{a}%
}^{i}$ \\ 
$\kappa _{M}^{u}$ & $\equiv $ & $\vartheta _{M}^{\underline{m}}\,k_{ 
\underline{m}}^{u}$%
\end{tabular}%
\end{equation}%
with ($k_{\underline{a}}^{i},k_{\underline{m}}^{u}$) ungauged Killing
vectors, and%
\begin{equation}
\begin{tabular}{lll}
$\mathcal{P}_{M}^{0}$ & $=$ & $\vartheta _{M}^{\underline{a}}\,P_{\underline{
a}}$ \\ 
$\mathcal{P}_{M}^{a}$ & $=$ & $\vartheta _{M}^{\underline{m}}\,P_{\underline{
m}}^{a}$%
\end{tabular}%
\end{equation}%
with ($P_{\underline{a}},P_{\underline{m}}$) ungauged moments. These
quantities satisfy the following algebras \textrm{\cite{Andrianopoli:2015rpa}%
}%
\begin{equation}
i\mathcal{G}_{i\bar{\jmath}}\, \kappa _{\lbrack M}^{i}\, \kappa _{N]}^{\bar{
\jmath}}=\frac{1}{2}\,X_{MN}^{P}\, \mathcal{P}_{P}^{0}  \label{gq1}
\end{equation}%
and%
\begin{equation}
2\,K_{uv}^{a}\, \kappa _{M}^{u}\, \kappa _{N}^{v}+\varepsilon ^{abc}\, 
\mathcal{P}_{M}^{b}\, \mathcal{P}_{N}^{c}=X_{MN}^{P}\mathcal{P}_{P}^{a}
\label{gq2}
\end{equation}%
where $K_{uv}^{a}$ is the hyperkahler 2-forms.

\  \  \  \ 

$\bullet $ $\mathcal{N}=2$ \emph{gauged supergravity Lagrangian}\newline
The bosonic part of the lagrangian density of the $\mathcal{N}=2$ gauged
supergravity, with bosonic degrees of freedom as described previously, is
obtained by covariantizing space time derivatives as follows \textrm{\cite%
{Trigiante:2016mnt}}%
\begin{equation}
\begin{tabular}{lll}
$e^{-1}\mathcal{L}_{gauged}^{bos}$ & $=$ & $-\frac{1}{2}R+\mathcal{G}_{i\bar{%
\jmath}}\mathcal{D}_{\mu }z^{i}\mathcal{D}^{\mu }\bar{z}^{\bar{\jmath}%
}+h_{uv}\mathcal{D}_{\mu }Q^{u}\mathcal{D}^{\mu }Q^{v}$ \\ 
&  & $+\frac{1}{4}\, \func{Im}\mathcal{N}_{\Lambda \Sigma }\,{\mathcal{\ 
\tilde{F}}}_{\mu \nu }^{\Sigma }{\mathcal{\tilde{F}}}^{\Lambda \mu \nu }+%
\frac{1}{8e}\func{Re}\mathcal{N}_{\Lambda \Sigma }\; \varepsilon ^{\mu \nu
\rho \sigma }{\mathcal{\tilde{F}}}_{\mu \nu }^{\Sigma }{\mathcal{\tilde{F}}}%
_{\rho \sigma }^{\Lambda }$ \\ 
&  & $+\mathcal{L}_{top}+\mathcal{L}_{CS}-\mathcal{V}\left( z,\bar{z}%
,Q\right) $%
\end{tabular}%
\end{equation}%
Here, $\mathcal{L}_{top}$ and $\mathcal{L}_{CS}$ are topological terms
required to maintain the gauge invariance; they are given by%
\begin{equation}
\mathcal{L}_{top}=-\frac{1}{8e}\varepsilon ^{\mu \nu \rho \sigma }\vartheta
^{\Lambda \underline{\alpha }}B_{\mu \nu \, \underline{\alpha }}\, \left(
2\partial _{\rho }A_{\sigma \Lambda }+X_{MN\Lambda }A_{\rho }^{M}A_{\sigma
}^{N}-\frac{1}{4}\, \vartheta _{\Lambda }^{\underline{\beta }}B_{\rho \sigma 
\underline{\beta }}\right)
\end{equation}%
and%
\begin{equation}
\begin{tabular}{lll}
$\mathcal{L}_{CS}$ & $=$ & $-\frac{1}{3e}\, \varepsilon ^{\mu \nu \rho
\sigma }X_{MN\, \Lambda }\,A_{\mu }^{M}A_{\nu }^{N}\left( \partial _{\rho
}A_{\sigma }^{\Lambda }+\frac{1}{4}X_{PQ}^{\Lambda }A_{\rho }^{P}A_{\sigma
}^{Q}\right) $ \\ 
&  & $-\frac{1}{6e}\, \varepsilon ^{\mu \nu \rho \sigma }X_{MN}^{\Lambda
}\,A_{\mu }^{M}A_{\nu }^{N}\left( \partial _{\rho }A_{\sigma \Lambda }+\frac{%
1}{4}\,X_{PQ\Lambda }A_{\rho }^{P}A_{\sigma }^{Q}\right) $%
\end{tabular}%
\end{equation}%
with%
\begin{equation}
\begin{tabular}{lll}
$\mathcal{D}_{\mu }z^{i}$ & $=$ & $\partial _{\mu }z^{i}+\vartheta _{M}^{%
\underline{a}}A_{\mu }^{M}k_{\underline{a}}^{i}$ \\ 
$\mathcal{D}_{\mu }Q^{u}$ & $=$ & $\partial _{\mu }Q^{u}+\vartheta _{M}^{%
\underline{m}}A_{\mu }^{M}k_{\underline{m}}^{u}$%
\end{tabular}%
\end{equation}%
and%
\begin{equation*}
A_{\mu }^{M}=\left( 
\begin{array}{c}
A_{\mu }^{\Lambda } \\ 
A_{\mu \Lambda }%
\end{array}%
\right) \qquad ,\qquad \Lambda =0,I,\quad I=1,...,n_{v}
\end{equation*}%
We also have 
\begin{equation}
\begin{tabular}{lll}
${\mathcal{F}}_{\mu \nu }^{\Lambda }$ & $=$ & $\partial _{\mu }A_{\nu
}^{\Lambda }-\partial _{\nu }A_{\mu }^{\Lambda }$ \\ 
$\mathcal{\tilde{F}}_{\mu \nu }^{\Lambda }$ & $=$ & ${\mathcal{F}}_{\mu \nu
}^{\Lambda }+\frac{1}{2}\vartheta ^{\Lambda \underline{m}}B_{\mu \nu 
\underline{m}}$%
\end{tabular}%
\end{equation}%
The $B_{\mu \nu \, \underline{\alpha }}$ are massless antisymmetric tensor
fields which must be introduced in order to construct gauge covariant field
strengths $\mathcal{\tilde{F}}_{\mu \nu }^{\Lambda }$\textrm{\footnote{%
\ The $B_{\mu \nu \underline{m}}$'s can be also interpreted as the dualized
scalars used in gauging isometries as in sec 2.}}. The scalar potential $%
\mathcal{V}(z,\bar{z},Q)$ is given by%
\begin{equation}
{V}(z,\bar{z},Q)=(\mathcal{G}_{i\bar{\jmath}}\kappa _{M}^{i}\kappa _{N}^{%
\bar{\jmath}}+4\,h_{uv}\kappa _{M}^{u}\kappa _{N}^{v})\overline{V}%
^{M}\,V^{N}+(\mathcal{U}^{MN}-3\,V^{M}\overline{V}^{N})\mathcal{P}_{N}^{a}%
\mathcal{P}_{M}^{a}  \label{gpv}
\end{equation}%
where $V^{M}$ is the covariantly holomorphic section \ of the special Kahler
manifold $\boldsymbol{M}_{SK}$ and 
\begin{equation}
\mathcal{U}^{MN}=U_{i}^{M}\mathcal{G}^{i\bar{j}}\bar{U}_{\bar{j}}^{N}
\label{mn}
\end{equation}%
with $U_{i}^{M}$ is holomorphic section of $\boldsymbol{M}_{SK}.$

\section{Appendix B: Scalar manifold in $\mathcal{N}=2$ gauged supergravity}

\label{appendixB}

In this appendix, we collect useful geometrical features on the scalar
manifold $\boldsymbol{M}_{scal}$ in $\mathcal{N}=2$ gauged supergravity. As
this manifold factorises like $\boldsymbol{M}_{scal}=\boldsymbol{M}%
_{SK}\times \boldsymbol{M}_{QK}$, we first describe properties of the
special K\"{a}hler $\boldsymbol{M}_{SK}$, and turn after to those of the
quaternionic Kahler $\boldsymbol{M}_{QK}$.

\subsection{Special Kahler Manifolds $\boldsymbol{M}_{SK}$}

{\small A }special K\"{a}hler manifold $\boldsymbol{M}_{SK}$ is a Hodge-
Kahler manifold endowed with a holomorphic flat vector bundle with structure
group $SP(2n,\mathbb{R})$ satisfying special properties \textrm{\cite%
{Strominger:1990pd,D'Auria:1990fj,deWit:1995tf,Ceresole:1995jg,Fre:1995dw,Fre:1995bc,Andrianopoli:1996cm,Craps:1997gp,Maruyoshi:2006te,Freedman:2012zz,Andrianopoli:2015rpa,Trigiante:2016mnt}%
}.. In what follows, we describe properties of sections on $\boldsymbol{M}%
_{SK}$ and the decomposition of eq(\ref{mn}) as in (\ref{tmn}).{\small \ }

\  \  \ 

$\bullet $ \emph{holomorphic section }$\Omega ^{M}(z)$ \emph{and Kahler
metric}\newline
The holomorphic section of $\boldsymbol{M}_{SK}$ is given by eq(\ref{rg})
that we recall hereafter%
\begin{equation}
\Omega ^{M}(z)=\left( 
\begin{array}{c}
X^{\Lambda }(z) \\ 
F_{\Lambda }(z)%
\end{array}%
\right) \, \, \qquad ,\, \qquad \, \, \Lambda =0,\dots ,n
\end{equation}%
It transforms as a global $SP(2n+2,\mathbb{R})$ symplectic vector;
holomorphy and symplectic symmetries act on $\Omega ^{M}(z)$ like \textrm{\ }%
\begin{equation}
\Omega _{\left( \alpha \right) }^{M}=e^{f_{(\alpha ,\beta )}}\, \left[ 
\mathfrak{R}_{(\alpha ,\beta )}\right] _{N}^{M}\, \Omega _{(\beta )}^{N}
\label{tomg}
\end{equation}%
where:\textrm{\ }$\left( i\right) $ the\textrm{\ }$f_{(\alpha ,\beta
)}=f(z^{\left( \alpha \right) },z^{\left( \beta \right) })$ is a holomorphic
transition function connecting two \emph{overlapping} patches\textrm{\ }$%
U_{\left( \alpha \right) }$\textrm{\ }and\textrm{\ }$U_{\left( \beta \right)
}$\textrm{\ }on the special Kahler manifold\textrm{\ }$\boldsymbol{M}_{SK}$
with local coordinates $z^{\left( \alpha \right) }$ and $z^{\left( \beta
\right) }$ ($U_{\left( \alpha \right) }\cap U_{\left( \beta \right)
}=U_{(\alpha ,\beta )}\neq \varnothing $)\textrm{; }and\textrm{\ }$\left(
ii\right) $ $\left[ \mathfrak{R}_{(\alpha ,\beta )}\right] _{N}^{M}$ is a
constant\textrm{\ }$SP(2n+2,R)$\textrm{\ }matrix. \newline
From the holomorphic section $\Omega ^{M}$, one can define other objects
like the K\"{a}hler potential $\mathcal{K}=\mathcal{K}(z,\bar{z})$, the
covariantly holomorphic section $V^{M}=V^{M}\left( z,\bar{z}\right) $ and
its covariant derivatives $U_{i}^{M}=D_{i}V^{M}\left( z,\bar{z}\right) $.
For the Kahler potential $\mathcal{K}$, it is related to $\Omega ^{M}$ and
its complex conjugate like\textrm{\ }%
\begin{equation}
\mathcal{K}=-\log [i\, \overline{\Omega }^{M}\mathcal{C}_{MN}\Omega ^{N})]\,
\label{kpl}
\end{equation}%
where $\mathcal{C}_{MN}$ is the invariant $SP(2n+2,\mathbb{R})$ metric 
\begin{equation}
\mathcal{C}_{MN}=%
\begin{pmatrix}
0 & \mathcal{I}_{n+1} \\ 
-\mathcal{I}_{n+1} & 0%
\end{pmatrix}%
\,.  \label{24}
\end{equation}%
With eq(\ref{kpl}), we can also build the metric $\mathcal{G}_{i\bar{\jmath}%
} $ of $\boldsymbol{M}_{SK}$ and the closed Kahler 2-form $\boldsymbol{K}%
_{2} $ as given hereafter%
\begin{equation}
\mathcal{G}_{i\bar{\jmath}}=\partial _{i}\partial _{\bar{\jmath}}\mathcal{\
K\qquad },\mathcal{\qquad K}=\mathcal{K}\left( z,\bar{z}\right)  \label{km}
\end{equation}%
which is invariant under Kahler transformation%
\begin{equation}
\mathcal{K}_{(\alpha )}=\mathcal{K}_{(\beta )}-f_{(\alpha ,\beta )}-\bar{f}%
_{(\alpha ,\beta )}  \label{ktr}
\end{equation}%
and 
\begin{equation}
\boldsymbol{K}_{2}=i\mathcal{G}_{i\bar{\jmath}}dz^{i}\wedge d\bar{z}^{\bar{%
\jmath}}\, \qquad ,\, \qquad \,d\boldsymbol{K}_{2}=0\,
\end{equation}%
From $d\boldsymbol{K}_{2}=0,$ we learn that 
\begin{equation}
\boldsymbol{K}_{2}=d\boldsymbol{\omega }_{1}^{0}  \label{dq}
\end{equation}%
where $\boldsymbol{\omega }_{1}^{0}$ is the $U(1)_{R}$ Kahler connection
1-form 
\begin{equation}
\boldsymbol{\omega }_{1}^{0}=-\frac{i}{2}\left[ \partial \mathcal{K}-\bar{%
\partial}\mathcal{K}\, \right] \qquad ,\, \qquad \partial \mathcal{K}%
=\partial _{i}\mathcal{K}\,dz^{i}  \label{cnx}
\end{equation}

\  \  \  \ 

$\bullet $ \emph{Sections }$V^{M}$ \emph{and} $U_{i}^{M}$\emph{\ }\newline
The covariantly holomorphic section is defined as $V^{M}=e^{\frac{\mathcal{K}
}{2}}\, \Omega ^{M}$ with symplectic components like 
\begin{equation}
V^{M}=\left( 
\begin{array}{c}
\Upsilon ^{\Lambda } \\ 
\Gamma _{\Lambda }%
\end{array}%
\right) \qquad ,\, \qquad V^{M}=V^{M}(z,\bar{z})  \label{chs2}
\end{equation}%
it satisfies the properties%
\begin{equation}
\bar{D}_{\bar{\imath}}V^{M}\equiv \left( \partial _{\bar{\imath}}-\frac{
\partial _{\bar{\imath}}\mathcal{K}}{2}\right) V^{M}=0
\end{equation}%
and 
\begin{equation}
-iV^{M}\mathcal{C}_{MN}\bar{V}^{N}=1\quad ,\quad i\bar{V}^{M}\mathcal{C}%
_{MN}V^{N}=1\quad ,\quad V^{M}\mathcal{C}_{MN}V^{N}=0  \label{vbv}
\end{equation}%
Eqs(\ref{vbv}) can be obtained from the definition of $V^{M}$ and (\ref{kpl}%
). By using (\ref{tomg}) and (\ref{ktr}), we obtain the transformation of $%
V^{M}$ under the U$\left( 1\right) $ Kahler and SP$\left( 2n+2,\mathbb{R}%
\right) $ symmetries namely 
\begin{equation}
V_{(\alpha )}=e^{i\,Imf_{(\alpha ,\beta )}}\, \left[ \mathfrak{R}_{(\alpha
,\beta )}\right] _{N}^{M}\,V_{(\beta )}^{N}  \label{tchs}
\end{equation}%
Regarding the $U(1)$- covariant derivatives $U_{i}=D_{i}V$; it reads
explicitly like 
\begin{equation}
U_{i}=\left( \partial _{i}+\frac{\partial _{i}\mathcal{K}}{2}\right) V\qquad
,\, \qquad \partial _{i}=\frac{\partial }{\partial z^{i}}  \label{gchs}
\end{equation}%
In a special Kahler manifold, the section $V^{M}$ and its covariant
derivative $U_{i}^{M}$ need to satisfy the following properties%
\begin{equation}
\begin{tabular}{lll}
$D_{i}U_{j}^{M}$ & $\equiv $ & $\partial _{i}U_{j}^{M}+\frac{\partial _{i} 
\mathcal{K}}{2}\,U_{j}^{M}-\Gamma _{ij}^{k}\,U_{k}^{M}$ \\ 
& $=$ & $i\,C_{ijk}\,g^{k\bar{k}}\, \bar{U}_{\bar{k}}^{M}$%
\end{tabular}%
\end{equation}%
and 
\begin{equation}
\begin{tabular}{lll}
$D_{i}\overline{U}_{\bar{\jmath}}^{M}$ & $=$ & $\mathcal{G}_{i\bar{\jmath}}\,%
\overline{V}^{M}$ \\ 
$V^{M}\mathcal{C}_{MN}U_{i}^{N}$ & $=$ & $0$ \\ 
$V^{M}\mathcal{C}_{MN}\overline{U}_{\bar{k}}^{N}$ & $=$ & $0$%
\end{tabular}
\label{vuv}
\end{equation}%
From above equation, one has \textrm{\cite{deWit:1995tf}}%
\begin{equation}
\mathcal{G}_{k\bar{l}}=iU_{k}^{M}\mathcal{C}_{MN}\bar{U}_{\bar{l}}^{N}\quad
,\quad U^{iM}\mathcal{C}_{MN}\bar{U}_{\bar{j}}^{N}=-i\delta _{\bar{j}%
}^{i}\quad ,\quad \bar{U}^{\bar{j}M}\mathcal{C}_{MN}U_{i}^{N}=i\delta _{i}^{ 
\bar{j}}  \label{ucu}
\end{equation}

\  \  \  \ 

$\bullet $ \emph{Decomposing the} \emph{factor} $\mathcal{U}^{MN}$\newline
The symplectic tensor, given by%
\begin{equation}
\mathcal{U}^{MN}=\mathcal{G}^{i\bar{\jmath}}\,U_{i}^{M}U_{\bar{\jmath}}^{N},
\label{umn}
\end{equation}%
appears in the study of Ward identities and the induced scalar potential (%
\ref{vscc}); its reducibility is interesting in the derivation of its rigid
limit. Following,\textrm{\cite{Andrianopoli:2015rpa}, }by using $V^{M}$ and $%
U_{i}^{M}$, we can construct a $\left( 2n+2\right) \times \left( 2n+2\right) 
$ matrix $\mathbb{L}=\mathbb{L}(z,\bar{z})$ whose entries as follows%
\begin{equation}
\mathbb{L}=%
\begin{pmatrix}
V^{M} & \boldsymbol{\bar{U}}_{I^{\prime }}^{M} & \bar{V}^{M} & \boldsymbol{U}%
_{I^{\prime }}^{M}%
\end{pmatrix}%
\qquad ,\qquad \mathbb{L}=\left( \mathbb{L}_{N^{\prime }}^{M}\right)
\end{equation}%
where $M$, $N^{\prime }=0,I^{\prime }$ are indices of $SP(2n+2,\mathbb{R})$
and $USP(n+1,n+1)$ respectively, and 
\begin{equation}
\boldsymbol{U}_{I^{\prime }}^{M}=e_{I^{\prime }}^{i}U_{i}^{M}\qquad ,\qquad 
\boldsymbol{\bar{U}}_{I^{\prime }}^{M}=\bar{e}_{I^{\prime }}^{\bar{\imath}}%
\bar{U}_{\bar{\imath}}^{M}
\end{equation}%
where $e_{I}^{i}{}$is the (inverse) vielbein on the $\boldsymbol{M}_{SK}$
related to the Kahler metric $\mathcal{G}_{i\bar{\jmath}}$ as 
\begin{equation}
\mathcal{G}_{i\bar{\jmath}}=\sum_{I^{\prime }=1}^{n_{V}}e_{i}^{I^{\prime }}%
\text{ }\bar{e}_{\bar{\jmath}I^{\prime }}\quad ,\quad e_{i}^{I^{\prime }}%
\bar{e}_{J^{\prime }}^{i}=\delta _{J^{\prime }}^{I^{\prime }}\quad ,\quad
e_{I^{\prime }}^{i}\text{ }e_{j}^{I^{\prime }}=\delta _{j}^{i}
\end{equation}%
Eqs(\ref{vuv},\ref{ucu}) imply the matrix $\mathbb{L}$ has remarkable
properties; in particular its transforms in the bifundamental of $%
SP(2n+2)\times USP(n+1,n+1)$ and obeys \textrm{\cite%
{Andrianopoli:1996ve,Andrianopoli:2015rpa}} 
\begin{equation}
\begin{tabular}{lll}
$\mathbb{L}^{\dagger }\mathcal{C}\mathbb{L}$ & $=$ & $\varpi $ \\ 
$\mathbb{L}\varpi \mathbb{L}^{\dagger }$ & $=$ & $\mathcal{C}$%
\end{tabular}%
\,  \label{lcl}
\end{equation}%
with%
\begin{equation}
\varpi =-i\,%
\begin{pmatrix}
\mathcal{I}_{n+1} & 0 \\ 
0 & -\mathcal{I}_{n+1}%
\end{pmatrix}%
\end{equation}%
and where $\mathcal{C}$ as in (\ref{24}). Explicitly, we have%
\begin{equation}
\begin{tabular}{lll}
$\left( \mathbb{L}^{\dagger }\right) _{M}^{M^{\prime }}\left( \mathcal{C}%
\right) _{N}^{M}\left( \mathbb{L}\right) _{N^{\prime }}^{N}$ & $=$ & $\left(
\varpi \right) _{N^{\prime }}^{M^{\prime }}$ \\ 
$\left( \mathbb{L}\right) _{N^{\prime }}^{N}\left( \varpi \right)
_{M^{\prime }}^{N^{\prime }}\left( \mathbb{L}^{\dagger }\right)
_{M}^{M^{\prime }}$ & $=$ & $\mathcal{C}_{M}^{N}$%
\end{tabular}
\label{lm}
\end{equation}%
Notice that the index $N^{\prime }\ $of $USP(n+1,n+1)$ can be brought into
an index of $SP(2n+2,\mathbb{R})$ by using the following $\left( 2n+2\right)
\times \left( 2n+2\right) $ Cayley matrix $\mathcal{A}=\left( \mathcal{A}%
_{N}^{N^{\prime }}\right) $ giben by\textrm{\cite%
{Fre:1995dw,Andrianopoli:1996cm}}%
\begin{equation}
\mathcal{A}=\frac{1}{\sqrt{2}}%
\begin{pmatrix}
\mathcal{I}_{n+1} & i\mathcal{I}_{n+1} \\ 
\mathcal{I}_{n+1} & -i\mathcal{I}_{n+1}%
\end{pmatrix}
\label{ml}
\end{equation}%
with the properties%
\begin{equation}
\mathcal{AA}^{\dagger }=\mathcal{I}_{2n+2}\qquad ,\, \qquad \varpi =\mathcal{%
ACA}^{\dagger }\qquad ,\, \qquad \mathcal{C}=\mathcal{A}^{\dagger }\varpi 
\mathcal{A}
\end{equation}%
Multiplying the matrix $\mathbb{L}_{N^{\prime }}^{M}$ from the right by $%
\mathcal{A}_{N}^{N^{\prime }}$, we can bring $\mathbb{L}_{N^{\prime }}^{M}$
to a symplectic matrix as follows%
\begin{equation}
\begin{tabular}{lllllll}
$\boldsymbol{S}$ & $=$ & $\mathbb{L}\mathcal{A}$ & $\qquad ,\, \qquad $ & $%
\boldsymbol{S}_{N}^{M}$ & $=$ & $\mathbb{L}_{N^{\prime }}^{M}\mathcal{A}%
_{N}^{N^{\prime }}$ \\ 
$\mathbb{L}$ & $=$ & $\boldsymbol{S}\mathcal{A}^{\dagger }$ & $\qquad ,\,
\qquad $ & $\mathbb{L}_{N^{\prime }}^{M}$ & $=$ & $\boldsymbol{S}%
_{N}^{M}\left( \mathcal{A}^{\dagger }\right) _{N^{\prime }}^{N}$%
\end{tabular}
\label{la}
\end{equation}%
Moreover, multiplying both sides of eq(\ref{lcl}) on left by $\mathcal{A}$
and on right by $\mathcal{A}^{\dagger }$, \ and using the relation $\mathcal{%
\ C}=\mathcal{A}^{\dagger }\varpi \mathcal{A}$, we obtain 
\begin{equation}
\left( \mathcal{A}^{\dagger }\mathbb{L}^{\dagger }\right) \mathcal{C}\left( 
\mathbb{L}\mathcal{A}\right) =\mathcal{C}
\end{equation}%
showing that $\boldsymbol{S}=\mathbb{L}\mathcal{A}$ is indeed a symplectic
matrix. In other words%
\begin{equation}
\mathbb{L}^{\dagger }\mathcal{C}\mathbb{L}=\varpi \qquad \Leftrightarrow \,
\qquad \boldsymbol{S}^{\dagger }\mathcal{C}\boldsymbol{S}=\mathcal{C}
\label{cs}
\end{equation}%
Notice also that $\boldsymbol{S}$ is a real matrix; by using eqs(\ref{lm}-%
\ref{ml}), we have%
\begin{equation}
\boldsymbol{S}_{N}^{M}=\frac{1}{\sqrt{2}}\left[ V^{M}+\bar{V}^{M},%
\boldsymbol{\bar{U}}_{K}^{M}+\boldsymbol{U}_{K}^{M},i\left( V^{M}-\bar{V}%
^{M}\right) ,i\left( \boldsymbol{\bar{U}}_{K}^{M}-\boldsymbol{U}%
_{K}^{M}\right) \right]
\end{equation}%
With the matrix $\mathbb{L}$, its adjoint $\mathbb{L}^{\dagger }$ and the
metric $\mathcal{C}$, we can build a \textrm{hermitian, symmetric} matrix $%
\mathcal{M}=\mathcal{M}(z,\bar{z})$ encoding all information about the
coupling of the vector fields to the scalars; this matrix defined like 
\begin{equation}
\mathcal{M}=\mathcal{C}\mathbb{L}\mathbb{L}^{\dagger }\mathcal{C\quad },%
\mathcal{\quad M}^{-1}=-\mathbb{LL}^{\dagger }
\end{equation}%
has several remarkable properties that we describe below: First, given $%
\mathcal{M}=\mathcal{C}\mathbb{L}\mathbb{L}^{\dagger }\mathcal{C}$, one can
directly check that $\mathcal{M}^{-1}=-\mathbb{LL}^{\dagger }$. Indeed, let
us compute the product $\mathcal{MM}^{-1}$ explicitly; we have after
substitution%
\begin{equation}
\mathcal{MM}^{-1}=\mathcal{C}\mathbb{L}\mathbb{L}^{\dagger }\mathcal{C}%
\left( -\mathbb{LL}^{\dagger }\right) =-\mathcal{C}\mathbb{L}\mathbb{L}%
^{\dagger }\mathcal{C}\mathbb{LL}^{\dagger }
\end{equation}%
By help of the first relation of eq(\ref{lcl}), we also have%
\begin{equation}
\mathcal{MM}^{-1}=-\mathcal{C}\mathbb{L\varpi L}^{\dagger }=-\mathcal{C}^{2}=%
\mathcal{I}
\end{equation}%
Moreover, using the factorisation $\mathbb{L}=\boldsymbol{S}\mathcal{A}%
^{\dagger }$ (\ref{la}), the matrices $\mathcal{M}$ and its inverse $%
\mathcal{M}^{-1}$ can be expressed in terms of the symplectic matrix $%
\boldsymbol{S}$ as follows 
\begin{equation}
\mathcal{M}=\mathcal{C}\boldsymbol{SS}^{T}\mathcal{C}\qquad ,\, \qquad 
\mathcal{M}^{-1}=-\boldsymbol{SS}^{T}
\end{equation}%
Another property of the matrix $\mathcal{M}$ is that it is a symplectic
matrix satisfying the usual property 
\begin{equation}
\mathcal{MCM}=\mathcal{C}  \label{cm}
\end{equation}%
This feature can be checked by substituting $\mathcal{M}=\mathcal{C}\mathbb{L%
}\mathbb{L}^{\dagger }\mathcal{C}$ and using the properties of $\mathbb{L}$.
In doing so, we first have%
\begin{equation}
\mathcal{MCM}=-\mathcal{C}\mathbb{L}\mathbb{L}^{\dagger }\mathcal{C}\mathbb{L%
}\mathbb{L}^{\dagger }\mathcal{C}
\end{equation}%
where we have used $\mathcal{C}^{2}=-I$; then using $\mathbb{L}^{\dagger }%
\mathcal{C}\mathbb{L=\varpi }$, we can reduce the expression of $\mathcal{MCM%
}$ down to $-\mathcal{C}\mathbb{L\varpi L}^{\dagger }\mathcal{C}$ which, by
help of $\mathbb{L\varpi L}^{\dagger }=\mathcal{C}$ given by eq(\ref{lcl}),
can be further reduced to (\ref{cm}). \newline
The third property concerns the change of the local matrix $\mathcal{M}(z,%
\bar{z})$ with respect to Kahler U$\left( 1\right) $ and symplectic SP$%
\left( 2n+2,\mathbb{R}\right) $ symmetries on $\boldsymbol{M}_{SK}$. Under a
symmetry transformation, the matrix $\mathcal{M}(z,\bar{z})$ transforms into 
$\mathcal{M}^{\left( g\right) }(z^{\prime },\bar{z}^{\prime })$ given by 
\textrm{\ }%
\begin{equation}
\mathcal{M}^{\left( g\right) }(z^{\prime },\bar{z}^{\prime })=\left[ 
\mathfrak{R}^{\left( g\right) }\right] ^{T}\mathcal{M}(z,\bar{z})\mathfrak{R}%
^{\left( g\right) }
\end{equation}%
where $z^{\prime }=z^{\prime }\left( z\right) $ a holomorphic transformation
and $\mathfrak{R}^{\left( g\right) }$ a symplectic representation as in eq(%
\ref{tomg}). The last feature we give deals with the relation between $%
\mathcal{M}^{MN}$ and the $\mathcal{U}^{MN}$ of eq(\ref{umn}). We have $%
\mathcal{M}^{MN}=-2\mathcal{U}^{MN}-i\, \mathcal{C}^{MN}-2\overline{V}%
^{M}V^{N}$ which reads also like \textrm{\cite{Andrianopoli:2015rpa} }%
\begin{equation}
\mathcal{U}^{MN}=-\frac{1}{2}\mathcal{M}^{MN}-\frac{i}{2}\, \mathcal{C}^{MN}-%
\overline{V}^{M}V^{N}\,  \label{uu}
\end{equation}

\subsection{Quaternionic Kahler Manifolds}

Here we give properties of the quaternionic Kahler manifold $\boldsymbol{M}%
_{QK}$ useful in the study of the $\mathcal{N}=2$ gauged supergravity 
\textrm{\cite%
{Bagger:1983tt,Hitchin:1986ea,Galicki:1986ja,D'Auria:1990fj,Andrianopoli:1996cm,D'Auria:2001kv,Andrianopoli:2015rpa,Trigiante:2016mnt}%
}; this is a real $4n_{H}$- dimensional Riemanian variety having an $%
SU(2)_{R}\times G^{\prime }$ holonomy group with $G^{\prime }$\ contained in
a symplectic group as follows\textrm{\ }%
\begin{equation}
G^{\prime }\subset SP(2n_{H},\mathbb{R})  \label{hgp}
\end{equation}%
Notice that the combination of the $SU(2)_{R}$ symmetry of $\boldsymbol{M}%
_{QK}$ with the $U(1)_{R}$\textrm{\ }Kahler group of $\boldsymbol{M}_{SK}$
define the usual $U(2)_{R}=U(1)_{R}\times SU(2)_{R}$\ R-symmetry of the $%
\mathcal{N}=2$\ supersymmetric\textrm{\ }algebra%
\begin{equation}
\left \{ Q^{A},Q_{B}\right \} \sim \delta _{B}^{A}\sigma ^{\mu }P_{\mu }
\end{equation}

\  \  \ 

$\bullet $ \emph{hyperKahler metric and hyperKahler 2- form}\newline
On the quaternionic Kahler manifold $\boldsymbol{M}_{QK}$ of the matter
sector, the metric (infinitesimal length) $ds_{H}^{2}$ and the hermitian
Kahler 2-form triplet $\boldsymbol{K}_{2}^{a}$ can be expressed in terms of
covariantly constant 1-form veilbeins $\mathcal{E}^{AM}$ as follows 
\begin{equation}
ds_{H}^{2}=\sum_{A,B=1}^{2}\varepsilon _{AB}\left( \sum_{M=1}^{2n_{H}}%
\mathcal{C}_{MN}\mathcal{E}^{AM}\otimes \mathcal{E}^{BN}\right)  \label{B1}
\end{equation}%
and%
\begin{equation}
\boldsymbol{K}_{2}^{a}=\sum_{A,B=1}^{2}\left( \tau ^{a}\varepsilon \right)
_{AB}\left( \sum_{M=1}^{2n_{H}}\mathcal{E}^{AM}\wedge \mathcal{E}%
^{BN}C_{MN}\right)  \label{2k}
\end{equation}%
where the three $\tau ^{a}$'s are the usual 2$\times $2 Pauli matrices with
the symmetric property $\left( \tau ^{a}\varepsilon \right) _{AB}=\left(
\tau ^{a}\varepsilon \right) _{BA}$. The $ds_{H}^{2}$ is invariant under $%
SU\left( 2\right) _{R}\times SP\left( 2n_{H}\right) $ isotropy group with
metric $\varepsilon _{AB}C_{MN}$; the hyperkahler 2-form $\boldsymbol{K}%
_{2}^{a}$ is invariant under $SP\left( 2n_{H}\right) $ but behaves as a
triplet under $SU\left( 2\right) _{R}$. By using the real $4n_{H}$ local
coordinates field variables $Q^{u}$, we can express the 1-form vielbeins $%
\mathcal{E}^{AM}$ in above equations like 
\begin{equation}
\mathcal{E}^{AM}=\mathcal{E}_{u}^{AM}dQ^{u}  \label{vi}
\end{equation}%
with $\mathcal{E}_{u}^{AM}$ satisfying the reality condition $\left( 
\mathcal{E}_{u}^{AM}\right) ^{\dagger }=\mathcal{E}_{uAM}$ with%
\begin{equation}
\mathcal{E}_{uAM}=\varepsilon _{AB}\mathcal{C}_{MN}\mathcal{E}_{u}^{BN}
\label{3k}
\end{equation}%
By substituting (\ref{3k}) back into eqs(\ref{1k}\ref{2k}), we can bring the
hyperkahler metric and the hyperKahler 2-form to 
\begin{equation}
ds^{2}=h_{uv}dQ^{u}dQ^{v}\qquad ,\qquad \boldsymbol{K}_{2}^{a}=\mathcal{K}%
_{uv}^{a}dQ^{u}\wedge dQ^{v}
\end{equation}%
with symmetric $h_{uv}=h_{vu}$ and antisymmetric \QTR{cal}{K}$_{uv}^{a}=-%
\mathcal{K}_{vu}^{a}$ as follows%
\begin{equation}
\begin{tabular}{lll}
$h_{uv}$ & $=$ & $\varepsilon _{AB}\mathcal{E}_{u}^{AM}C_{MN}\mathcal{E}%
_{v}^{BN}$ \\ 
$\mathcal{K}_{uv}^{a}$ & $=$ & $-i\left( \tau ^{a}\right) _{AB}\mathcal{E}%
_{u}^{AM}\mathcal{C}_{MN}\mathcal{E}_{v}^{BN}$%
\end{tabular}%
\end{equation}

$\bullet $ \emph{three complex structures}\newline
In the above relations, the $SP\left( 2n_{H}\right) $ symplectic invariance
of the metric $h_{uv}$ and the closed hyperkahler 2-forms $\mathcal{K}%
_{uv}^{a}$ is manifestly exhibited. By using the hermiticity property of the
vielbeins (\ref{3k}) expressing $\mathcal{E}_{uAM}$ as $\varepsilon _{AB}%
\mathcal{C}_{MN}\mathcal{E}_{u}^{BN}$; we can rewrite the hermitian metric
like $h_{uv}=\mathcal{E}_{u}^{AM}\mathcal{E}_{vAM}$ and the hyperkahler
2-form like $\mathcal{K}_{uv}^{a}=i\mathcal{E}_{u}^{AM}\left( \tau
^{a}\right) _{A}^{B}\mathcal{E}_{vBM}$. From the SU$\left( 2\right) $ R-
symmetry view, the metric is nothing but the trace of the matrix $\mathcal{E}%
_{u}^{AM}\mathcal{E}_{vBM}$. The last object can in general be decomposed as
the sum of a singlet and a triplet as follows%
\begin{equation}
\mathcal{E}_{u}^{AM}\mathcal{E}_{vBM}=\frac{1}{2}\delta _{B}^{A}h_{uv}-\frac{
i}{2}\mathcal{K}_{uv}^{a}\left( \tau _{a}\right) _{B}^{A}
\end{equation}%
Notice also that the quaternionic Kahler $\boldsymbol{M}_{QK}$ has three
complex structures $J_{1},$ $J_{2},$ $J_{3}$, generating $SU(2)_{R}$, acting
on the \textrm{\ }tangent space like $\left( J_{a}\right) _{v}^{u}$ and
satisfying the quaternionic algebra%
\begin{equation}
J^{a}J^{b}=-\delta ^{ab}+\varepsilon ^{abc}J^{c}  \label{qst}
\end{equation}%
In terms of the quaternionic structure, the three $4n_{H}\times 4n_{H}$
matrices $\mathcal{K}_{uv}^{a}$ are related to the hyperkahler metric $%
h_{uv} $ as follows 
\begin{equation}
\mathcal{K}_{uv}^{a}=h_{uw}\left( J^{a}\right) _{v}^{w}
\end{equation}%
Notice moreover that the hyperkahler 2-forms $\boldsymbol{K}_{2}^{a}$ obey
other special features; in particular the following ones: First, by using $%
h^{uw}$, the inverse of the metric, we can reexpress $\left( J^{a}\right)
_{v}^{w}$ like $h^{uw}\mathcal{K}_{wv}^{a}$. Putting it back into eq(\ref%
{qst}) we obtain the property 
\begin{equation}
\mathcal{K}_{uw}^{a}h^{ws}\mathcal{K}_{sv}^{b}=-\delta
^{ab}h_{uv}+\varepsilon ^{abc}\mathcal{K}_{uv}^{c}  \label{hkf}
\end{equation}%
By multiplying both sides of this relation by $h^{tu}$ and setting $\left( 
\mathcal{K}^{b}\right) ^{tw}=h^{tv}h^{ws}\mathcal{K}_{sv}^{b}$, we have 
\begin{equation}
\mathcal{K}_{uw}^{a}\left( \mathcal{K}^{b}\right) ^{tw}=-\delta ^{ab}\delta
_{u}^{t}+\varepsilon ^{abc}\mathcal{K}_{uv}^{c}h^{tv}
\end{equation}%
from which we deduce that for a given value of index a, $\left( \mathcal{K}%
^{a}\right) ^{wt}$ is the inverse of $\mathcal{K}_{uw}^{a}$; i.e: 
\begin{equation}
\mathcal{K}_{uw}^{a}\left( \mathcal{K}^{a}\right) ^{wt}=\delta _{u}^{t}
\end{equation}%
Second, the 2-forms $\boldsymbol{K}_{2}^{a}$ are covariantly constant $%
\nabla \boldsymbol{K}_{2}^{a}=0$ with respect to the SU$\left( 2\right) _{R}$
connection given by the three 1-forms $\mathbf{\omega }_{1}^{a}=\omega
_{u}^{a}dQ^{u}$ living in the cotangent space bundle of the scalar manifold.
This condition reads explicitly as follows 
\begin{equation}
\nabla \boldsymbol{K}_{2}^{a}=d\boldsymbol{K}_{2}^{a}+\varepsilon ^{abc}%
\mathbf{\omega }_{1}^{b}\boldsymbol{K}_{2}^{c}=0  \label{dek}
\end{equation}%
This is a remarkable condition as it allows to get more insight into the
properties of $\boldsymbol{K}_{2}^{a}$. Indeed, the gauge curvature 2-forms $%
\boldsymbol{\Omega }_{2}^{a}=d\boldsymbol{\omega }_{1}^{a}+\frac{1}{2}%
\varepsilon ^{abc}\boldsymbol{\omega }_{1}^{b}\boldsymbol{\omega }_{1}^{c}$
of the SU$\left( 2\right) $ connection $\boldsymbol{\omega }_{1}^{a}$ obey a
similar constraint following from the Jacobi identity of the SU$\left(
2\right) $ bracket namely $d\boldsymbol{\Omega }_{2}^{a}+\varepsilon ^{abc}%
\mathbf{\omega }_{1}^{b}\boldsymbol{\Omega }_{2}^{c}=0$. By comparing this
relation with (\ref{dek}), it follows that the 2-forms $\boldsymbol{K}%
_{2}^{a}$ and $\boldsymbol{\Omega }_{2}^{a}$ should be proportional like%
\begin{equation}
\boldsymbol{\Omega }_{2}^{a}=\lambda \, \boldsymbol{K}_{2}^{a}  \label{olk}
\end{equation}%
where $\lambda $\ is a real coefficient, depending on the normalization of
the metric, taken here $\lambda =-1$\textrm{.}

\  \  \  \  \  \textrm{\ }

$\bullet $ \emph{tri-holomorphic}\textrm{\ }\emph{moment maps}\newline
On the quaternionic Kahler manifold $\boldsymbol{M}_{QK}$, the moment maps $%
\mathcal{P}_{n}^{a}\left( Q\right) $, associated with the Killing vector
fields $k_{n}=k_{n}^{u}\left( Q\right) \frac{\partial }{\partial Q^{u}}$,
carry a quantum number of the $SU\left( 2\right) $ R-symmetry; they behave
as hermitian triplets. To get the relationship between $\mathcal{P}_{n}^{a}$
and $\iota _{n}\left( \boldsymbol{\omega }_{1}^{a}\right) $, the contraction
of the 1-forms isotriplet $\boldsymbol{\omega }_{1}^{a}$, we consider
infinitesimal isometries generated by $t_{m}$, whose action on the scalar
fields given by the Killing vectors $k_{m}$\textrm{. }These generators obey
the isometry algebra\textrm{\ }%
\begin{equation}
\begin{tabular}{lll}
$\lbrack t_{m},\,t_{n}]$ & $=$ & $f_{mn}{}^{p}\,t_{p}$ \\ 
$\lbrack k_{m},\,k_{n}]$ & $=$ & $-f_{mn}{}^{p}\,k_{p}$%
\end{tabular}%
\end{equation}%
living invariant the 4-form\textrm{\ }$\sum \boldsymbol{K}_{2}^{a}\wedge 
\boldsymbol{K}_{2a}$ \textrm{\cite{D'Auria:1990fj,Andrianopoli:2015rpa}}.
This property leads to\textrm{\ }%
\begin{equation}
\mathcal{L}_{n}\boldsymbol{K}_{2}^{a}=\varepsilon ^{abc}\boldsymbol{K}%
_{2}^{b}W_{n}^{c}  \label{lk}
\end{equation}%
where $W_{n}^{a}$\ is an $SU(2)_{R}$- compensator. Eq(\ref{lk}) is solved as
follows \textrm{\cite{D'Auria:1990fj,Andrianopoli:2015rpa}}%
\begin{equation}
\iota _{n}\left( \boldsymbol{K}_{2}^{a}\right) =-\nabla \mathcal{P}_{n}^{a}
\end{equation}%
with covariant derivative 1-form as follows%
\begin{equation}
\nabla \mathcal{P}_{n}^{a}=d\mathcal{P}_{n}^{a}+\varepsilon ^{abc}%
\boldsymbol{\omega }_{1}^{b}\, \mathcal{P}_{n}^{c})
\end{equation}%
which gives the following expression of the moment maps \textrm{\cite%
{Galicki:1986ja, D'Auria:1990fj}}%
\begin{equation}
\mathcal{P}_{n}^{a}=\lambda ^{-1}(\left[ \iota _{n}\left( \boldsymbol{\omega 
}_{1}^{a}\right) -W_{n}^{a}\right]
\end{equation}%
By taking $\lambda =-1$, we have%
\begin{equation}
\mathcal{P}_{n}^{a}=W_{n}^{a}-\iota _{n}\left( \boldsymbol{\omega }%
_{1}^{a}\right)
\end{equation}%
Notice that for those isometries with vanishing compensator\textrm{, }$%
W_{n}^{a}=0$\textrm{, }the moment maps take the simple expression\textrm{\ }$%
P_{n}^{a}=-k_{n}^{u}\, \omega _{u}^{a}$\textrm{. }In this case, the
quaternionic manifold is globally isometric to a solvable Lie group $%
e^{Solv} $ generated by a solvable Lie algebra $Solv$\textrm{\  \cite%
{Andrianopoli:2015rpa}.}

\section{Appendix C: Coupling matrices $\mathcal{N}_{\Lambda \Sigma }$ and $%
\mathcal{U}^{MN}$}

\textrm{\label{appendixC}}

In this appendix, we determine the rigid limit of the coupling matrix $%
\mathcal{N}_{\Lambda \Sigma }$ appearing in the supergravity Lagrangian
density (\ref{act},\ref{tca}) as well as the rigid limit $\mathcal{\mathring{%
U}}^{MN}$ of the factor 
\begin{equation}
\mathcal{U}^{MN}=U_{j}^{M}\mathcal{G}^{i\bar{j}}\bar{U}_{\bar{j}}^{N}
\end{equation}%
involved in the structure of the scalar potential (\ref{vscc}).

\subsection{Rigid limit of period matrix $\mathcal{N}_{\Lambda \Sigma }$}

Using the SP$\left( 2n+2\right) $ symplectic structure on the special Kahler
manifold $\boldsymbol{M}_{SK}$, the $2\left( n+1\right) $ components of $%
V^{M}$ can be decomposed into a$\  \left( n+1\right) $ electric part and a $%
\left( n+1\right) $ magnetic dual like 
\begin{equation}
V^{M}=\left( 
\begin{array}{c}
\Upsilon ^{\Lambda } \\ 
\Gamma _{\Lambda }%
\end{array}%
\right)
\end{equation}%
with $\Upsilon ^{\Lambda }=\Upsilon ^{\Lambda }\left( z,\bar{z}\right) $ and 
$\Gamma _{\Lambda }=\Gamma _{\Lambda }\left( z,\bar{z}\right) $ splitting in
turn in terms of graviphoton components and Coulomb branch ones as follows%
\begin{equation}
\Upsilon ^{\Lambda }=\left( 
\begin{array}{c}
\Upsilon ^{0} \\ 
\Upsilon ^{i}%
\end{array}%
\right) \qquad ,\qquad \Gamma _{\Lambda }=\left( 
\begin{array}{c}
\Gamma _{0} \\ 
\Gamma _{i}%
\end{array}%
\right)
\end{equation}%
where the Coulomb branch index $i$ runs from 1 to n. Following \textrm{\cite%
{Ceresole:1995jg, Ceresole:1995ca, Craps:1997gp}}, the blocks $\Gamma
_{\Lambda }$ and $\Upsilon ^{\Lambda }$ are related like 
\begin{equation}
\Gamma _{\Lambda }=\mathcal{N}_{\Lambda \Sigma }\Upsilon ^{\Sigma }
\label{permat}
\end{equation}%
with complex coupling and symmetric $\left( n+1\right) \times \left(
n+1\right) $ matrix $\mathcal{N}_{\Lambda \Sigma }$ defined as follows 
\begin{equation}
\mathcal{N}_{\Lambda \Sigma }=\mathcal{\bar{F}}_{\Lambda \Sigma }+\mathcal{N}%
_{\Lambda \Sigma }^{loc}
\end{equation}%
The first term in above relation is given by $\mathcal{\bar{F}}_{\Lambda
\Sigma }=\frac{\partial ^{2}\mathcal{\bar{F}}}{\partial \bar{X}^{\Lambda
}\partial \bar{X}^{\Sigma }}$ with leading terms in the $\frac{1}{\mathrm{%
\mu }}$-expansion as 
\begin{equation}
\mathcal{\bar{F}}_{\Lambda \Sigma }=%
\begin{pmatrix}
\frac{i}{2}+\mathfrak{O}(\frac{1}{\mathrm{\mu }^{2}}) & \frac{i}{2\mathrm{%
\mu }}\eta _{j}+\mathfrak{O}(\frac{1}{\mathrm{\mu }^{2}}) \\ 
\frac{i}{2\mathrm{\mu }}\eta _{i}+\mathfrak{O}(\frac{1}{\mathrm{\mu }^{2}})
& \frac{i}{2\mathrm{\mu }^{2}}\partial _{ji}\bar{\phi}+\mathfrak{O}(\frac{1}{%
\mathrm{\mu }^{3}})%
\end{pmatrix}%
\end{equation}%
The extra matrix term $\mathcal{N}_{\Lambda \Sigma }^{loc}$ is given by 
\begin{equation}
\mathcal{N}_{\Lambda \Sigma }^{loc}=2i\frac{\left[ \left( \func{Im}\mathcal{F%
}\right) _{\Lambda \Delta }X^{\Delta }\right] \left[ \left( \func{Im}%
\mathcal{F}\right) _{\Sigma \Gamma }X^{\Gamma }\right] }{X^{\Delta }\left( 
\func{Im}\mathcal{F}\right) _{\Delta \Gamma }X^{\Gamma }}
\end{equation}%
its leading terms in the $\frac{1}{\mathrm{\mu }}$-expansion read as follows 
\begin{equation}
\mathcal{N}_{\Lambda \Sigma }^{loc}=%
\begin{pmatrix}
-i+\mathfrak{O}(\frac{1}{\mathrm{\mu }^{2}}) & -\frac{i}{\mathrm{\mu }}\eta
_{j}+\mathfrak{O}(\frac{1}{\mathrm{\mu }^{2}}) \\ 
-\frac{i}{\mathrm{\mu }}\eta _{i}+\mathfrak{O}(\frac{1}{\mathrm{\mu }^{2}})
& -\frac{i}{\mathrm{\mu }^{2}}\eta _{i}\eta _{j}+\mathfrak{O}(\frac{1}{%
\mathrm{\mu }^{3}})%
\end{pmatrix}%
\end{equation}%
Using the symplectic transformation (\ref{spf}), one can show that the
period matrix $\mathcal{N}_{\Lambda \Sigma }$ may be diagonalised by help of
the two matrices $\mathbb{A}$ and $\mathbb{D}$ (\ref{spf}) like 
\begin{equation}
\widetilde{\mathcal{N}}_{\Lambda \Sigma }=\left( \mathbb{D}\mathcal{N}%
\mathbb{A}^{-1}\right) _{\Lambda \Sigma }
\end{equation}%
and can as well be splited into the following form \textrm{\cite%
{Ceresole:1995ca, Ceresole:1995jg, deWit:1995jd, Sabra:1996xg,
DAuria:1996lbq}}%
\begin{equation}
\widetilde{\mathcal{N}}_{\Lambda \Sigma }=\widetilde{\bar{F}}_{\Lambda
\Sigma }+\widetilde{\mathcal{N}}_{\Lambda \Sigma }^{loc}
\end{equation}%
By using eqs(\ref{10}-\ref{ef}), we get%
\begin{equation}
\begin{tabular}{lll}
$\widetilde{\bar{F}}_{\Lambda \Sigma }$ & $=$ & $%
\begin{pmatrix}
\frac{i}{2} & \mathbf{0}_{1\times n} \\ 
\mathbf{0}_{n\times 1} & -\frac{i}{2}\left( \eta _{i}\eta _{j}-\partial _{ji}%
\bar{\phi}\right)%
\end{pmatrix}%
+\mathfrak{O}\left( \frac{1}{\mathrm{\mu }}\right) $ \\ 
&  &  \\ 
$\widetilde{\mathcal{N}}_{\Lambda \Sigma }^{loc}$ & $=$ & $%
\begin{pmatrix}
-i & \mathbf{0}_{1\times n} \\ 
\mathbf{0}_{n\times 1} & \mathbf{0}_{n\times n}%
\end{pmatrix}%
+\mathfrak{O}\left( \frac{1}{\mathrm{\mu }}\right) $%
\end{tabular}%
\end{equation}%
and then%
\begin{equation}
\widetilde{\mathcal{N}}_{\Lambda \Sigma }=%
\begin{pmatrix}
-\frac{i}{2} & \mathbf{0}_{1\times n} \\ 
\mathbf{0}_{n\times 1} & -\frac{i}{2}\left( \eta _{i}\eta _{j}-\partial _{ji}%
\bar{\Phi}\right)%
\end{pmatrix}%
+\mathfrak{O}\left( \frac{1}{\mathrm{\mu }}\right)  \label{pm}
\end{equation}%
We notice that because of this nontrivial form of $\widetilde{\mathcal{N}}%
_{\Lambda \Sigma }$; which contains contribution from both the graviphoton
and rigid parts having the same zero order of $\frac{1}{\mathrm{\mu }}$; the
graviphoton will play an important role in the rigid theory as we will see
in \textrm{\autoref{sec5}}. From eq(\ref{pm}), we learn the expressions of
the prepotential $\mathcal{\mathring{F}}$ of the rigid theory and the
prepotential $\mathcal{F}^{grav}$ associated with the graviphoton part; they
read as 
\begin{equation}
\begin{tabular}{lll}
$\mathcal{\mathring{F}}$ & $=$ & $\frac{i}{4}\left[ \frac{1}{2\mathrm{\mu }}%
\left( \mathbf{\eta .z}\right) ^{2}-2\Phi \right] +\mathfrak{O}\left( \frac{1%
}{\mathrm{\mu }}\right) $ \\ 
$\mathcal{F}^{grav}$ & $=$ & $c\text{ }X^{0}-\frac{i}{4}(X^{0})^{2}+%
\mathfrak{O}\left( \frac{1}{\mathrm{\mu }}\right) $%
\end{tabular}%
\end{equation}%
where $c$ is independent of $X^{0}$; it can be set to zero by performing the
particular Kahler transformation (\ref{1k}). So the leading term of the pure
supergravity prepotential $\mathcal{F}^{grav}$ in the $\frac{1}{\mathrm{\mu }%
}$-expansion reduces to the following relation which is in agreement with
the usual term of the literature \textrm{\cite{Ortin:2015hya}}%
\begin{equation}
\mathcal{F}^{grav}=-\frac{i}{4}(X^{0})^{2}+\mathfrak{O}\left( \frac{1}{%
\mathrm{\mu }}\right)
\end{equation}

\subsection{Rigid limit of\emph{\ }$\mathcal{U}^{MN}$}

Recall that in $\mathcal{N}=2$ supergravity theory, we have the following
decomposition property of the rank 2 symplectic tensor $\mathcal{U}%
^{MN}=U_{j}^{M}\mathcal{G}^{i\bar{j}}\bar{U}_{\bar{j}}^{N}$ \textrm{\cite%
{Andrianopoli:2015rpa}, for a its derivation see eq(B.37) also appendix B,} 
\begin{equation}
\mathcal{U}^{MN}=-\frac{1}{2}\mathcal{M}^{MN}-\frac{i}{2}\mathcal{C}^{MN}-%
\bar{V}^{M}V^{N}  \label{tmn}
\end{equation}%
where the antisymmetric $\mathcal{C}^{MN}$ is the symplectic metric of SP$%
\left( 2n+2\right) $ and $\mathcal{M}^{MN}=\left( \mathcal{M}_{MN}\right)
^{-1}$, the inverse of the symmetric matrix built of $\func{Re}\mathcal{N}%
_{\Lambda \Sigma }$ and $\func{Im}\mathcal{N}_{\Lambda \Sigma }$ as follows 
\begin{equation}
\mathcal{M}_{MN}=%
\begin{pmatrix}
\func{Im}\mathcal{N}+\func{Re}\mathcal{N}\left( \func{Im}\mathcal{N}\right)
^{-1}\func{Re}\mathcal{N} & -\func{Re}\mathcal{N}\left( \func{Im}\mathcal{N}%
\right) ^{-1} \\ 
-\left( \func{Im}\mathcal{N}\right) ^{-1}\func{Re}\mathcal{N} & \left( \func{%
Im}\mathcal{N}\right) ^{-1}%
\end{pmatrix}
\label{sm}
\end{equation}%
with $\mathcal{N}_{\Lambda \Sigma }$ standing for the period matrix of eq(%
\ref{permat}). The rigid limit of $\mathcal{U}^{MN}$ is obtained by
determining the rigid limit of $\mathcal{M}^{MN}$ and $\bar{V}^{M}V^{N}$. To
obtain the rigid limit of $\mathcal{M}_{MN}$, we use eq(\ref{pm}) which
tells us that the rigid theory is described by $\widetilde{\mathcal{%
\mathring{N}}}_{ij}=\overline{\mathcal{\mathring{F}}}_{ij}$ with%
\begin{equation}
\overline{\mathcal{\mathring{F}}}_{ij}=-\frac{i}{2}\left( \eta _{i}\eta
_{j}-\partial _{ji}\bar{\Phi}\right) +\mathfrak{O}\left( \frac{1}{\mathrm{%
\mu }}\right)
\end{equation}%
Putting $\func{Im}\widetilde{\mathcal{\mathring{N}}}_{ij}=$ $-\func{Im}%
\mathcal{\mathring{F}}_{ij}$ back into (\ref{sm}), we get 
\begin{equation}
\mathcal{M}^{MN}=-\delta _{\underline{M}}^{M}\delta _{\underline{N}}^{N}%
\mathcal{\mathring{M}}^{\underline{M}\underline{N}}+\mathfrak{O}\left( \frac{%
1}{\mathrm{\mu }}\right)
\end{equation}%
with $\mathcal{\mathring{M}}^{\underline{M}\underline{N}}\equiv (\mathcal{%
\mathring{M}}_{\underline{M}\underline{N}})^{-1}$ and%
\begin{equation}
\mathcal{\mathring{M}}_{\underline{M}\underline{N}}=%
\begin{pmatrix}
\func{Im}\mathcal{\mathring{F}}+\func{Re}\mathcal{\mathring{F}}\left( \func{%
Im}\mathcal{\mathring{F}}\right) ^{-1}\func{Re}\mathcal{\mathring{F}} & -%
\func{Re}\mathcal{\mathring{F}}\left( \func{Im}\mathcal{\mathring{F}}\right)
^{-1} \\ 
-\left( \func{Im}\mathcal{\mathring{F}}\right) ^{-1}\func{Re}\mathcal{%
\mathring{F}} & \left( \func{Im}\mathcal{\mathring{F}}\right) ^{-1}%
\end{pmatrix}
\label{mmn}
\end{equation}%
where $\func{Im}\mathcal{\mathring{F}}\equiv \func{Im}\mathcal{\mathring{F}}%
_{IJ}$ and $\func{Re}\mathcal{\mathring{F}}\equiv \func{Re}\mathcal{%
\mathring{F}}_{IJ}$. Moreover, we learn from eq(\ref{rd}) that the
components $\bar{V}^{\underline{M}}V^{\underline{N}}=\delta _{M}^{\underline{%
M}}\delta _{N}^{\underline{N}}\bar{V}^{M}V^{N}$ in eq(\ref{tmn}) do not
contribute; so we have the two following following expressions%
\begin{equation}
U_{i}^{M}\mathcal{G}^{i\bar{j}}\bar{U}_{\bar{j}}^{N}\qquad \rightarrow
\qquad \delta _{\underline{M}}^{M}\delta _{\underline{N}}^{N}\left[ \frac{1}{%
2}\mathcal{\mathring{M}}^{\underline{M}\underline{N}}-\frac{i}{2}\mathcal{C}%
^{\underline{M}\underline{N}}\right]
\end{equation}%
and%
\begin{equation}
U_{i}^{M}\mathcal{G}^{i\bar{j}}\bar{U}_{\bar{j}}^{N}\qquad \rightarrow
\qquad \frac{1}{\mathrm{\mu }}\mathring{U}_{i}^{\underline{M}}\left( \mathrm{%
\mu }^{2}\mathcal{\mathring{G}}^{i\bar{j}}\right) \frac{1}{\mathrm{\mu }}%
\overline{\mathring{U}}_{\bar{j}}^{\underline{N}}
\end{equation}%
from which we deduce the relation 
\begin{equation}
\mathring{U}_{i}^{\underline{M}}\mathcal{\mathring{G}}^{i\bar{j}}\overline{%
\mathring{U}}_{\bar{j}}^{\underline{N}}=\frac{1}{2}\mathcal{\mathring{M}}^{%
\underline{M}\underline{N}}-\frac{i}{2}\mathcal{C}^{\underline{M}\underline{N%
}}  \label{rlm}
\end{equation}%
which coincides with the rigid theory relation given in \textrm{\cite%
{Andrianopoli:2015wqa, Andrianopoli:2006ub}}. Notice also that the
components associated with the graviphoton direction in (\ref{41}) is
trivially satisfied at the $\mathfrak{O}\left( \frac{1}{\mathrm{\mu }}%
\right) $ order as shown here below%
\begin{equation}
-\frac{1}{2}\mathcal{\mathring{M}}_{grav}^{\underline{\rho }\underline{%
\sigma }}-\frac{i}{2}\mathcal{C}^{\underline{\rho }\underline{\sigma }}-\bar{%
V}_{grav}^{\underline{\rho }}V_{grav}^{\underline{\sigma }}=\mathbf{0}%
_{2\times 2}  \label{grad}
\end{equation}%
with $\mathcal{C}^{\underline{\rho }\underline{\sigma }}$ is the symplectic $%
SP(2)$ metric, $V_{grav}^{\underline{\sigma }}$ as in\ eq(\ref{vgra}) and%
\begin{equation}
\mathcal{\mathring{M}}_{grav}^{\underline{\rho }\underline{\sigma }}=%
\begin{pmatrix}
-2 & 0 \\ 
0 & -\frac{1}{2}%
\end{pmatrix}
\label{mgra}
\end{equation}

\section{Appendix D: Solving Ward identities}

\label{appendixD}

In this appendix, we study the Ward identities (\ref{cst}) and a class of
solutions in terms of the geometric objects of the scalar manifold of $%
\mathcal{N}=2$ supergravity. We also give an explicit check of these
solutions.

\subsection{Ward identities for quaternionic gauging}

First, we recall the Ward identities of gauging isometries; then we give the
solution for the particular case of Abelian quaternionic isometries.

\  \  \ 

$\bullet $ \emph{Ward identities}\newline
Following \textrm{\cite{Andrianopoli:2015rpa}}, the Ward identities of $%
\mathcal{N}=2$ gauged supergravity%
\begin{equation}
-12\bar{S}_{C}^{A}S_{B}^{C}+2\bar{N}_{\dot{C}}^{A}N_{B}^{\dot{C}%
}+\sum_{i,j=1}^{n_{V}}\mathcal{G}_{i\bar{j}}\left( W^{i}\right)
_{C}^{A}\left( \bar{W}^{\bar{j}}\right) _{B}^{C}=\delta _{B}^{A}.\mathcal{V}%
_{\text{sugra}}^{\mathcal{N}=2}  \label{wi}
\end{equation}%
By setting%
\begin{equation}
\begin{tabular}{lll}
$X_{B}^{A}$ & $=$ & $\sum_{i,j=1}^{n_{V}}\mathcal{G}_{i\bar{j}}\left(
W^{i}\right) _{C}^{A}\left( \bar{W}^{\bar{j}}\right) _{B}^{C}$ \\ 
$Y_{B}^{A}$ & $=$ & $2\bar{N}_{\dot{C}}^{A}N_{B}^{\dot{C}}$ \\ 
$Z_{B}^{A}$ & $=$ & $-12\bar{S}_{C}^{A}S_{B}^{C}$%
\end{tabular}%
\end{equation}%
the matrix eq(\ref{wi}) becomes%
\begin{equation}
X_{B}^{A}+Y_{B}^{A}+Z_{B}^{A}=\delta _{B}^{A}\mathcal{V}_{\text{sugra}}^{ 
\mathcal{N}=2}
\end{equation}%
and splits as%
\begin{equation}
tr\left[ \tau ^{a}\left( X_{B}^{A}+Y_{B}^{A}+Z_{B}^{A}\right) \right] =0
\label{wj}
\end{equation}%
and%
\begin{equation}
tr\left( X_{B}^{A}+Y_{B}^{A}+Z_{B}^{A}\right) =2\mathcal{V}_{\text{sugra}}^{ 
\mathcal{N}=2}  \label{wk}
\end{equation}

\  \  \  \ 

$\bullet $ \emph{Solutions of Ward identities}\newline
For the gauging of Abelian quaternionic isometries, the $\left( W^{i}\right)
_{B}^{A}$, $N_{B}^{A}$ and $S_{B}^{A}$ matrices can be expressed like%
\begin{equation}
\begin{tabular}{lll}
$\left( W^{i}\right) _{B}^{A}$ & $=$ & $W^{ia}\left( \tau _{a}\right) ^{AB}$
\\ 
$N_{B}^{A}$ & $=$ & $N^{a}\left( \tau _{a}\right) _{B}^{A}$ \\ 
$S_{B}^{A}$ & $=$ & $S^{a}\left( \tau _{a}\right) _{B}^{A}$%
\end{tabular}
\label{wiid1}
\end{equation}%
and the above Ward identities have a solution in terms of the geometric
objects of the scalar manifold $\boldsymbol{M}_{SK}\times \boldsymbol{M}%
_{QK} $ as follows%
\begin{equation}
\begin{tabular}{lll}
$W^{ia}$ & $=$ & $-i\mathcal{P}_{M}^{a}\mathcal{G}^{i\bar{j}}\bar{U}_{\bar{j}%
}^{M}$ \\ 
$N^{a}$ & $=$ & $i\mathcal{P}_{M}^{a}\bar{V}^{M}$ \\ 
$S^{a}$ & $=$ & $\frac{i}{2}\mathcal{P}_{M}^{a}V^{M}$%
\end{tabular}
\label{wiid2}
\end{equation}

\subsection{Checking the solution of Ward identities}

Here, we want to explicitly check that the solution (\ref{wiid2}) satisfies
indeed the Ward identities (\ref{wi}); the sum of contributions coming from
gauge and matter sectors are compensated by a negative contribution coming
from gravity sector.

\  \  \  \ 

$\bullet $ \emph{Computing }$X_{B}^{A}$\newline
Substituting $W^{iAC}$ by its expression,\ we have%
\begin{equation}
X_{B}^{A}=\left( \tau _{a}\right) _{C}^{A}\left( \tau _{b}\right) _{B}^{C} 
\left[ \mathcal{G}^{i\bar{k}}\bar{U}_{\bar{k}}^{N}U_{k}^{M}\right] \mathcal{P%
}_{N}^{a}\mathcal{P}_{M}^{b}
\end{equation}%
Expanding $\tau _{a}\tau _{b}$ as $\frac{1}{2}\left \{ \tau _{a},\tau
_{b}\right \} +\frac{1}{2}\left[ \tau _{a},\tau _{b}\right] $, we find that
the $2\times 2$ matrix $X_{B}^{A}$ is a sum of two contributions, an
isosinglet, coming from $\left \{ \tau _{a},\tau _{b}\right \} ,$ and an
isotriplet, associated with $\left[ \tau _{a},\tau _{b}\right] $, as shown
on the following relation%
\begin{equation}
X_{B}^{A}=\delta _{ab}\delta _{B}^{A}\left[ \mathcal{G}^{i\bar{k}}\bar{U}_{ 
\bar{k}}^{M}U_{k}^{N}\right] \mathcal{P}_{M}^{a}\mathcal{P}%
_{N}^{b}+i\varepsilon _{abc}\left( \tau ^{c}\right) _{B}^{A}\mathcal{P}%
_{N}^{a}\mathcal{P}_{M}^{b}\left[ V^{M}\bar{V}^{N}\right]  \label{xab2}
\end{equation}

\  \  \ 

$\bullet $ \emph{Computing }$Y_{B}^{A}$\newline
By using eqs(\ref{wiid1},\ref{wiid2}), we obtain%
\begin{equation}
Y_{B}^{A}=2\left[ -i\left( \tau _{a}\right) _{C}^{A}\mathcal{P}_{M}^{a}V^{M}%
\right] \left[ i\left( \tau _{b}\right) _{B}^{C}\mathcal{P}_{N}^{b}\bar{V}%
^{N}\right]
\end{equation}%
having as well two contributions as 
\begin{equation}
Y_{B}^{A}=2\delta _{ab}\delta _{B}^{A}\mathcal{P}_{M}^{a}\mathcal{P}%
_{N}^{b}V^{M}\bar{V}^{N}+2i\varepsilon _{abc}\mathcal{P}_{M}^{a}\mathcal{P}%
_{N}^{b}V^{M}\bar{V}^{N}\left( \tau ^{c}\right) _{B}^{A}  \label{yab2}
\end{equation}

\  \  \ 

$\bullet $ \emph{Computing }$Z_{B}^{A}$\newline
For the gravitini shifts $S_{BC}$, we have a quite similar expression given
by%
\begin{equation}
Z_{B}^{A}=-12\left[ -\frac{i}{2}\left( \tau _{a}\right) _{C}^{A}\mathcal{P}%
_{N}^{a}\bar{V}^{N}\right] \left[ \frac{i}{2}\left( \tau _{b}\right) _{B}^{C}%
\mathcal{P}_{N}^{b}V^{N}\right]
\end{equation}%
and splitting as%
\begin{equation}
Z_{B}^{A}=-3\delta _{ab}\delta _{B}^{A}\mathcal{P}_{N}^{a}\mathcal{P}%
_{M}^{b}V^{M}\bar{V}^{N}-3i\varepsilon _{abc}\mathcal{P}_{N}^{a}\mathcal{P}%
_{M}^{b}V^{M}\bar{V}^{N}\left( \tau ^{c}\right) _{B}^{A}  \label{zab2}
\end{equation}%
Combining eqs(\ref{xab2},\ref{yab2},\ref{zab2}), we obtain%
\begin{equation}
X_{B}^{A}+Y_{B}^{A}+Z_{B}^{A}=\delta _{B}^{A}\mathcal{V}_{\text{sugra}}^{%
\mathcal{N}=2}
\end{equation}%
with%
\begin{equation}
\mathcal{V}_{\text{sugra}}^{\mathcal{N}=2}=-3\mathcal{P}_{N}^{a}\mathcal{P}%
_{M}^{a}V^{M}\bar{V}^{N}+2\mathcal{P}_{M}^{a}\mathcal{P}_{N}^{a}V^{M}\bar{V}%
^{N}+\left[ \mathcal{G}^{i\bar{k}}\bar{U}_{\bar{k}}^{M}U_{k}^{N}\right] 
\mathcal{P}_{M}^{a}\mathcal{P}_{N}^{a}
\end{equation}%
reducing to%
\begin{equation}
\mathcal{V}_{\text{sugra}}^{\mathcal{N}=2}=\left[ \mathcal{G}^{i\bar{k}}\bar{%
U}_{\bar{k}}^{M}U_{k}^{N}\right] \mathcal{P}_{N}^{a}\mathcal{P}_{M}^{a}-%
\mathcal{P}_{N}^{a}\mathcal{P}_{M}^{a}V^{M}\bar{V}^{N}
\end{equation}%
and%
\begin{equation}
\left( \tau ^{c}\right) _{B}^{A}\mathcal{P}_{N}^{a}\mathcal{P}_{M}^{b}V^{M}%
\bar{V}^{N}\left[ i\varepsilon _{abc}+2i\varepsilon _{abc}-3i\varepsilon
_{abc}\right] =0
\end{equation}

\section{Appendix E: Rigid limit of the scalar potential}

\textrm{\label{appendixE}}

The aim of this appendix is to compute the rigid limit of the scalar
potential $\mathcal{V}_{\text{sugra}}^{\mathcal{N}=2}$. To this end, we
first give the $\frac{1}{\mathrm{\mu }}$-expansion of the fermion shift
matrices, namely\ $W^{ia},$\emph{\ }$N^{a}$\emph{\ }and $S^{a}$.

\subsection{$\frac{1}{\mathrm{\protect \mu }}$\emph{- }expansion of \ fermion
shift matrices}

To derive the rigid limit of (\ref{prf}), we have to determine the $\frac{1}{%
\mathrm{\mu }}$- expansions of $W^{ia},$ $N^{a}$ and $S^{a}$ by proceeding
in three steps as follows: First, we compute the $\frac{1}{\mathrm{\mu }}$%
\emph{- }expansion of the $W^{ia}$ describing the contribution coming from
the Coulomb branch. Then, we determine the expansion of the term $N^{a}$ of
hypermultiplet sector; and after that we turn to the $\frac{1}{\mathrm{\mu }}
$- development of the factor $S^{a}$ of gravity branch.

\  \  \  \  \  \ 

$\bullet $ $\frac{1}{\mathrm{\mu }}$\emph{- expansion of the term} $W^{ia}$%
\newline
To obtain the $\frac{1}{\mathrm{\mu }}$- expansion of $W^{ia}$, we use
results derived in previous sections concerning developments in power series
of $\frac{1}{\mathrm{\mu }}$ of symplectic sections and moment maps on the
scalar manifold $\boldsymbol{M}_{SK}\times \boldsymbol{M}_{QK}\mathrm{;}$ in
particular the following things: $\left( i\right) $ eq(\ref{met}) giving the 
$\frac{1}{\mathrm{\mu }}$- expansion of the metric $\mathcal{G}_{i\bar{j}}$; 
$\left( ii\right) $ eq(\ref{uurig}) determining the development of the $%
U_{i}^{M}$ sections; $\left( iii\right) $ eq(\ref{mp}) giving the expression
of the moment maps $\mathcal{P}_{M}^{a}$ in terms of the embedding tensor $%
\vartheta _{M}^{a}$; and $\left( iv\right) $ eqs(\ref{ts}-\ref{ss})
regarding the $\frac{1}{\mathrm{\mu }}$- expansion of the embedding tensor.
For convenience, we recall these relationships here below 
\begin{equation}
\begin{tabular}{lll}
$\mathcal{G}_{i\bar{j}}$ & $=$ & $\frac{1}{\mathrm{\mu }^{2}}\mathcal{%
\mathring{G}}_{i\bar{j}}+\mathfrak{O}\left( \frac{1}{\mathrm{\mu }^{3}}%
\right) $ \\ 
$U_{i}^{M}$ & $\simeq $ & $\frac{1}{\mathrm{\mu }}\mathring{U}_{i}^{M}+%
\mathfrak{O}(\frac{1}{\mathrm{\mu }^{2}})$ \\ 
$\mathcal{P}_{M}^{m}$ & $=$ & $\delta _{M}^{\underline{\tau }}\mathcal{P}_{%
\underline{\tau }}^{a}+\frac{1}{\mathrm{\mu }}\delta _{M}^{\underline{M}}%
\mathcal{P}_{\underline{M}}^{a}$ \\ 
& $=$ & $e^{\frac{\varphi }{M_{pl}^{2}}}\left( \underline{\mathbf{\vartheta }%
}_{M}^{m}+\frac{1}{\mathrm{\mu }}\mathring{\vartheta}_{M}^{m}\right) $%
\end{tabular}%
\end{equation}%
Substituting these $\frac{1}{\mathrm{\mu }}$- developments into%
\begin{equation}
W^{ia}=-i\mathcal{G}^{i\bar{j}}\bar{U}_{\bar{j}}^{M}\mathcal{P}_{M}^{a}
\end{equation}%
with $\mathcal{G}^{i\bar{j}}$ the inverse of $\mathcal{G}_{i\bar{j}}$, we
have%
\begin{equation}
W^{ia}=-i\mathrm{\mu }^{2}\mathcal{G}^{i\bar{j}}\frac{1}{\mathrm{\mu }}\left[
\delta _{\underline{M}}^{M}\overline{\mathring{U}}_{\bar{j}}^{\underline{M}}+%
\mathfrak{O}(\frac{1}{\mathrm{\mu }})\right] \left[ \delta _{M}^{\underline{%
\tau }}\mathcal{P}_{\underline{\tau }}^{a}\oplus \frac{1}{\mathrm{\mu }}%
\delta _{M}^{\underline{N}}\mathcal{P}_{\underline{N}}^{a}\right]
\label{wiaa}
\end{equation}%
By expanding the product on right hand side of above expression in a power
series of the parameter $\mathrm{\mu }$, we obtain the leading contributions 
\begin{equation}
W^{ia}=\mathring{W}^{ia}+\mathfrak{O}(\frac{1}{\mathrm{\mu }})  \label{wia}
\end{equation}%
with%
\begin{equation}
\begin{tabular}{lll}
$\mathring{W}^{ia}$ & $=$ & $-i\mathcal{G}^{i\bar{j}}\overline{\mathring{U}}%
_{\bar{j}}^{\underline{M}}\mathcal{P}_{\underline{M}}^{a}$ \\ 
& $=$ & $-ie^{\frac{\varphi }{M_{pl}^{2}}}\mathcal{\mathring{G}}^{i\bar{k}}%
\overline{\mathring{U}}_{\bar{k}}^{M}\mathring{\vartheta}_{M}^{m}$%
\end{tabular}%
\end{equation}%
where $\underline{\mathbf{\vartheta }}_{M}^{m}$ and $\mathring{\vartheta}%
_{M}^{m}$ scale as mass$^{2}$, in same manner as the Fayet-Iliopoulos
coupling constants. The rigid limit of $W^{ia}$ has no contribution along
the graviphoton direction since the $\mathring{W}^{ia}$'s are proportional
to the $\mathring{\vartheta}_{M}^{m}$'s having components in the Coulomb
branch as given below 
\begin{equation}
\mathring{\vartheta}_{M}^{m}=\left( 
\begin{array}{c}
0 \\ 
\theta _{i}^{m} \\ 
0 \\ 
\tilde{\theta}^{mi}%
\end{array}%
\right) \equiv \mathring{\vartheta}_{\underline{M}}^{m}
\end{equation}%
Here also the $\theta _{i}^{m}$ and $\tilde{\theta}^{mi}$ components of the
embedding tensor scale as mass$^{2}$ and can be interpreted in terms of
electric and magnetic FI coupling constants$.$

\  \  \  \ 

$\bullet $ $\frac{1}{\mathrm{\mu }}$-\emph{\ expansion of the term} $N^{a}$%
\newline
By using the $\frac{1}{\mathrm{\mu }}$- expansion of the covariantly
holomorphic section\emph{\ }(\ref{rd}) namely 
\begin{equation}
V^{M}=V_{grav}^{M}+\frac{1}{\mathrm{\mu }}\mathring{\Omega}^{M}+\mathfrak{O}(%
\frac{1}{\mathrm{\mu }^{2}})
\end{equation}%
and the expression of $\mathcal{P}_{M}^{a}$ in terms of the embedding tensor
(\ref{ts}) 
\begin{equation}
\mathcal{P}_{M}^{a}=e^{\text{\texttt{$\QTR{sc}{\lambda }$}}\varphi }\left( 
\underline{\mathbf{\vartheta }}_{M}^{m}+\frac{1}{\mathrm{\mu }}\mathring{%
\vartheta}_{M}^{m}\right)
\end{equation}%
where we have set\textrm{\footnote{%
\ Notice that in the limit where $M_{pl}$ is thought of very large with
respect to $\Lambda $, $\left( M_{pl}\rightarrow \infty \right) $; then 
\texttt{$\QTR{sc}{\lambda }$}$=\frac{1}{M_{pl}^{2}}\rightarrow 0$ and
therefore $e^{\text{\texttt{$\QTR{sc}{\lambda }$}}\varphi }\mapsto 1$.}} 
\begin{equation*}
\text{\texttt{$\QTR{sc}{\lambda }$}}=\frac{1}{M_{pl}^{2}}
\end{equation*}%
the $\frac{1}{\mathrm{\mu }}$- expansion of the term $N^{a}=i\mathcal{P}%
_{M}^{a}\bar{V}^{M}$ reads as follows%
\begin{equation}
N^{a}=i\left[ e^{\text{\texttt{$\QTR{sc}{\lambda }$}}\varphi }\left( 
\underline{\mathbf{\vartheta }}_{M}^{a}+\frac{1}{\mathrm{\mu }}\mathring{%
\vartheta}_{M}^{a}\right) \right] \left[ \bar{V}_{grav}^{M}+\frac{1}{\mathrm{%
\mu }}\overline{\mathring{\Omega}}^{M}+\mathfrak{O}(\frac{1}{\mathrm{\mu }%
^{2}})\right]
\end{equation}%
with leading $\mathrm{\mu }^{-n}$- terms as%
\begin{equation}
N^{a}=\mathring{N}^{a}+\mathfrak{O}(\frac{1}{\mathrm{\mu }^{2}})
\end{equation}%
In this relation, $\mathring{N}^{a}$ is the rigid limit whose expression in
terms of the components of the embedding tensor reads as $\mathring{N}^{a}=i%
\mathcal{P}_{\underline{\tau }}^{a}\bar{V}_{grav}^{\underline{\tau }};$ by
substituting $\mathcal{P}_{\underline{\tau }}^{a}$ and $\bar{V}_{grav}^{%
\underline{\tau }}$ by their expressions, we also have%
\begin{equation}
\mathring{N}^{a}=-\frac{1}{2}e^{\text{\texttt{$\QTR{sc}{\lambda }$}}\varphi
}\left( \tilde{\theta}^{a0}-2i\theta _{0}^{a}\right)  \label{na}
\end{equation}%
The expansion of $N^{a}$ has a gravity- like contribution in the sense that
it involves only the embedding tensor components $\theta _{0}^{a}$ and $%
\tilde{\theta}^{a0}$ associated with the graviphoton direction. Notice that
this particular dependence only in $\left( \theta _{0}^{a},\tilde{\theta}%
^{a0}\right) $ does not mean that $\mathring{N}^{a}$ is free from $\theta
_{i}^{a},\tilde{\theta}^{ai}$; this is because $\theta _{0}^{a}$ and $\tilde{%
\theta}^{a0}$ are related to $\theta _{i}^{a}$ and $\tilde{\theta}^{ai}$
through the constraint relations (\ref{13}-\ref{ctt}).

\  \  \  \  \ 

$\bullet $ $\frac{1}{\mathrm{\mu }}$-\emph{\ expansion of} $S^{a}$\newline
The expansion of $S^{a}$ has a quite similar behaviour as the $N^{a}$
associated with the transformations of the hyperini (\ref{stt}). By using
eqs(\ref{rd}, \ref{ts}) the leading term $\mathring{S}^{a}$ of the $\frac{1}{%
\mathrm{\mu }}$- expansion of $S^{a}=\frac{i}{2}\mathcal{P}_{M}^{a}V^{M}$
associated with the transformations of the two gravitini is given by $%
\mathring{S}^{a}=\frac{i}{2}\mathcal{P}_{\underline{\tau }}^{a}V_{grav}^{%
\underline{\tau }}$ and reads in terms of the embedding tensor components as
follows 
\begin{equation}
\mathring{S}^{a}=\frac{1}{2}e^{\text{\texttt{$\QTR{sc}{\lambda }$}}\varphi
}\left( \tilde{\theta}^{a0}+2i\theta _{0}^{a}\right)  \label{an}
\end{equation}%
Like $\mathring{N}^{a}$, the rigid $\mathring{S}^{a}$ has a gravity- like
contribution with no manifest contribution from the embedding tensor
components $\theta _{i}^{a}$ and $\tilde{\theta}^{ai}$ associated with the
Coulomb branch dimensions. By using the constraint eqs(\ref{13}-\ref{ctt}),
one can also express $\mathring{S}^{a}$ in terms of $\theta _{i}^{a}$ and $%
\tilde{\theta}^{ai}$; see also the comment given in the conclusion regarding
this matter.

\subsection{Rigid limit of the supergravity scalar potential}

Using eqs (\ref{wia}-\ref{na}), we can expand the induced potential $%
\mathcal{V}_{\text{sugra}}^{\mathcal{N}=2}$ as a power series of the
parameter $\mathrm{\mu }$. The obtained development is as given below%
\begin{equation}
\mathcal{V}_{\text{sugra}}^{\mathcal{N}=2}=\mathcal{\mathring{V}}_{\left(
0\right) }^{\mathcal{N}=2}+\mathfrak{O}\left( \frac{1}{\mathrm{\mu }}\right)
\label{2m}
\end{equation}%
with%
\begin{equation}
\mathcal{\mathring{V}}_{\left( 0\right) }^{\mathcal{N}=2}=\mathcal{\mathring{%
V}}_{\text{kah}}^{\mathcal{N}=2}+\mathcal{\mathring{V}}_{\text{hyper}}^{%
\mathcal{N}=2}-\mathcal{\mathring{V}}_{\text{gra}}^{\mathcal{N}=2}
\label{kh}
\end{equation}%
where%
\begin{equation}
\begin{tabular}{lll}
$\mathcal{\mathring{V}}_{\text{kah}}^{\mathcal{N}=2}$ & $=$ & $\mathcal{%
\mathring{G}}_{i\bar{j}}\mathring{W}^{ai}\overline{\mathring{W}}^{a\bar{j}}$
\\ 
$\mathcal{\mathring{V}}_{\text{hyper}}^{\mathcal{N}=2}$ & $=$ & $2\overline{%
\mathring{N}}^{a}\mathring{N}^{a}$ \\ 
$\mathcal{\mathring{V}}_{\text{gra}}^{\mathcal{N}=2}$ & $=$ & $12\overline{%
\mathring{S}}^{a}\mathring{S}^{a}$%
\end{tabular}
\label{vo}
\end{equation}%
\begin{equation*}
\text{ }
\end{equation*}%
By substituting the expressions of $\mathring{W}^{ai},$ $\mathring{N}^{a}$
and $\mathring{S}^{a}$ back into (\ref{vo}), we first have $\mathcal{%
\mathring{V}}_{\text{kah}}^{\mathcal{N}=2}=\frac{1}{2}\mathcal{\mathring{M}}%
^{\underline{M}\underline{N}}\mathcal{P}_{\underline{N}}^{a}\mathcal{P}_{%
\underline{M}}^{a};$ and by replacing the moment maps by their expressions
in terms of the embedding tensor, we obtain%
\begin{equation}
\mathcal{\mathring{V}}_{\text{kah}}^{\mathcal{N}=2}=\frac{1}{2}\mathcal{%
\mathring{M}}^{\underline{M}\underline{N}}\mathring{\vartheta}_{\underline{N}%
}^{a}\mathring{\vartheta}_{\underline{M}}^{a}  \label{vkr}
\end{equation}%
with $\mathcal{\mathring{M}}^{\underline{M}\underline{N}}$ the inverse of
the matrix $\mathcal{\mathring{M}}_{\underline{M}\underline{N}}$ given by eq(%
\ref{mmn}). Similarly, we have for the extra term $\Delta \mathcal{\mathring{%
V}}=\mathcal{\mathring{V}}_{\text{hyper}}^{\mathcal{N}=2}-\mathcal{\mathring{%
V}}_{\text{gra}}^{\mathcal{N}=2}$ in (\ref{kh})%
\begin{equation}
\Delta \mathcal{\mathring{V}}=2\overline{\mathring{N}}^{a}\mathring{N}^{a}-12%
\overline{\mathring{S}}^{a}\mathring{S}^{a}
\end{equation}%
the following expression%
\begin{equation}
\Delta \mathcal{\mathring{V}}=2\left( -i\mathcal{P}_{\underline{\sigma }%
}^{a}V_{grav}^{\underline{\sigma }}\right) \left( i\mathcal{P}_{\underline{%
\rho }}^{a}\bar{V}_{grav}^{\underline{\rho }}\right) -12\left( -\frac{i}{2}%
\mathcal{P}_{\underline{\rho }}^{a}\bar{V}_{grav}^{\underline{\rho }}\right)
\left( \frac{i}{2}\mathcal{P}_{\underline{\sigma }}^{a}V_{grav}^{\underline{%
\sigma }}\right)
\end{equation}%
By expanding, we have 
\begin{equation}
\Delta \mathcal{\mathring{V}}=-\left( \mathcal{P}_{\underline{\sigma }%
}^{a}V_{grav}^{\underline{\sigma }}\right) \left( \mathcal{P}_{\underline{%
\rho }}^{a}\bar{V}_{grav}^{\underline{\rho }}\right)
\end{equation}%
and by substituting $\mathcal{P}_{\underline{\sigma }}^{a}$ and $V_{grav}^{%
\underline{\sigma }}$ by their values in terms of $\theta _{0}^{a}$ and $%
\tilde{\theta}^{a0}$, we end with%
\begin{equation}
\Delta \mathcal{\mathring{V}}=-\frac{1}{4}\left( \tilde{\theta}^{a0}\tilde{%
\theta}^{a0}+4\theta _{0}^{a}\theta _{0}^{a}\right)  \label{vhid}
\end{equation}%
where we have dropped out the factor $e^{\text{\texttt{$\QTR{sc}{\lambda }$}}%
\varphi }$ which, in the limit $M_{pl}$ very large, reduces to 1. By adding (%
\ref{vkr}) and (\ref{vhid}), we obtain the explicit expression of the rigid
potential (\ref{kh}) in terms of the components of the embedding tensor
namely 
\begin{equation}
\mathcal{\mathring{V}}_{\left( 0\right) }^{\mathcal{N}=2}=\frac{1}{2}%
\mathcal{\mathring{M}}^{\underline{M}\underline{N}}\mathring{\vartheta}_{%
\underline{N}}^{a}\mathring{\vartheta}_{\underline{M}}^{a}-\frac{1}{4}\left( 
\tilde{\theta}^{a0}\tilde{\theta}^{a0}+4\theta _{0}^{a}\theta _{0}^{a}\right)
\label{vrig1}
\end{equation}%
Notice that eq(\ref{vrig1}) can be derived directly from (\ref{sps}) by
taking its rigid limit as follows:%
\begin{equation}
\mathcal{\mathring{V}}_{\left( 0\right) }^{\mathcal{N}=2}=\frac{1}{2}%
\mathcal{\mathring{M}}^{\underline{M}\underline{N}}\mathcal{P}_{\underline{N}%
}^{a}\mathcal{P}_{\underline{M}}^{a}-\frac{1}{2}\mathcal{\mathring{M}}%
_{grav}^{\underline{\rho }\underline{\sigma }}\mathcal{P}_{\underline{\rho }%
}^{a}\mathcal{P}_{\underline{\sigma }}^{a}-2V^{\underline{\rho }}\bar{V}^{%
\underline{\sigma }}\mathcal{P}_{\underline{\rho }}^{a}\mathcal{P}_{%
\underline{\sigma }}^{a}  \label{vrig2}
\end{equation}%
where $\mathcal{\mathring{M}}_{grav}^{\underline{\rho }\underline{\sigma }}$
as in (\ref{mgra}). From eq(\ref{grad}), we have%
\begin{equation}
-\frac{1}{2}\mathcal{\mathring{M}}_{grav}^{\underline{\rho }\underline{%
\sigma }}\mathcal{P}_{\underline{\rho }}^{a}\mathcal{P}_{\underline{\sigma }%
}^{a}=\left( \frac{i}{2}\mathcal{C}^{\underline{\rho }\underline{\sigma }}+%
\bar{V}_{grav}^{\underline{\rho }}V_{grav}^{\underline{\sigma }}\right) 
\mathcal{P}_{\underline{\rho }}^{a}\mathcal{P}_{\underline{\sigma }}^{a}
\end{equation}%
which reduces to%
\begin{equation}
-\frac{1}{2}\mathcal{\mathring{M}}_{grav}^{\underline{\rho }\underline{%
\sigma }}\mathcal{P}_{\underline{\rho }}^{a}\mathcal{P}_{\underline{\sigma }%
}^{a}=\left( \mathcal{P}_{\underline{\rho }}^{a}\bar{V}_{grav}^{\underline{%
\rho }}\right) \left( \mathcal{P}_{\underline{\sigma }}^{a}V_{grav}^{%
\underline{\sigma }}\right)
\end{equation}%
Putting back into (\ref{vrig2}), we obtain%
\begin{equation}
\mathcal{\mathring{V}}_{\left( 0\right) }^{\mathcal{N}=2}=\frac{1}{2}%
\mathcal{\mathring{M}}^{\underline{M}\underline{N}}\mathcal{P}_{\underline{N}%
}^{a}\mathcal{P}_{\underline{M}}^{a}-V^{\underline{\rho }}\bar{V}^{%
\underline{\sigma }}\mathcal{P}_{\underline{\rho }}^{a}\mathcal{P}_{%
\underline{\sigma }}^{a}
\end{equation}%
which coincides with eq(\ref{vrig1}).

\end{document}